\documentclass[longauth]{aa} 

\usepackage{silence}
\usepackage{graphicx}
\usepackage{lipsum}

\usepackage{txfonts}
\usepackage{natbib}
\usepackage{hyperref}
\usepackage{longtable}
\usepackage{supertabular}
\usepackage{placeins}

\usepackage[normalem]{ulem}

\usepackage{caption}
\usepackage{subcaption}

\usepackage{lscape}

\usepackage[utf8]{inputenc}
\usepackage{booktabs}
\usepackage{hyperref}
\usepackage{multirow}
\usepackage{enumitem}

\usepackage{lscape}

\usepackage[switch]{lineno}

\usepackage{lineno}

\usepackage[dvipsnames]{xcolor}
\definecolor{myblue}{HTML}{1F77B4}
\definecolor{mygreen}{HTML}{2CA02C}
\definecolor{m}{HTML}{D62728}
\definecolor{mymagenta}{HTML}{D33682}
\definecolor{codepurple}{HTML}{C42043}

\newcommand{\program}[1]{\textsc{#1}}

\usepackage{xspace}
\newcommand{\xga}{SN~2020xga}
\newcommand{\xgc}{SN~2022xgc}

\hypersetup{
unicode=true,           
colorlinks=true,        
citecolor=myblue,       
urlcolor=black        
}
\begin{document}

   \title{Eruptive mass loss less than a year before the explosion of superluminous supernovae}

   \subtitle{I. The cases of SN\,2020xga and SN\,2022xgc}

   \author{A.~Gkini\inst{1} \href{https://orcid.org/0009-0000-9383-2305}{\includegraphics[scale=0.5]{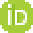}}\and
   C.~Fransson\inst{1} \href{https://orcid.org/0000-0001-8532-3594}{\includegraphics[scale=0.5]{Images/ORCIDiD_icon16x16.eps}}\and
   R.~Lunnan\inst{1} \href{https://orcid.org/0000-0001-9454-4639}{\includegraphics[scale=0.5]{Images/ORCIDiD_icon16x16.eps}}\and
    S.~Schulze\inst{2} \href{https://orcid.org/0000-0001-6797-1889}{\includegraphics[scale=0.5]{Images/ORCIDiD_icon16x16.eps}} \and
   F.~Poidevin\inst{3,4}\and \href{https://orcid.org/0000-0002-5391-5568}{\includegraphics[scale=0.5]{Images/ORCIDiD_icon16x16.eps}}
   N.~Sarin\inst{5,6} \href{https://orcid.org/0000-0003-2700-1030}{\includegraphics[scale=0.5]{Images/ORCIDiD_icon16x16.eps}} \and
   R.~K\"onyves-T\'oth\inst{7,8} \href{https://orcid.org/0000-0002-8770-6764}{\includegraphics[scale=0.5]{Images/ORCIDiD_icon16x16.eps}}\and
   J.~Sollerman\inst{1} \href{https://orcid.org/0000-0003-1546-6615}{\includegraphics[scale=0.5]{Images/ORCIDiD_icon16x16.eps}}\and 
   C.~M.~B.~Omand\inst{9} \href{https://orcid.org/0000-0002-9646-8710}{\includegraphics[scale=0.5]{Images/ORCIDiD_icon16x16.eps}} \and
   S. J.~Brennan \inst{1} \href{https://orcid.org/0000-0003-1325-6235}{\includegraphics[scale=0.5]{Images/ORCIDiD_icon16x16.eps}}\and
   K.~R.~Hinds \inst{9} \href{https://orcid.org/0000-0002-0129-806X}{\includegraphics[scale=0.5]{Images/ORCIDiD_icon16x16.eps}} \and
   J.~P.~Anderson\inst{10,11} \href{https://orcid.org/0000-0003-0227-3451}{\includegraphics[scale=0.5]{Images/ORCIDiD_icon16x16.eps}}\and
   M.~Bronikowski\inst{12}  \href{https://orcid.org/0000-0002-1537-6911}{\includegraphics[scale=0.5]{Images/ORCIDiD_icon16x16.eps}}\and
   T.-W.~Chen\inst{13} \href{https://orcid.org/0000-0002-1066-6098}{\includegraphics[scale=0.5]{Images/ORCIDiD_icon16x16.eps}} \and
   R.~Dekany  \inst{14} \href{https://orcid.org/[0000-0002-5884-7867}{\includegraphics[scale=0.5]{Images/ORCIDiD_icon16x16.eps}}\and
   M.~Fraser \inst{15}  \href{https://orcid.org/0000-0003-2191-1674}{\includegraphics[scale=0.5]{Images/ORCIDiD_icon16x16.eps}} \and
   C.~Fremling \inst{14,16} \href{https://orcid.org/[0000-0002-4223-103X}{\includegraphics[scale=0.5]{Images/ORCIDiD_icon16x16.eps}}\and
   L.~Galbany \inst{17,18} \href{https://orcid.org/0000-0002-1296-6887}{\includegraphics[scale=0.5]{Images/ORCIDiD_icon16x16.eps}}\and
   A.~Gal-Yam \inst{19} \href{https://orcid.org/0000-0002-3653-5598}{\includegraphics[scale=0.5]{Images/ORCIDiD_icon16x16.eps}}\and
   A.~Gangopadhyay \inst{1} \href{https://orcid.org/0000-0002-3884-5637}{\includegraphics[scale=0.5]{Images/ORCIDiD_icon16x16.eps}}\and
   S.~Geier \inst{3,20} \and
   E.~P.~Gonzalez \inst{21,22} \and
   M.~Gromadzki\inst{23} \href{https://orcid.org/0000-0002-1650-1518}{\includegraphics[scale=0.5]{Images/ORCIDiD_icon16x16.eps}}\and
   S.~L.~Groom \inst{24} \href{https://orcid.org/0000-0001-5668-3507}{\includegraphics[scale=0.5]{Images/ORCIDiD_icon16x16.eps}}\and
   C.~P.~Guti\'errez \inst{18,17} \href{https://orcid.org/0000-0003-2375-2064}{\includegraphics[scale=0.5]{Images/ORCIDiD_icon16x16.eps}}\and
   D.~Hiramatsu \inst{25,26} \href{https://orcid.org/0000-0002-1125-9187}{\includegraphics[scale=0.5]{Images/ORCIDiD_icon16x16.eps}}\and
   D.~A.~Howell   \inst{21,22}  \href{https://orcid.org/0000-0003-4253-656X}{\includegraphics[scale=0.5]{Images/ORCIDiD_icon16x16.eps}}\and
   Y.~Hu \inst{1} \and
   C.~Inserra\inst{27} \href{https://orcid.org/0000-0002-3968-4409}{\includegraphics[scale=0.5]{Images/ORCIDiD_icon16x16.eps}}\and
   M.~Kopsacheili \inst{17,18} \href{https://orcid.org/0000-0002-3563-819X}{\includegraphics[scale=0.5]{Images/ORCIDiD_icon16x16.eps}}\and
   L.~Lacroix \inst{28,6} \href{https://orcid.org/0000-0003-0629-5746}{\includegraphics[scale=0.5]{Images/ORCIDiD_icon16x16.eps}}\and
   F.~J.~Masci \inst{24} \href{https://orcid.org/0000-0002-8532-9395}{\includegraphics[scale=0.5]{Images/ORCIDiD_icon16x16.eps}}\and
   K.~Matilainen\inst{29} \href{https://orcid.org/0000-0002-8111-4581}{\includegraphics[scale=0.5]{Images/ORCIDiD_icon16x16.eps}}\and
   C.~McCully   \inst{21} \href{https://orcid.org/0000-0001-5807-7893}{\includegraphics[scale=0.5]{Images/ORCIDiD_icon16x16.eps}}  \and
   T.~Moore \inst{10,30} \href{https://orcid.org/ 0000-0001-8385-372}{\includegraphics[scale=0.5]{Images/ORCIDiD_icon16x16.eps}}\and
   T.~E.~M\"uller-Bravo\inst{17,18} \href{https://orcid.org/0000-0003-3939-7167}{\includegraphics[scale=0.5]{Images/ORCIDiD_icon16x16.eps}}\and
   M.~Nicholl\inst{30} \href{https://orcid.org/0000-0002-2555-3192}{\includegraphics[scale=0.5]{Images/ORCIDiD_icon16x16.eps}}\and
   C.~Pellegrino \inst{21,22} \href{https://orcid.org/0000-0002-7472-1279}{\includegraphics[scale=0.5]{Images/ORCIDiD_icon16x16.eps}} \and
   I.~P\'erez-Fournon \inst{3,4} \href{https://orcid.org/0000-0002-2807-6459}{\includegraphics[scale=0.5]{Images/ORCIDiD_icon16x16.eps}}\and 
   D.~A.~Perley \inst{9} \href{https://orcid.org/0000-0001-8472-1996}{\includegraphics[scale=0.5]{Images/ORCIDiD_icon16x16.eps}}\and 
   P.~J.~Pessi\inst{1} \href{https://orcid.org/0000-0002-8041-8559}{\includegraphics[scale=0.5]{Images/ORCIDiD_icon16x16.eps}}\and 
   T.~Petrushevska\inst{12} \href{https://orcid.org/0000-0003-4743-1679}{\includegraphics[scale=0.5]{Images/ORCIDiD_icon16x16.eps}}\and 
   G.~Pignata\inst{31} \href{https://orcid.org/0000-0003-0006-0188}{\includegraphics[scale=0.5]{Images/ORCIDiD_icon16x16.eps}}\and
   F.~Ragosta\inst{32,33}  \href{https://orcid.org/0000-0003-2132-3610}{\includegraphics[scale=0.5]{Images/ORCIDiD_icon16x16.eps}}\and
   A.~Sahu \inst{34} \href{https://orcid.org/0009-0007-6825-3230}{\includegraphics[scale=0.5]{Images/ORCIDiD_icon16x16.eps}}\and
   A.~Singh \inst{1} \href{https://orcid.org/0000-0003-2091-622X}{\includegraphics[scale=0.5]{Images/ORCIDiD_icon16x16.eps}}\and
   S.~Srivastav \inst{35} \href{https://orcid.org/0000-0003-4524-6883}{\includegraphics[scale=0.5]{Images/ORCIDiD_icon16x16.eps}}\and
   J.~L.~Wise \inst{9} \href{https://orcid.org/0000-0003-0733-2916}{\includegraphics[scale=0.5]{Images/ORCIDiD_icon16x16.eps}}\and
   L.~Yan \inst{14} \href{https://orcid.org/0000-0003-1710-9339}{\includegraphics[scale=0.5]{Images/ORCIDiD_icon16x16.eps}}\and
   D.~R.~Young \inst{30} \href{https://orcid.org/0000–0002–1229–2499}{\includegraphics[scale=0.5]{Images/ORCIDiD_icon16x16.eps}}
}

   \institute{The Oskar Klein Centre, Department of Astronomy, Stockholm University, Albanova University Center, 106 91 Stockholm, Sweden
   \and
   Center for Interdisciplinary Exploration and Research in Astrophysics (CIERA), Northwestern University, 1800 Sherman Ave, Evanston, IL 60201, USA
   \and
   Instituto de Astrof\'{\i}sica de Canarias, V\'{\i}a   L\'actea, 38205 La Laguna, Tenerife, Spain
   \and
   Universidad de La Laguna, Departamento de Astrof\'{\i}sica,  38206 La Laguna, Tenerife, Spain
  \and
    The Oskar Klein Centre, Department of Physics, Stockholm University, Albanova University Center, SE-106 91 Stockholm, Sweden
  \and
    Nordita,Stockholm University and KTH Royal Institute of Technology, Hannes Alfv\'ens v\"ag 12, 106 91, Stockholm, Sweden
  \and
  Konkoly Observatory, Research Center for Astronomy and Earth Sciences, H-1121 Budapest Konkoly Th. M. út 15-17., Hungary; MTA Centre of Excellence 
  \and
  Department of Experimental Physics, Institute of Physics, University of Szeged, D\'om t\'er 9, Szeged, 6720 Hungary
   \and
  Astrophysics Research Institute, Liverpool John Moores University, Liverpool Science Park, 146 Brownlow Hill, Liverpool L3 5RF, UK
 \and
  European Southern Observatory, Alonso de C\'ordova 3107, Casilla 19, Santiago, Chile
  \and
  Millennium Institute of Astrophysics MAS, Nuncio Monsenor Sotero Sanz 100, Off.104, Providencia, Santiago, Chile
  \and
  Center for Astrophysics and Cosmology, University of Nova Gorica, Vipavska 11c, 5270 Ajdov\v{s}\v{c}ina, Slovenia 
  \and
  Graduate Institute of Astronomy, National Central University, 300 Jhongda Road, 32001 Jhongli, Taiwan
  \and
  Caltech Optical Observatories, California Institute of Technology, Pasadena, CA 91125, USA 
  \and
  School of Physics, University College Dublin, LMI Main Building, Beech Hill Road, Dublin 4, D04 P7W1
  \and
  Division of Physics, Mathematics and Astronomy, California Institute of Technology, Pasadena, CA 91125, USA
  \and
  Institute of Space Sciences (ICE, CSIC), Campus UAB, Carrer de Can Magrans, s/n, E-08193 Barcelona, Spain 
  \and
  Institut d'Estudis Espacials de Catalunya (IEEC), 08860  
  Castelldefels (Barcelona), Spain 
  \and
  Department of Particle Physics and Astrophysics, Weizmann Institute of Science, 234 Herzl St, 76100 Rehovot, Israel 
    \and
  GRANTECAN, Cuesta de San Jos\'{e} s/n, 38712 Bre{\~n}a Baja, La Palma, Spain
    \and
  Las Cumbres Observatory, 6740 Cortona Dr. Suite 102, Goleta, CA, 93117, USA 
  \and
  Department of Physics, University of California, Santa Barbara, CA 93106-9530, USA 
  \and
  Astronomical Observatory, University of Warsaw, Al. Ujazdowskie 4,00-478 Warszawa, Poland
  \and
  IPAC, California Institute of Technology, 1200 E. California Blvd, Pasadena, CA 91125, USA 
  \and
  Center for Astrophysics, Harvard \& Smithsonian, 60 Garden Street, Cambridge, MA 02138-1516, USA 
  \and
  The NSF AI Institute for Artificial Intelligence and Fundamental Interactions, USA 
  \and
  Cardiff Hub for Astrophysics Research and Technology, School of Physics \& Astronomy, Cardiff University, Queens Buildings, The Parade, Cardiff, CF24 3AA, UK
  \and
  LPNHE, CNRS/IN2P3, Sorbonne Université, Université Paris-Cité, Laboratoire de Physique Nucléaire et de Hautes Énergies, 75005 Paris, France 
   \and
  Department of Physics and Astronomy, University of Turku, 20014 Turku, Finland
    \and
  Astrophysics Research Centre, School of Mathematics and Physics, Queens University Belfast, Belfast BT7 1NN, UK
   \and
  Instituto de Alta Investigación, Universidad de Tarapacá, Casilla 7D, Arica, Chile
  \and
  Dipartimento di Fisica “Ettore Pancini”, Università di Napoli Federico II, Via Cinthia 9, 80126 Naples, Italy
  \and
  INAF - Osservatorio Astronomico di Capodimonte, Via Moiariello 16, I-80131 Naples, Italy
  \and
  Department of Physics, University of Warwick, Gibbet Hill Road, Coventry CV4 7AL, UK 
  \and
  Astrophysics sub-Department, Department of Physics, University of Oxford, Keble Road, Oxford, OX1 3RH, UK
  }

   \date{}

\abstract
{We present photometric and spectroscopic observations of \xga\ and \xgc, two hydrogen-poor superluminous supernovae (SLSNe-I) at $z =  0.4296$ and $z = 0.3103$, respectively, which show an additional set of broad \ion{Mg}{II} absorption lines, blueshifted by a few thousands kilometer ~second$^{-1}$ with respect to the host galaxy absorption system. Previous work interpreted this as due to resonance line scattering of the SLSN continuum by rapidly expanding circumstellar material (CSM) expelled shortly before the explosion. The peak rest-frame $g$-band magnitude of \xga\ is $-22.30 \pm 0.04$~mag and of \xgc\ is $-21.97 \pm 0.05$~mag, placing them among the brightest SLSNe-I. We used high-quality spectra from ultraviolet to near-infrared wavelengths to model the \ion{Mg}{II} line profiles and infer the properties of the CSM shells. We find that the CSM shell of \xga\ resides at $\sim 1.3 \times 10^{16}~\rm cm$, moving with a maximum velocity of $4275~\rm km~s^{-1}$, and the shell of \xgc\ is located at $\sim 0.8 \times 10^{16}~\rm cm$, reaching up to $4400~\rm km~s^{-1}$. These shells were expelled $\sim 11$ and $\sim 5$~months before the explosions of \xga\ and \xgc, respectively, possibly as a result of luminous-blue-variable-like eruptions or pulsational pair instability (PPI) mass loss. We also analyzed optical photometric data and modeled the light curves, considering powering from the magnetar spin-down mechanism. The results support very energetic magnetars, approaching the mass-shedding limit, powering these SNe with ejecta masses of $\sim 7-9~\rm M_\odot$. The ejecta masses inferred from the magnetar modeling are not consistent with the PPI scenario pointing toward stars $> 50~\rm M_\odot$ He-core; hence, alternative scenarios such as fallback accretion and CSM interaction are discussed. Modeling the spectral energy distribution of the host galaxy of \xga\ reveals a host mass of 10$^{7.8}$~$\rm M_\odot$, a star formation rate of $0.96^{+0.47}_{-0.26}$~$\rm M_\odot$~yr$^{-1}$, and a metallicity of $\sim$ 0.2~$\rm Z_\odot$.} 


\authorrunning{Gkini et al.}
\titlerunning{Eruptive mass loss less than a year before the explosion of superluminous supernovae}

   \maketitle
%

\section{Introduction}
Superluminous supernovae (SLSNe; \citealt{Quimby2011,GalYam2012}) constitute a rare class of massive star explosions \citep{Perley2020} that reach absolute magnitudes between $-20$ and $-23$~mag at peak \citep{DeCia2018,Lunnan2018b,Chen2023a}. Today, more than 200\footnote{reported in Transient Name Server} SLSNe have been detected out to $z=2$ \citep{Angus2019}. They are frequently found in low-metallicity dwarf host galaxies with high specific star formation rates (SFRs) \citep{Neill2011,Chen2013,Lunnan2014,Leloudas2015,Angus2016,Perley2016,Chen2017b,Schulze2018,Taggart2021}.

There are two types of SLSNe, which are differentiated by the presence or absence of hydrogen in their spectra \citep{GalYam2012}: hydrogen-poor (type I; SLSNe-I hereafter) and hydrogen-rich (type II; SLSNe-II hereafter). The early spectra of the majority of SLSNe-I could show a prominent blue continuum and a series of \ion{O}{II} features at $3500 - 5000$~\AA\,, with the feature at $4350 - 4650$~\AA\ being the most dominant \citep{Quimby2011,Mazzali2016,Quimby2018}.  Studies \citep{Mazzali2016,Dessart2019,Konyves2022,Saito2024} have shown that the \ion{O}{II} features require either nonthermal excitation and/or temperatures higher than 12\,000 -- 14\,000~K; but in a few SLSNe-I \citep{Nicholl2014,Gutierrez2022,Schulze2024} this feature has not been detected in their spectra. 

The high luminosities observed in SLSNe-I cannot be explained by the amount of radioactive $^{56}$\ion{Ni}{} generated in the normal core-collapse process, which is the major power source of type I SNe, and thus alternative scenarios have been proposed. A popular scenario, which could potentially explain the majority of the observed properties in SLSNe-I \citep[e.g.,][]{Inserra2013,Nicholl2017,Liu2017,Blanchard2020,Hsu2021,Chen2023b}, is the spin-down of a newly formed rapidly rotating highly magnetized neutron star (NS) known as a magnetar \citep{Ostriker1971,Kasen2010,Woosley2010,Vurm2021}. Other proposed scenarios are the long-term fallback accretion of material onto a black hole \citep{Dexter2013,Moriya2018},
the thermonuclear explosion of $140$ -- $260$~$\rm M_\odot$ zero-age main sequence (ZAMS) metal-poor stars referred to as pair-instability supernovae (PISNe; \citealt{Barkat1967,Rakavy1967,Woosley2002,Heger2002}) and interaction of the SN ejecta with circumstellar material (CSM) formed by material previously expelled from the star \citep{Chatzopoulos2012,Sorokina2016,Wheeler2017,Chen2023b}.

The fate of the stars, their powering mechanism, and the type of the resulting explosion are closely related to the final years of their stellar lives before the core collapse. During their lifetime, stars can lose a substantial part of their initial mass due to stellar winds \citep[e.g.,][]{Lucy1970,Lamers1999,Puls2008}, binary interactions \citep[e.g.,][]{Petrovic2005,Smith2014,Gotberg2017,Yoon2017,Petrovic2020,Laplace2020}, or eruptive mass loss \citep[e.g.,][]{Heger2002,Woosley2007,Quataert2012,Shiode2014,Smith2014b,Smith2014,Woosley2017,Fuller2018,Leung2019,Renzo2020,Leung2021}. Mass loss in the form of violent outbursts becomes critical in the late stages of stellar evolution and, in extreme situations, can remove tens of solar masses. Such eruptive mass loss has been observed in $\eta$ Carinae  \citep{Westphal1969} and is thought to come from a group of post-main-sequence stars called luminous blue variables (LBVs; \citealt{Humphreys1999}). 

Eruptive mass loss can also be achieved in the case of pulsational pair instability (PPI; \citealt{Woosley2007,Woosley2017,Leung2019}) in which the formation of positron–electron pairs in the CO core of a star with mass as low as $40~\rm M_\odot$ (if metallicity and rotationally induced mixing is taken into account; \citealt{Chatzopoulos2012c,Chatzopoulos2012b}) ZAMS results in explosive O-burning, and the energy released drives a series of mass ejections. The more massive the star is, the more energetic the pulses are, and thus the more mass will be ejected in the pulses \citep{Renzo2020}. \cite{Woosley2017} and \cite{Renzo2020} note that the time interval between the mass ejection and the core collapse in PPI could be between a few hours to 10\,000 years, which, along with the ejection velocity, could determine the distance of the ejected material.

Eruptive mass loss, and especially PPI, can generate CSM shell(s) around the progenitor stars, which potentially can be seen in the spectra of the SNe. There are a few SLSNe-I in the literature with evidence of late-time mass loss, such as them showing late-time broad \ion{H}{} emission \citep{Yan2015,Yan2017,Fiore2021,Pursiainen2022,Gkini2024} or early forbidden emission of [\ion{O}{II}] and [\ion{O}{III}] \citep{Lunnan2016,Inserra2017,Aamer2024,Schulze2024} in their spectra. The former has been explained by interaction of the ejecta with H-rich CSM located at $\sim 10^{15} - 10^{16}$~cm \citep[e.g.,][]{Yan2015}, and the latter by the interaction with low-density matter moving at a few $10^{3}$~km~s$^{-1}$. However, recently, two SLSNe-I, iPTF16eh \citep{Lunnan2018} and SN\,2018ibb \citep{Schulze2024}, were discovered that show a unique spectroscopic feature, a second \ion{Mg}{II} absorption system blueshifted by $\sim 3000$~km~s$^{-1}$ with respect to the \ion{Mg}{II} absorption lines originating in the interstellar medium of the host galaxy. This feature has been associated with the photoionization of a rapidly expanding CSM shell expelled decades before the explosion. In the case of iPTF16eh, \cite{Lunnan2018} also detected a \ion{Mg}{II} emission line that moved from $-1600~\rm km~s^{-1}$ to $2900~\rm km~s^{-1}$ between $100$ and $300$ days after maximum light, and this was attributed to a light echo from that shell. The CSM was located at $\sim 10^{17}$~cm and matched with theoretical predictions of shell ejections due to PPI. However, the detections of these shells were both serendipitous, and so it is not known whether these properties are typical, or how common this phenomenon is.

We present results from a dedicated study using the X-shooter spectrograph \citep{Vernet2011a} on the ESO Very Large Telescope (VLT) in Paranal, Chile to search for a second \ion{Mg}{II} absorption system. The full sample will be presented in a follow-up paper; here, we focus on the analysis of the two detections found in the X-shooter sample indicating the presence of a fast-moving CSM. An extensive dataset for \xga\ and \xgc\ enable us to extract the CSM shell properties and give insights into the late stages of the stellar evolution.

This paper is structured as follows. In Sect.~\ref{sec:obs_data}, we present photometric and spectroscopic data for \xga\ and \xgc\ along with photometric measurements of their host galaxies, and imaging polarimetry data for \xgc. In Sect.~\ref{sec:light_curve}, we analyze the light-curve properties of \xga\ and \xgc, derive their blackbody temperatures and radii, construct bolometric light curves, and compare them with a homogeneous sample of SLSNe-I as well as with the photometric properties of SN\,2018ibb and iPTF16eh. We also model the light curves of \xga\ and \xgc\ under the assumption that they are powered by a magnetar. In Sect.~\ref{sec:spectroscopy}, we present the spectroscopic sequences of \xga\ and \xgc, analyze the spectral properties of these two objects, and compare them with those of well-studied SLSNe-I, and with SN\,2018ibb and iPTF16eh. The modeling of the \ion{Mg}{II} lines to extract information about the CSM shell is done in Sect.~\ref{sec:mgii}. In Sect.~\ref{sec:host_galaxy}, we discuss the properties of the two host galaxies. We discuss our findings and provide possible mass loss scenarios and alternative powering mechanisms in Sect.~\ref{sec:discussion}, and we summarize our results in Sect.~\ref{sec:conclusions}. 

Throughout the paper, the photometric measurements are reported in the AB system and the uncertainties are provided with 1$\sigma$ confidence. We assume a flat Lambda cold dark matter cosmology with $H_{0} = 67.4$~km~s$^{-1}$~Mpc$^{-1}$, $\Omega_{m} = 0.31$, and $\Omega_{\Lambda} = 0.69$ \citep{Planck2020}.

\section{Observations} \label{sec:obs_data}

\subsection{Our X-shooter sample}

Motivated by the discovery of iPTF16eh and SN\,2018ibb, we collected a sample of 19 SLSNe with the medium-resolution X-shooter spectrograph ($\rm program~IDs$: $\rm 105.20PN$, $\rm 106.21L3$, $108.2262$ and $\rm 110.247C$). The triggering criteria of the program were objects that have been already classified as SLSNe-I, were observable from Paranal and have $z > 0.11$ so that the \ion{Mg}{II} $\lambda\lambda 2796, 2803$ resonance lines are observable with X-shooter. Our primary objectives are to constrain the occurrence of such mass ejections in SLSNe-I and determine the distribution of the CSM properties. This paper focuses on the analysis of two detections in the X-shooter sample, \xga\ and \xgc, which exhibit a second narrow \ion{Mg}{II} absorption system in their X-shooter spectra blueshifted by a few thousand km~s$^{-1}$ with respect to the \ion{Mg}{II} absorption lines originating in the interstellar medium of the host galaxy.

\subsection{Discovery and classification}

\subsubsection{SN\,2020xga}

\begin{figure}[!ht]
     \centering
     \begin{subfigure}[b]{0.5\textwidth} 
          \begin{subfigure}[b]{0.48\textwidth}
         \centering
         \includegraphics[width=\textwidth]{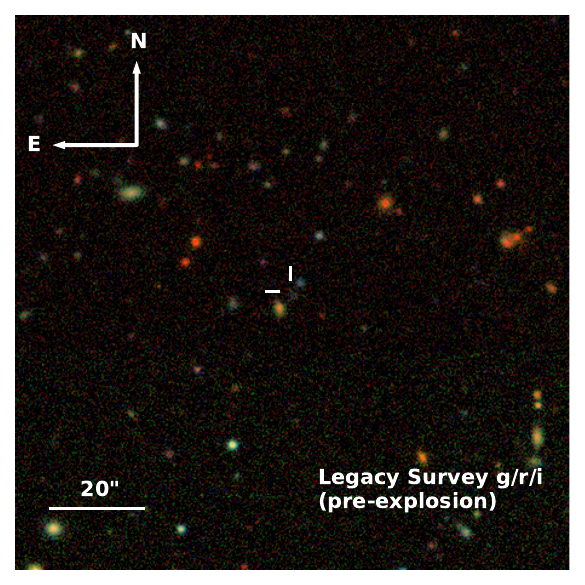}
     \end{subfigure}
     \begin{subfigure}[b]{0.48\textwidth}
         \centering
         \includegraphics[width=\textwidth]{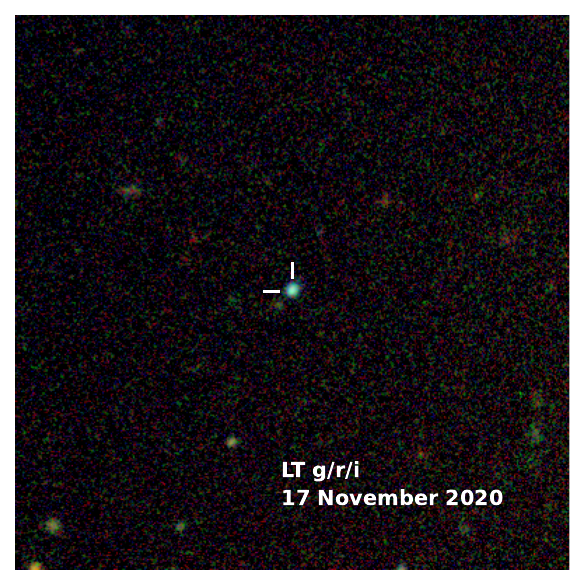}
    \end{subfigure}
    \caption{}
    \label{fig:image_xga}
    \end{subfigure}
     \begin{subfigure}[b]{0.5\textwidth} 
     \begin{subfigure}[b]{0.48\textwidth}
         \centering
         \includegraphics[width=\textwidth]{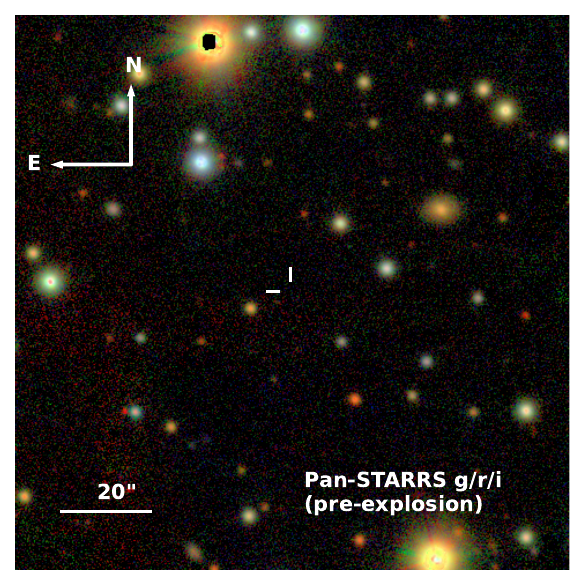}
     \end{subfigure}
     \begin{subfigure}[b]{0.48\textwidth}
         \centering
         \includegraphics[width=\textwidth]{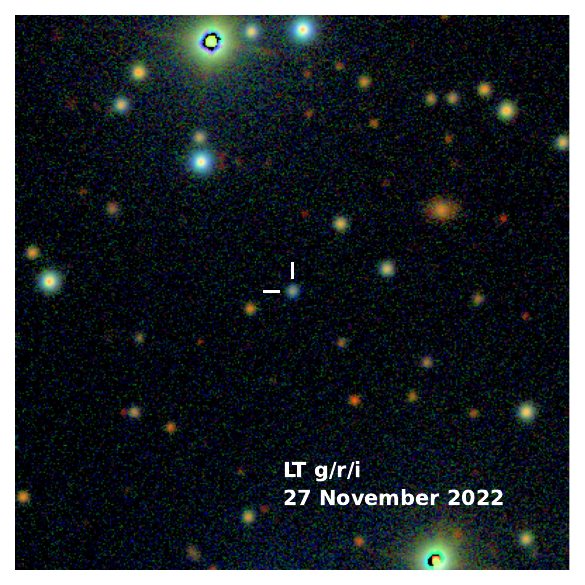}
    \end{subfigure}
    \caption{}
    \label{fig:image_xgc}
    \end{subfigure}

    \caption{Images of the fields of \xga\ (a) and \xgc\ (b). Panel a: Legacy Survey DR10 image of the field of \xga\ before explosion. A faint host galaxy at the SN position is visible, marked by the white crosshairs. Right: $gri$ composite image of the SN near peak from Liverpool Telescope (LT).  Panel b: Pan-STARRS image of the field of \xgc\ before explosion. The SN position is marked by the white crosshairs. Right: $gri$ composite image of the SN near peak from the LT. All images have a size of $2 \times 2$ arcminutes and have been combined following the algorithm in \citet{Lupton2004}. }

    \label{fig:images}
\end{figure}

\xga\ was discovered by the Panoramic Survey Telescope and Rapid Response System (Pan-STARRS1; \citealt{Kaiser2010a}) as PS20jxm on the rise on October 4, 2020, at a $w$-band magnitude of 19.8~mag at right ascension, declination (J2000.0) 03$^{h}$46$^{m}$39.37$^{s}$, $-11^{\circ}14'33.90''$ \citep{Chambers2020a}. It was classified by the extended Public ESO Spectroscopic Survey for Transient Objects (ePESSTO+; \citealt{Smartt2015}) as a SLSN-I on November 6, 2020 \citep{Gromadzki2020a,Ihanec2020a}. An image of the field before and after the explosion is shown in Fig.~\ref{fig:image_xga}.

Spectroscopic follow-up showed a redshift of $z = 0.4296$ (see Sect.~\ref{sec:redshift}) corresponding to a distance modulus of $41.93$~mag. We corrected for the Milky Way (MW) extinction using the dust extinction model of \citet{Fitzpatric1999} based on $R_{V} = 3.1$ and $E(B-V) = 0.049$~mag \citep{Schlafly2011}. As for
the host galaxy extinction, we find that the host properties of \xga\ are consistent with no extinction within the uncertainties (see Sect.~\ref{sec:host_galaxy}). The estimated epoch of maximum light in the rest-frame $g$ band is  November 19, 2020, MJD = $59\,172.5$ (see Sect.~\ref{sec:lc_properties}).

\subsubsection{SN\,2022xgc}

\xgc\ was discovered by the Zwicky Transient Facility (ZTF; \citealt{Bellm2019,Graham2019,Dekany2020}) on October 9, 2022, as ZTF22abkbmob at a $g$-band magnitude of 20.8~mag at right ascension, declination (J2000.0) 07$^{h}$12$^{m}$41.81$^{s}$, $+07^{\circ}18'59.95''$ \citep{Fremling2022a}. ePESSTO+ classified it as a SLSN-I on December 2, 2022 \citep{Gromadzki2022a,Poidevin2022a,Grzesiak2022a}. 

To correct for the MW extinction, we again used the dust extinction model of \citet{Fitzpatric1999}, $R_{V} = 3.1$  and now with  $E(B-V) = 0.061$~mag \citep{Schlafly2011}. We adopted a spectroscopic redshift of $z = 0.3103$ (see Sect.~\ref{sec:redshift}) and computed the distance modulus to be $41.11$~mag. Since the host of \xgc\ is not detected in the photometric catalogs, we did not apply any host extinction. The epoch of maximum light in the rest-frame $g$ band is estimated to be November 18, 2022, MJD = $59\,901.9$ (see Sect.~\ref{sec:lc_properties}). An image of the field before and after the explosion is shown in Fig.~\ref{fig:image_xgc}.

\subsection{Photometry}

Photometric measurements of \xga\ and \xgc\ are available from sky surveys such as the Asteroid Terrestrial-impact Last Alert System (ATLAS; \citealt{Tonry2020a}), and the ZTF survey. We retrieved forced photometry from the ATLAS forced photometry server\footnote{\url{https://fallingstar-data.com/forcedphot/}} \citep{Tonry2018a,Smith2020a,Shingles2021a} for both $c$ and $o$ filters. The clipping and binning, with a bin size of 1 day, of the ATLAS data were done using the \texttt{plot\_atlas\_fp.py}\footnote{\url{https://gist.github.com/thespacedoctor/86777fa5a9567b7939e8d84fd8cf6a76}} python script \citep{Young_plot_atlas_fp}. We removed the measurements with < 3$\sigma$ significance and converted the resulting fluxes to the AB magnitude system using the 3631 Jy zeropoint. The ZTF forced point spread function (PSF)-fit photometry was requested from the Infrared Processing and Analysis Center \citep{Masci2019a} for the $gri$ bands. To obtain the rest-frame light curve, we followed the ZTF data processing procedure\footnote{\url{https://irsa.ipac.caltech.edu/data/ZTF/docs/ZTF_zfps_userguide.pdf}} including baseline correction, validation of the flux uncertainties, combining measurements obtained the same night and converting the differential fluxes to the AB magnitude system. Similarly to ATLAS data, a quality cut of 3$\sigma$ was performed to the data.

In addition, both objects were monitored with the 2m Liverpool Telescope (LT; \citealt{Steele2004a}) using the IO:O imager at the Roque de los Muchachos Observatory in the $griz$ bands. The images were retrieved from the LT data archive\footnote{\url{https://telescope.livjm.ac.uk/cgi-bin/lt_search}} and were processed through a PSF photometry script developed by Hinds and Taggart et al. (in prep). Each measurement was calibrated using stars from the Pan-STARRS \citep{Flewelling2020a} catalog and a cut of 3$\sigma$ was performed.

\xga\ was also monitored between November 2020 and July 2021 by ePESSTO+ using the Las Cumbres Observatory  in the $griz$ bands. We performed photometry using the AUTOmated Photometry of Transients\footnote{\url{https://github.com/Astro-Sean/autophot}} pipeline developed by \cite{Brennan2022a}. The instrumental magnitude of the SN is measured through PSF fitting and the zero point in each image is calibrated with stars from the Pan-STARRS \citep{Flewelling2020a} catalog. We do not discuss the $z$-band photometry because of the poor quality of these images.

The photometric dataset of \xgc\ is complemented with four epochs obtained with the Rainbow Camera at the Spectral Energy Distribution Machine (SEDM; \citealt{Blagorodnova2018a,Rigault2019,Kim2022}) at Palomar Observatory in the $gri$ bands. The data were reduced with the \texttt{FPipe} pipeline described in \cite{Fleming2016}. Finally, one epoch of photometry in the $gr$ bands was obtained with the Alhambra Faint Object Spectrograph and Camera (ALFOSC) at the 2.56m Nordic Optical Telescope (NOT). For the reduction the \texttt{PyNOT}\footnote{\url{https://github.com/jkrogager/PyNOT}} data processing pipeline was utilized. For nights with multiple exposures, we computed the weighted average and we kept only the data with > 3$\sigma$ significance. 

Overall, for \xga\ we obtained 68 epochs of photometry spanning from $-44$ and $+59$ days post maximum in the $gcroiz$ bands with a cadence of 1.5 days in the best-covered $r$ band. For \xgc\ we obtained 86 photometric epochs between $-59$ and $+110$ days after the peak in the $gcroiz$ bands with an average cadence of 2 days in the $g$ and $r$ bands, which were best covered (for the photometric tables of \xga\ and \xgc\ see Sect.~\ref{data}.)

\subsection{Spectroscopy}

We acquired five low-resolution spectra of \xga\ between November 6, 2020, and December 30, 2020, and six low-resolution spectra of \xgc\ between December 1, 2022, and February 12, 2023, with the ESO Faint Object Spectrograph and Camera 2 \citep[EFOSC2;][]{Buzzoni1984a} on the 3.58m ESO New Technology Telescope (NTT) at the La Silla Observatory in Chile under the ePESSTO+ program \citep{Smartt2015}. Additional medium-resolution spectra were obtained for both \xga\ between November 2020 and January 2021, and \xgc\ between December 2022 and March 2023, with the X-shooter spectrograph.

The spectroscopic data for \xgc\ was supplemented with three low-resolution spectra obtained with ALFOSC between November and December 2021, one spectrum obtained on November 22, 2022, with the Kast double spectrograph mounted on the Shane 3m telescope at Lick Observatory, and two spectra taken in November 2022 with the SEDM. We acquired one additional epoch of spectroscopy for \xga\ with the DouBle-SPectrograph (DBSP; \citealt{Oke1982}) mounted on Palomar 200-inch telescope on Palomar Observatory on January 07, 2021. Observations using the SEDM and DBSP were coordinated using the FRITZ data platform \citep{vanderWalt2019,Coughlin2023}. 

The NTT spectra were reduced with the \texttt{PESSTO}\footnote{\url{https://github.com/svalenti/pessto}} pipeline. The observations were performed with grisms \#11, \#13, and \#16 using a 1\farcs0 wide slit. The integration times varied between $1500$ and $5400$~s for \xga\ and between $900$ and $4800$~s for \xgc. The spectra of \xga\ on November 16 and 17, 2020, were combined to boost the signal-to-noise (S/N).

The X-shooter observations were performed for the ultra-violet (UV), visible (VIS), and near-infrared (NIR) arms in nodding mode using 1\farcs0, 0\farcs9, 0\farcs9 wide slits, respectively, and were reduced using the ESO X-shooter pipeline. We followed the following procedure; first, the tool \texttt{astroscrappy}\footnote{\url{https://github.com/astropy/astroscrappy}} was used for the removal of cosmic-rays based on the algorithm of \citet{vanDokkum2001a}, then the data were processed with the X-shooter pipeline v3.6.3 and the ESO workflow engine ESOReflex \citep{Goldoni2006a, Modigliani2010}. The UV and VIS-arm data were reduced in stare mode. The corrected two-dimensional spectra were co-added utilizing tools developed by \citet{Selsing2019a}\footnote{\url{https://github.com/jselsing/XSGRB_reduction_scripts}}. To achieve proper skyline subtraction, the NIR-arm data were processed in nodding mode. The wavelength calibration of all spectra was adjusted to account for barycentric motion. The spectra of the separate arms were combined by averaging the overlap areas. Since observations of SLSNe-I have shown that the spectra tend to evolve slower compared to other SNe \citep[e.g.,][]{Quimby2018}, we stitched the X-shooter spectra of \xga\ on January 10 and 14, 2021, to increase the S/N.

The spectroscopic data obtained with ALFOSC were reduced using the  \program{PypeIt}\footnote{\url{https://pypeit.readthedocs.io/en/release/}} pipeline \citep{Prochaska2020a,Prochaska2020b}. The observations were obtained with a 1\farcs3 wide slit and grism \#4, and the exposure times were between $3344$~s and $4000$~s. The spectrum on November 13, 2022, was observed under cloudy conditions and thus we do not consider it. The SEDM observations had an integration time of $2250$~s and were reduced using the pipeline described in \cite{Rigault2019}. The first SEDM spectrum of \xgc\ obtained on November 14, 2022, is of insufficient quality and is not presented in the paper. The epoch observed with the DBSP instrument was taken using the D-55 dichroic beam splitter, a blue grating with 600 lines per mm blazed at 4000~\AA, a red grating with 316 lines per mm blazed at 7500~\AA, and a 1\farcs5 wide slit. The data were reduced using the python package DBSP\_DRP4\footnote{\url{https://github.com/finagle29/dbsp_drp}} that is primarily based on \program{PypeIt}. Finally, the Kast observations utilized the 2\farcs0 wide slit, the 600/4310 grism, and the 300/7500 grating. The Kast data were reduced following standard techniques for CCD processing and spectrum extraction \citep{Silverman2012} utilizing IRAF routines and custom Python and IDL codes\footnote{\url{https://github.com/ishivvers/TheKastShiv}}. 
 
Each spectrum was flux calibrated against standard stars. The spectral logs for \xga\ and \xgc\ are presented in Table~\ref{tab:2020xga_spectra} and Table~\ref{tab:2022xgc_spectra}, respectively. 

\subsection{Polarimetry} \label{pol_data}

Linear polarimetry was obtained on \xgc\ at two epochs after maximum light at $+26.1$ (MJD 59928.0) days and at $+60.1$ (MJD 59962.0) days, observer-frame. 
A log of the observations is given in Table~\ref{tab:log_polarimetry}. The polarimetry was obtained using a half wave plate in the FAPOL unit and a calcite plate mounted in the aperture wheel of the ALFOSC instrument on the NOT. The calcite plate provides the simultaneous measurement of the ordinary and the extraordinary components of two orthogonal polarized beams. The half wave plate can be rotated over 16 angle positions in steps of $22.5^{\circ}$ from 0$^{\circ}$ to $337.5^{\circ}$.  As a standard, we used 4 angle positions ($0^{\circ},
22.5^{\circ}, 45^{\circ}$, and $67.5^{\circ}$) to sample the linear Stokes $Q-U$ parameters space. 

The pipeline used to reduce the data is the same as the one introduced in \citet{Poidevin2022a}. The photometry of the ordinary and extraordinary beams was done using aperture photometry of size $\sim$ 2 to 3 times the Full-Width at Half-Maximum (FWHM) of punctual sources in the images. For multiple sequences of 4 Half-Wave Plate angles the polarization was obtained by summing-up the fluxes from the ordinary and extra-ordinary beams to minimize the propagation of the uncertainties.
The instrumental polarization (IP) was first estimated using the unpolarized star HD\,14069. The IP degree is of order 0.1$\%$ in the $R$-band, and or order $0.2\%$ in the $V$-band (see Table~\ref{tab:pol_results}). These averaged Stokes $\overline{Q}$ and $\overline{U}$ values were subsequently removed from the Stokes parameters $Q-U$ estimates of the polarized calibration stars HD\,251204 and BD$+$59\,389 and of \xgc. The polarized stars were used to calculate the zero polarization angle (ZPA) used to rotate the Stokes $Q,U$ parameters from the ALFOSC FAPOL instrument reference frame to the sky reference frame in equatorial coordinates. The polarization angles are counted positively from north to east. When applicable, the polarization degree and polarization angle obtained at each of these steps are reported in Table~\ref{tab:pol_results}.

\subsection{Host galaxy observations}

We retrieved science-ready co-added images from the DESI Legacy Imaging Surveys \citep[LS;][]{Dey2018a} Data Release (DR) 10, and archival science-ready images obtained with MegaCAM at the 3.58\,m Canada-France-Hawaii Telescope (CFHT) for \xga. We measured the brightness with the aperture photometry tool presented in \citet{Schulze2018} using an aperture similar to the other images. The photometry was calibrated against stars from the Sloan Digital Sky Survey DR9 \citep{Ahn2012a} and Pan-STARRS1 \citep{Chambers2016a}. The host galaxy of \xgc\ is not detected in any catalog and thus, we provide the upper limits of the Dark Energy Survey images obtained with the Dark Energy Camera (DECam) at the Cerro Tololo Inter-American Observatory (CTIO). Table \ref{tab:phot:host} summarizes the measurements in the different bands.

\begin{table}[!ht]
\caption{Photometry of the host galaxies of \xga\ and \xgc.\label{tab:phot:host}}
\centering
\begin{tabular}{ccc}
\toprule
Survey or                  & Filter & Brightness \\
Telescope/Instrument       &        & (mag)      \\
\midrule
\multicolumn{3}{c}{SN 2020xga}\\
\midrule
LS             &$g $&$  23.56 \pm 0.09 $\\
LS             &$r $&$  23.26 \pm 0.12 $\\
CFHT/MegaCAM   &$i $&$  22.96 \pm 0.09 $\\
LS             &$z $&$  22.75 \pm 0.15 $\\
\midrule
\multicolumn{3}{c}{SN 2022xgc}\\
\midrule
CTIO/DECam &$ g $&$ >23.6$\\
CTIO/DECam & $r$ & $>23.4$\\
CTIO/DECam & $i$ & $>23.0$\\
CTIO/DECam & $z$ & $>22.7$\\
CTIO/DECam & $y$ & $>21.3$\\
\bottomrule
\end{tabular}
\tablefoot{All measurements are reported in the AB system and not corrected for reddening. Non-detections are reported with $3\sigma$ confidence.
}
\end{table}

\section{Photometry} \label{sec:light_curve}

\subsection{General light-curve properties} \label{sec:lc_properties}

\begin{figure*}[!ht]
     \centering
     \begin{subfigure}[b]{0.9\textwidth}
         \centering
         \includegraphics[width=\textwidth]{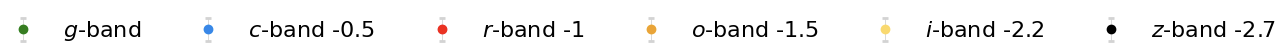}
     \end{subfigure}
          \begin{subfigure}[b]{0.49\textwidth}
         \centering
         \includegraphics[width=\textwidth]{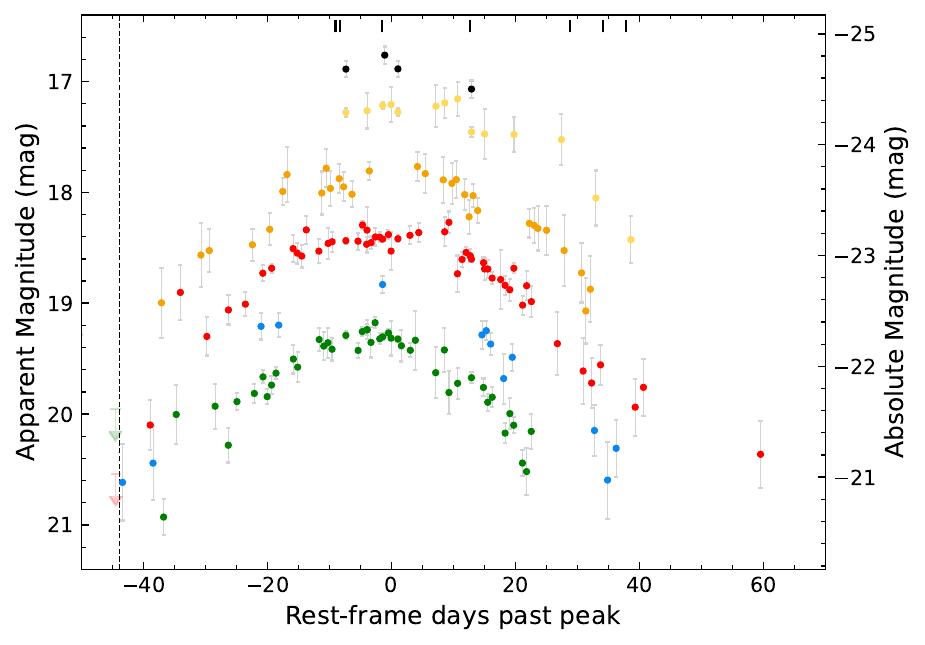}
     \end{subfigure}
     \begin{subfigure}[b]{0.49\textwidth}
         \centering
         \includegraphics[width=\textwidth]{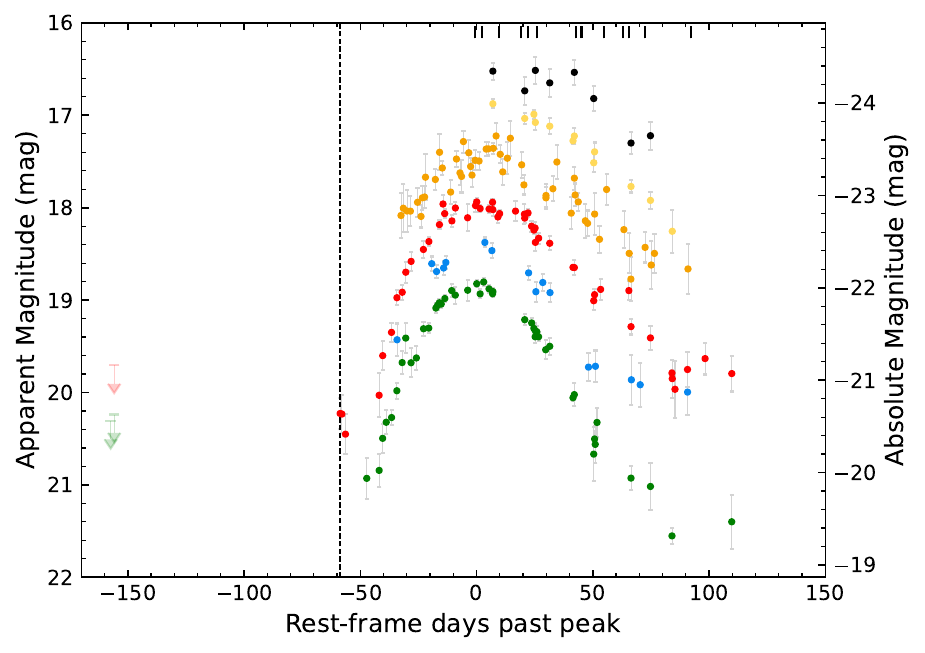}
     \end{subfigure}

    \caption{Optical light curves of \xga\ (left panel) and \xgc\ (right panel). The magnitudes are corrected for MW extinction and K-correction. Upper limits are presented as downward-pointing triangles in a lighter shade. The epochs of the spectra are marked as thick lines at the top of the figure. The dashed line represents the estimated time of the first light. The x axis is in rest-frame days with respect to the rest-frame $g$-band maximum.}
    \label{fig:LCs}
\end{figure*}

To estimate the absolute magnitudes of \xga\ and \xgc, we used $ M = m - \mu  - A_{\rm MW} - K_{\rm corr}$, where $m$ is the apparent magnitude, $\mu$ is the distance modulus, $A_{\rm MW}$ is the extinction caused by the MW, and the last term is the K-correction. For the last term, we used the expression $-2.5\log(1+z)$, which we found to be consistent within 0.1~mag with the full K-correction using the spectra near peak, as is suggested in \citet{Chen2023a} as well. This gave $K_{\rm corr}= -0.39 \pm 0.1$~mag for \xga\ and $K_{\rm corr}= -0.29 \pm 0.1$~mag for \xgc. The multiband light curves in apparent and absolute magnitude systems for \xga\ and \xgc\ are shown in Fig.~\ref{fig:LCs}.

To estimate the time of first light, we fit a baseline to the non-detection data points and a second-order polynomial to the rising part of the light curve, with the cross-point of the two fits being the time of first light. To estimate the uncertainty in the first-light epoch, we ran a Monte Carlo algorithm of randomly selected data points from a Gaussian distribution of the 1$\sigma$ uncertainties for each of the selected flux measurements. For \xga, the resulting dates are MJD $59109.8 \pm 0.2$ in the $g$ band and MJD $59110.5 \pm 1.0$ in the $r$ band. We adopted a weighted average of MJD $59109.8 \pm 0.2$, which is also before the first $c$-band detection (MJD $59110.5$). The uncertainty is statistical only, but the systematic error is likely a few days. This is shown for \xgc, where this method results in the dates of MJD $59836.5 \pm 0.4$ in the $g$ band and MJD $59842.2 \pm  2.6$ in the $r$ band. The weighted mean of MJD $59836.6 \pm 2.5$ is after the first three $r$-band detections, which could be associated with a pre-peak bump, as is seen in the light curves of some SLSNe-I \citep[e.g.,][]{Leloudas2012a,Nicholl2015b,Smith2016a,Vreeswijk2017,Angus2019}. These pre-bumps have been discussed in the context of shock breakout into a CSM \citep{Piro2015, Nicholl2015b,Smith2016a,Vreeswijk2017} or a shock generated by a central magnetar at early times \citep{Kasen2017}. Given that only three data points are shown in decline, we cannot conclusively favor one scenario over the other. However, since \xgc\ does not have stringent upper limits due to solar conjunction and the first $r$-band detections are real, we cannot exclude them and we instead took as time of the first light the first $r$-band data point MJD $59825$.

To estimate the light-curve properties, we used the $r$-band light curve, which falls into the rest-frame $g$ band at the redshifts of \xga\ and \xgc\ \citep{Chen2023a}. We used the method from \cite{Angus2019} for the light-curve interpolation and fit a Gaussian process (GP) regression, utilizing the \texttt{PYTHON} package \texttt{GEORGE} \citep{Ambikasaran2015} with a Matern 3/2 kernel. We used the interpolated $r$-band light curve to estimate the peak magnitude as well as to define the rise and decline timescales as a fraction of the maximum flux (e.g., $t_{1/2,\rm rise}$ is the time interval between $f_{\rm peak}/2$ and $f_{\rm peak}$) following \cite{Chen2023a}. To estimate the rest-frame $g-r$ color at the peak, we used the peak magnitudes inferred from the interpolated rest-frame $g$- and $r$-band light curves, K-corrected using the spectra closer to the peak. The photometric properties of \xga\ and \xgc\ obtained from this analysis are listed in Table~\ref{tab:lc_properties}. The timescales are reported in rest-frame days.

\begin{table}[!ht]
\centering
\caption{Light curve properties of \xga\ and SN\,2022xgc.} \label{tab:lc_properties}
\begin{tabular}{ccc}

\hline
\hline
Property & SN\,2020xga & SN\,2022xgc\\
\hline
$M_{\rm r,peak}$ (mag) & $-22.30 \pm 0.04$ & $-21.97 \pm 0.05$\\
Peak time (MJD) & $59172.5_{-11.7}^{+9.4}$        & $59901.9_{-12.0}^{+15.1}$\\
$t_{1/2,\rm rise}$ (day) & $26.2_{-8.3}^{+6.9}$       & $28.2_{-11.6}^{+9.2}$ \\
$t_{1/2,\rm decline}$ (day) &$23.2_{-6.7}^{+8.3}$        & $42.8_{-9.6}^{+11.9}$\\
$t_{1/e,\rm rise}$ (day) &  $32.4_{-8.4}^{+6.9}$      & $33.7_{-9.2}^{+11.6}$\\
$t_{1/e,\rm decline}$ (day) &$29.6_{-6.8}^{+8.4}$        & $58.6_{-12}^{+9.8}$\\
$(g-r)_{\rm peak}$ (mag) & $-0.23 \pm 0.05$        & $-0.24 \pm 0.05$\\

\hline
\end{tabular}
\end{table}

\begin{figure}[!ht]
   \centering
        \centering
     \begin{subfigure}[b]{0.45\textwidth}
         \centering
         \includegraphics[width=\textwidth]{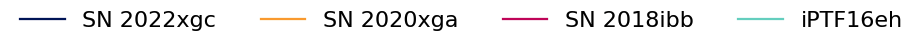}
     \end{subfigure}
          \begin{subfigure}[b]{0.5\textwidth}
         \centering
         \includegraphics[width=\textwidth]{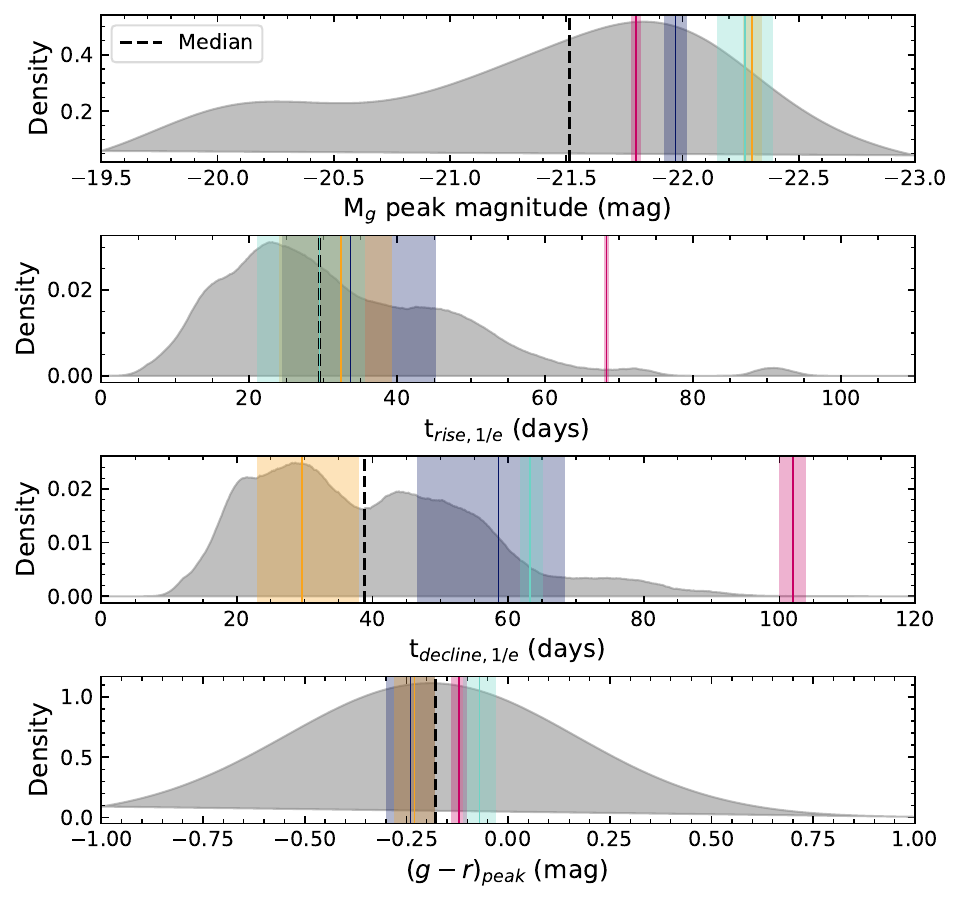}
     \end{subfigure}

   \caption{Comparison of the photometric properties of \xga, \xgc, SN\,2018ibb and iPTF16eh with the ZTF SLSN-I sample \citep{Chen2023a}. Top: KDE distribution of the M$_{g}$ peak magnitudes for 78 ZTF SLSNe-I. Second: KDE plot of the $e$-folding rise time for 69 ZTF SLSNe-I. Third: $e$-folding decline time distribution for 54 ZTF SLSNe-I. Bottom: Rest-frame peak $g-r$ color distribution for 39 ZTF SLSNe-I. The vertical colored lines along with the errors (shaded regions) illustrate the positions of \xga, \xgc, SN\,2018ibb and iPTF16eh and the vertical black lines the median values of the ZTF sample.}
      \label{fig:phot_comp}%
\end{figure}

In Fig.~\ref{fig:phot_comp}, we place the light curve properties of \xga\ and \xgc\ in the context of the homogeneous ZTF SLSN-I sample from \cite{Chen2023a}, which studied the
photometric properties of 78 H-poor SLSNe-I. In the four different panels, we show the kernel density estimates (KDEs) of the ZTF sample, which are an outcome of a Monte Carlo simulation accounting for the asymmetric errors. Both \xga\ and \xgc\ are placed in the bright side of the distribution while the decline times span across the whole distribution. The rise times and the $g-r$ peak magnitudes are rather average compared to the median values of the ZTF sample.

\subsection{Bolometric light curve} \label{sec:blackbody}

\begin{figure}[!ht]
   \centering
       \centering
     \begin{subfigure}[b]{0.45\textwidth}
         \centering
         \includegraphics[width=\textwidth]{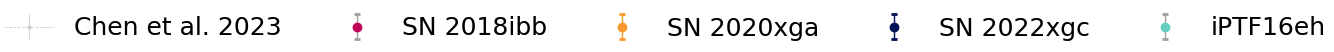}
        
     \end{subfigure}
          \begin{subfigure}[b]{0.5\textwidth}
         \centering
         \includegraphics[width=\textwidth]{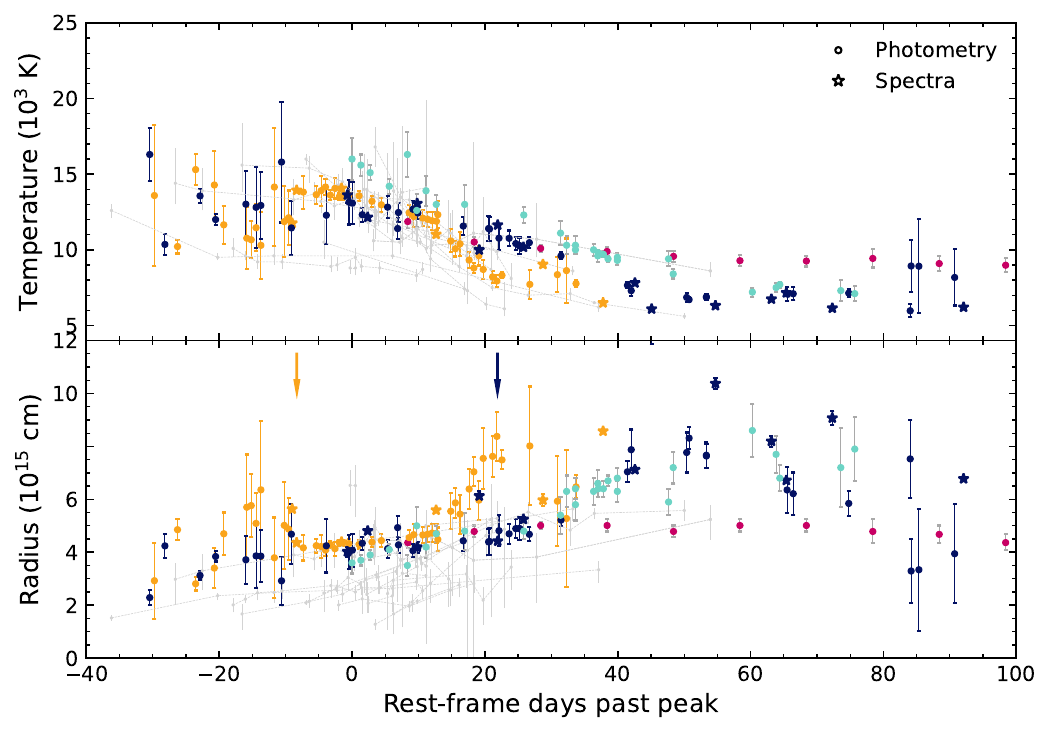}
     \end{subfigure}
   \caption{Blackbody temperatures and radii of \xga, \xgc, SN\,2018ibb and iPTF16eh. Top: Temperature evolution of \xga, \xgc, SN\,2018ibb and iPTF16eh derived from the blackbody fits to the photometric (circle symbols) and the spectroscopic (star symbols) data. The gray background points represent the temperature evolution of the ZTF sample \citep{Chen2023a}. Bottom: Blackbody radius evolution of \xga, \xgc, SN\,2018ibb and iPTF16eh utilizing photometry and spectra in comparison with the ZTF sample (gray). The arrows indicate the epochs where the X-shooter spectra of \xga\ and \xgc\ show the second \ion{Mg}{II} absorption system.}
              \label{fig:temperature}%
    \end{figure}

To construct the bolometric light curves of \xga\ and \xgc, and derive the blackbody temperatures and radii, we constructed the spectral energy distributions (SEDs); however, we note that the presence of nonthermal sources affecting the SN spectra could introduce additional errors that are not accounted for in this analysis. We interpolated all light curves using the GP method described in Sect.~\ref{sec:lc_properties} to match the epochs with $r$-band observations and converted the magnitudes to spectral luminosities $L_{\rm \lambda}$ for each band at each epoch. To extract information about the photospheric temperature and radius, we fit blackbody curves to each SED. The derived temperature and radius evolution for \xga\ and \xgc\ are plotted in Fig.~\ref{fig:temperature}. We compared only with events of the ZTF sample characterized as normal by \cite{Chen2023a}. We find that the temperatures of both objects are comparable to those of the ZTF SLSNe-I and evolve similarly to the ZTF sample, declining over time. On the other hand, while the radius evolution of \xga\ and \xgc\ follows the rising trend seen for the SLSNe-I in the ZTF sample, the photospheric radius of both objects expands to larger values than for the rest of the ZTF sample. This is why these objects are so luminous while their temperatures are typical. In \xga\ and SN\,2020xgc the photospheric radius decreases after $25$ and $50$ days, respectively. We caution that the quoted error bars are statistical only, and do not include any systematic effects; for example, from the fact that we are fitting to optical data only, while the peak of the blackbody is in the UV at early times \citep[e.g.,][]{Arcavi2022}.

\begin{figure}[!ht]
   \centering
   \includegraphics[width=0.5\textwidth]{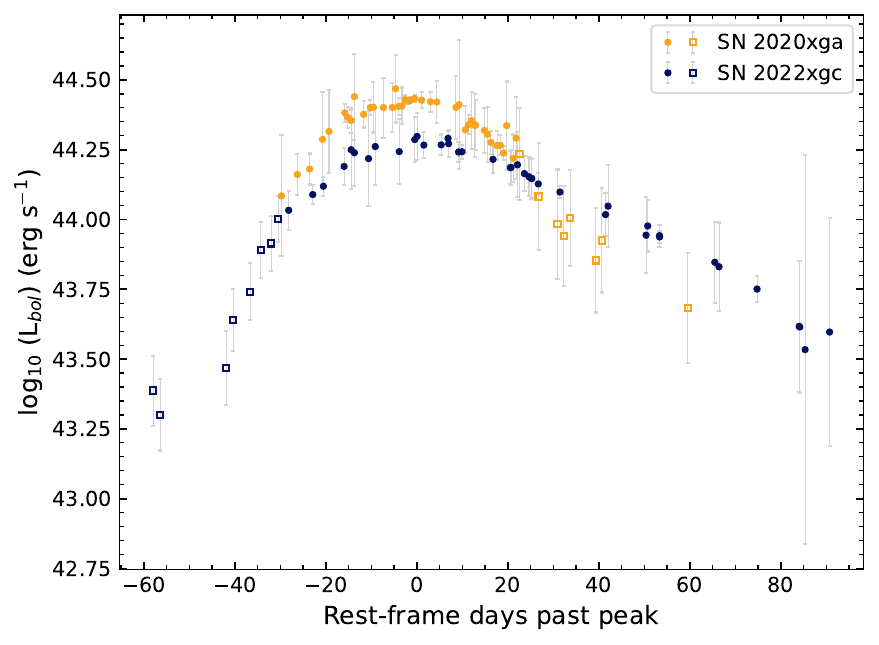}
   \caption{Bolometric light curves of \xga\ and \xgc\ with bolometric corrections applied. The circles correspond to the derived luminosities using $gro$ filters and the open square symbols illustrate the bolometric luminosity assuming the same bolometric correction as the nearest epochs with multiband data. The error bars represent statistical errors.}
              \label{fig:bolometric}%
    \end{figure}

To construct the bolometric light curves of \xga\ and \xgc, we started by integrating the SED using only the $gro$ filters, since the $i$ and $z$ light curves cover only a few epochs and the $c$ filter is already covered by the $g$ and $r$ bands. To account for the missing flux in the NIR we fit a blackbody to the $gro$ SED and integrated the blackbody tail up to 24\,400~\AA, beyond which the contribution to the bolometric light curve is negligible in the photospheric phase \citep[$\sim$ 1\%;][]{Ergon2013}. For the UV correction, we followed the approach of \cite{Lyman2014a} to capture the effect of the line blanketing commonly encountered in SLSNe \cite[e.g.,][]{Yan2017b}. To do this, we linearly extrapolated the SED from the observed $g$ band to 2000~\AA\ where the luminosity is set to zero. The total bolometric luminosity is the sum of the observed $gro$ luminosity and the UV and NIR corrections. The bolometric light curves of \xga\ and \xgc\ are shown in Fig.~\ref{fig:bolometric}. The peak bolometric luminosity is estimated to be $ L_{\rm bol,peak} \gtrsim 2.7 \pm 0.1 \times 10^{44}$~erg~s$^{-1}$ for \xga\ and $ L_{\rm bol,peak} \gtrsim 1.9 \pm 0.1 \times 10^{44}$~erg~s$^{-1}$  for \xgc. These values are typical for SLSNe-I being close to the median value of $2.00^{+1.97}_{-1.44} \times 10^{44}$~erg~s$^{-1}$ reported for 76 SLSNe-I in \cite{Chen2023a} using the $g$ and $r$ filters and the value of $2.00^{+1.98}_{-1.36} \times 10^{44}$~erg~s$^{-1}$ found in \cite{Gomez2024} studying a heterogeneous sample of 262 SLSNe-I.

To include the epochs for which we do not have complete $gro$ data and for which we therefore cannot construct the SED, we assumed a constant bolometric correction. For \xga, for the later epochs that only have $r$-band data available, we used the same ratio of the $r$-band flux to the total flux that we measured at the latest epoch with multiband data. Similarly, for the rising part of the light curve for \xgc\ for which we have only $g$- and $r$-band measurements, we applied the same bolometric correction measured in the first multiband epoch. The bolometric luminosity used is the average of the luminosities calculated for the $g$ and $r$ bands. The data points assuming bolometric corrections are shown as open squares in Fig.~\ref{fig:bolometric}. We note that this approach has two main caveats since at early phases the bolometric correction progressively underestimates the UV contribution while in the later phases the IR contribution gets more significant. By integrating the area below the bolometric light curves we estimated the total radiated energy to be $E_{\rm rad} \gtrsim 1.8 \pm 0.1 \times 10^{51}~{\rm erg}$ for \xga\ and $E_{\rm rad} \gtrsim 0.9 \pm 0.2 \times 10^{51}~{\rm erg}$ for \xgc. We note that these errors only account for the statistical errors in the fit and not for any systematic errors. These values are consistent within the uncertainties with the median $E_{\rm rad} = 1.3^{+1.2}_{-0.9} \times 10^{51}~\rm erg$ found in \cite{Gomez2024}.

\subsection{Photometric comparison to iPTF16eh and SN\,2018ibb}

In Fig.~\ref{fig:phot_comp}, we compare the light-curve properties of \xga\ and \xgc\ with the well-studied sample of SLSNe-I from the ZTF \citep{Chen2023a}. In this plot, we also include the photometric properties of SN\,2018ibb \citep{Schulze2024} and iPTF16eh \citep{Lunnan2018}. These two SLSNe-I, at $z = 0.166$ and $z = 0.427$, respectively, are the only other SLSNe-I in which the two \ion{Mg}{II} absorption-line system is detected in their spectra. To determine whether the objects with this remarkable spectroscopic similarity stand apart in photometry space, we also plot the rest-frame $g$-band peak magnitudes, rise and decline timescales, and $g-r$ magnitudes at the peak of SN\,2018ibb and iPTF16eh in Fig.~\ref{fig:phot_comp}. We note that since there are no data available in the rest-frame $g$ band in the rising part of iPTF16eh's light curve, we used the rest-frame $u$ band to estimate the rise time.

Similarly to \xga\ and \xgc, SN\,2018ibb and iPTF16eh are placed on the bright side of the ZTF luminosity distribution. The peak absolute magnitudes of these four objects span from $-21.8$~mag to $-22.3$~mag with a mean of $-22.1$~mag putting these objects among the most luminous SLSNe-I compared to the ZTF sample. The decline times of these four objects span across the whole distribution, while the $g-r$ colors at the peak are close to the median value of the ZTF sample. The rise times of iPTF16eh, \xga\ and \xgc\ are consistent within the errors with the median of the sample, whereas SN\,2018ibb is placed in the far slow end of the distribution as one of the longest rising SLSNe-I compared to the \cite{Chen2023a} sample.

The blackbody temperature and radius evolution of iPTF16eh and SN\,2018ibb are plotted in Fig.~\ref{fig:temperature}. Both objects follow the temperature evolution of the ZTF sample, with iPTF16eh having higher temperatures compared to the bulk of the population. The photospheric radius of SN\,2018ibb remains constant for almost 100 days after maximum light, while the radius evolution of iPTF16eh is increasing with time. However, similarly to \xga\ and \xgc\, the size of the photoshere in iPTF16eh is getting larger than that of most ZTF SLSNe-I. The photospheric radius of iPTF16eh appears to decline after $\sim$ 60 days.

\begin{figure}[!ht]
   \centering
  \includegraphics[width=0.5\textwidth]{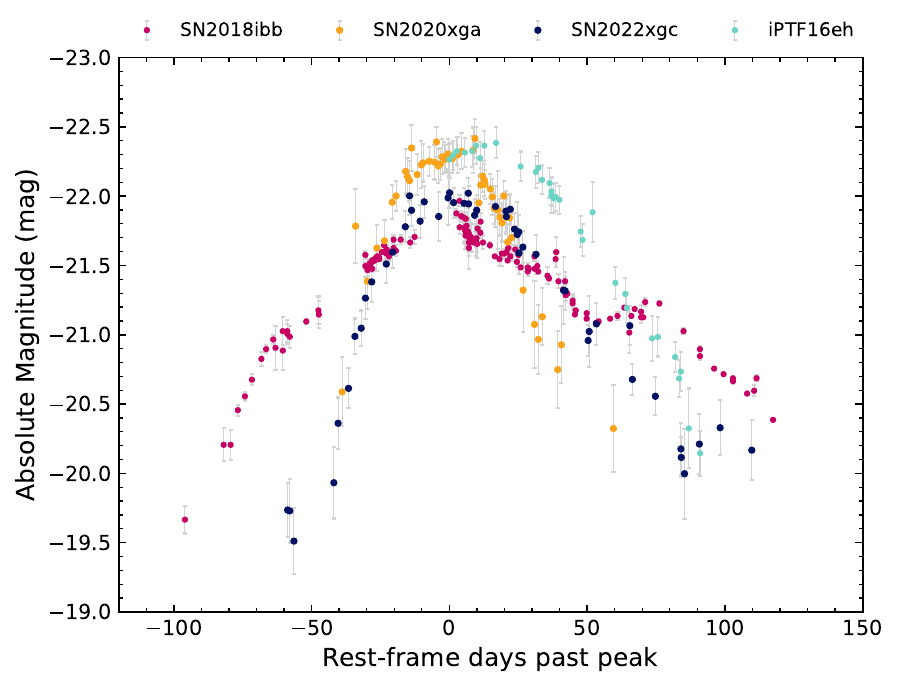}
   \caption{Rest-frame $g$-band absolute magnitude  light curves of \xga\ and \xgc\ in comparison with iPTF16eh and SN\,2018ibb. The magnitudes are K-corrected and corrected for MW extinction. 
   The x axis is in rest-frame days with respect to the $g$-band peak, with the exception of iPTF16eh, where the $u$ band was utilized for estimating the peak owing to the lack of data in the rising part of the $g$-band light curve.}
      \label{fig:phot_csm_comp}%
\end{figure}

To better illustrate the variety in the photometric properties of these four objects, we plot the rest-frame $g$-band light curves in Fig.~\ref{fig:phot_csm_comp}. The absolute magnitudes of all the objects are K-corrected and corrected for MW extinction. We see that the light curve of SN\,2018ibb differs significantly compared to the other three SLSNe-I with a second \ion{Mg}{II} system, by being very slow evolving and presenting bumps and undulations in its light curve. There are no signs of post-peak bumps or wiggles in the light curves of \xga, while the light curve of \xgc\ shows a possible flattening in the $gcr$ bands starting $\sim 80$ days after the peak (see Sect.~\ref{sec:discussion}). In \xgc, a pre-bump in the $r$ band was observed immediately after explosion, whereas in \xga\ a possible bump can be seen in the $g$ and $r$ band light curves at $\sim -30$ days. All four objects are very energetic with radiated energies of $E_{\rm rad} \gtrsim 1 \times 10^{51}~\rm erg~s^{-1}$.

\subsection{Light-curve modeling} \label{sec:lc_modeling}

\begin{table*} [!ht]
\centering
\caption{Priors and posterior of the parameters fit with \program{redback} for the generalized magnetar model.} 
\label{tab:redback_results}
\begin{tabular*}{.98\linewidth}[!ht]
{@{\extracolsep{\fill}}llll}

\hline
\hline

Parameters & Priors & Best-fit values & Best-fit values  \\
                
\hline

&  & \xga\ & \xgc \\ 
\hline

Initial magnetar spin-down luminosity  $L_{\rm 0}$ ($\rm erg~s^{-1}$) &$\mathcal{L}$ ($10^{40}$, $10^{50}$) & $1.2^{+0.1}_{-0.1} \times 10^{45}$ & $8.1^{+0.6}_{-0.4} \times 10^{44}$ \\
Magnetar spin-down time $t_{\rm SD}$ ($\rm s$) & $\mathcal{L}$ ($10^{2}$, $10^{8}$) & $1.9^{+0.7}_{-0.5} \times 10^{7}$ & $1.9^{+0.9}_{-0.5} \times 10^{7}$\\
Magnetar braking index $n$ & $\mathcal{U}$ (1.5, 10) & $1.9^{+0.5}_{-0.3}$  & $5.5^{+2.5}_{-1.9}$\\
Ejecta nickel mass fraction $f_{\rm ^{56}Ni}$ & $\mathcal{L}$ (10$^{-6}$, 1) & $5.1^{+49.8}_{-4.9} \times 10^{-4}$  & $1.9^{+44.9}_{-1.9} \times 10^{-3}$\\
Ejecta mass $M_{\rm ej}$ ($\rm M_\odot$) & $\mathcal{U}$ (0.1, 100) & $7.0^{+0.4}_{-0.4}$ & $9.4^{+0.7}_{-0.7}$\\
Supernova explosion energy $E_{\rm SN}$ ($\rm erg$) & $\mathcal{U}$ ($5 \times 10^{49}$, $3 \times 10^{51}$) & $2.97^{+0.02}_{-0.04} \times 10^{51}$  & $2.4^{+0.3}_{-0.3} \times 10 ^{51}$\\
Ejecta gamma-ray opacity $\rm \kappa_\gamma$ (cm$^{2}$ g$^{-1}$) & $\mathcal{L}$ ($10^{-4}$, $10^4$) & $8.3^{+1.2}_{-1.1} \times 10^{-3}$  & $3.6^{+0.9}_{-0.7} \times 10^{-3}$\\
Photospheric plateau temperature $T_{\rm floor}$ (K) & $\mathcal{U}$ (3~$\times$~10$^{3}$, 2~$\times$~10$^{4}$) & $11494^{+242}_{-232}$ & $6605^{+185}_{-205}$\\
Explosion date $t_{\rm exp}$ (MJD) & $\mathcal{U}$ ($\rm FD -100 $, $\rm FD - 0.1 $) & $59096.3^{+1.2}_{-1.2}$ & $59824.3^{+0.4}_{-0.6}$\\

\hline

\hline
\end{tabular*}
\tablefoot{The $\mathcal{U}$ stands for uniform, and $\mathcal{L}$ for log-uniform. The FD acronym stands for first detection and represents the date of the first real detection. The uncertainties are reported at $1\sigma$ significance.}
\end{table*}

\begin{figure*}[!ht]
     \centering
     \begin{subfigure}[b]{0.9\textwidth}
         \centering
         \includegraphics[width=\textwidth]{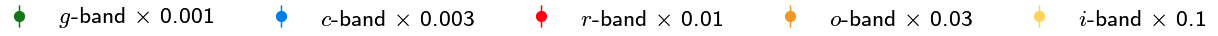}
        
     \end{subfigure}
          \begin{subfigure}[b]{0.49\textwidth}
         \centering
         \includegraphics[width=\textwidth]{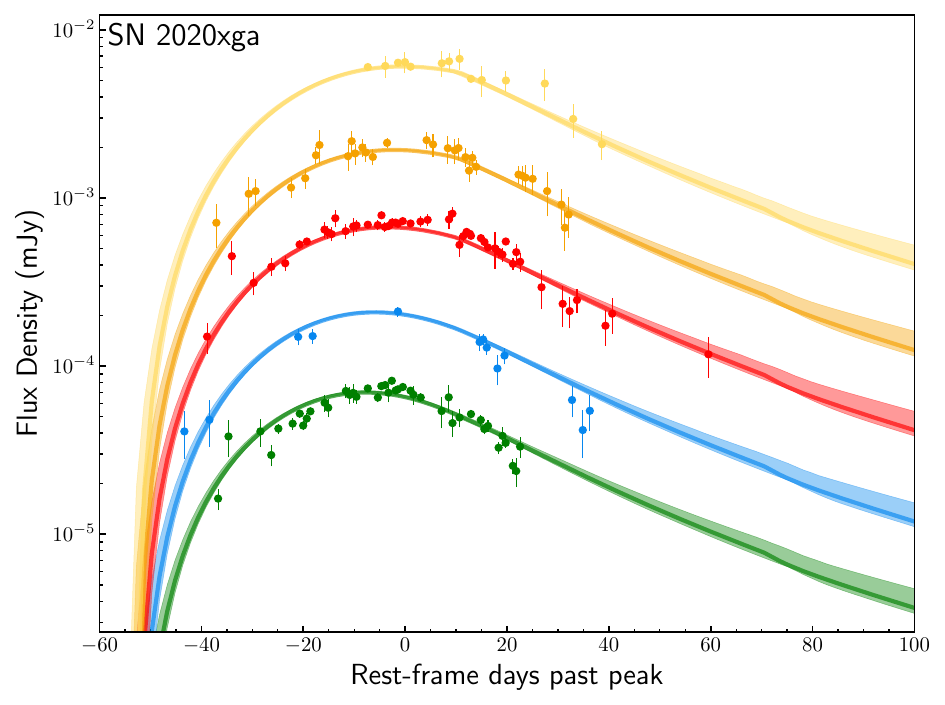}
     \end{subfigure}
     \begin{subfigure}[b]{0.49\textwidth}
         \centering
         \includegraphics[width=\textwidth]{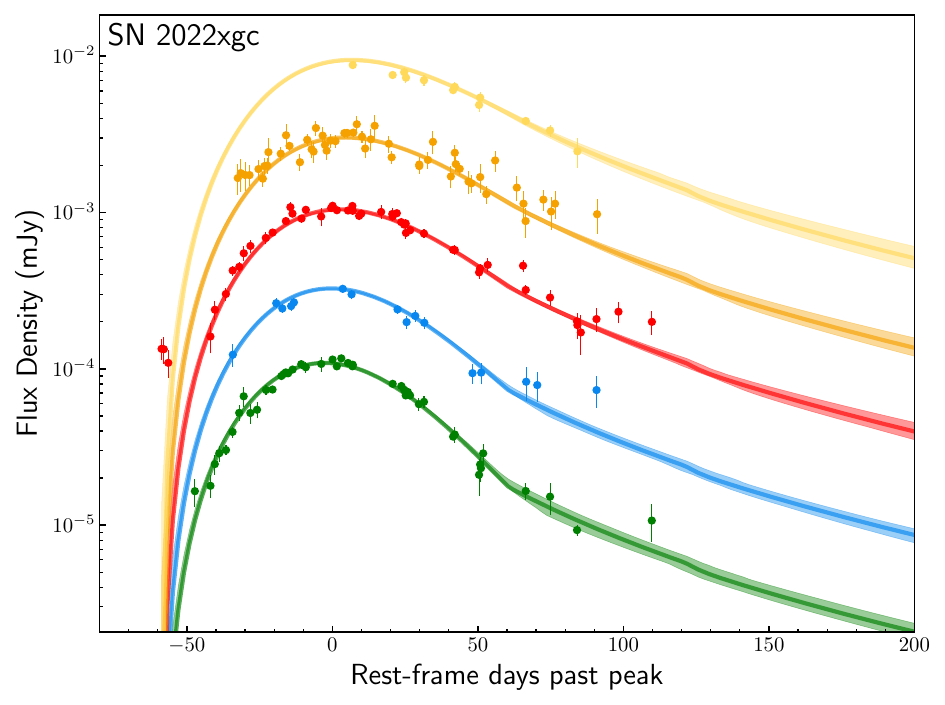}

     \end{subfigure}

    \caption{Multiband light curves of \xga\ (left panel) and \xgc\ (right panel) with their resulting fits from \program{redback}. The solid colored lines indicate the light curves from the model with the maximum likelihood, while the shaded areas depict the 90 percent credible interval. The x axis is in rest-frame days with respect to the rest-frame $g$-band maximum.}
    \label{fig:redback_slsn}
\end{figure*}

We modeled the observed multiband light curves of \xga\ and \xgc\ using the Bayesian inference software package for fitting electromagnetic transients, \program{redback} \citep{Sarin2024a}. We input the redshift of the SLSNe (see Sect.~\ref{sec:redshift}), the $gcroi$ photometric observations of \xga\ and \xgc\ corrected for extinction (we excluded the $z$ band due to the low number of datapoints), and a list of priors shown in Table~\ref{tab:redback_results}. We explored the parameter space with the nested sampling package \program{dynesty} \citep{Ashton2019,Speagle2020}. 

To fit the data, we selected the versatile \texttt{general magnetar-driven supernova} model described in \cite{Omand2024a} under the assumption that the light curves of \xga\ and \xgc\ are powered by the spin-down of a rapidly rotating newly formed magnetar \citep{Ostriker1971,Arnett1989,Kasen2010,Chatzopoulos2012,Inserra2013}. This model sets the magnetar braking index, $n$, as a variable, relaxing the assumption of a vacuum dipole spin-down mechanism and includes the dynamical evolution of the ejecta \citep{Sarin2022} coupling it to both the explosion energy and the spin-down luminosity of the magnetar itself. We used the default priors defined in \cite{Omand2024a} relaxing the priors for the explosion energy $E_{\rm SN}$ and the temperature floor $T_{\rm floor}$. The opacity, $\kappa$, was fixed at $0.04~\rm cm^{2}~g^{-1}$, which is a good approximation for type Ic SNe, as is shown in \cite{Kleiser2014} (see their Fig.~3) considering that the blackbody temperature tends to overestimate the temperature of the photosphere \citep{Dessart2019}. We note that if we set the $\kappa$ parameter free, our results do not change significantly, suggesting that the choice of opacity plays a minimal role in our inference. However, we note that the opacity is kept constant with time and a time-dependent opacity could yield different results. This model assumes a modified SED accounting for the line blanketing in the UV part of the SLSN spectra \citep{Chomiuk2011} similar to the one used in \cite{Nicholl2017}. The \program{redback} light curve fits are shown in Fig.~\ref{fig:redback_slsn}, and the resulting values of the posteriors are given in Table~\ref{tab:redback_results}. The corner plots are uploaded in \href{https://zenodo.org/records/14565605}{https://zenodo.org/records/14565605} (see Sect.~\ref{data}).

In \xga\ (Fig.~\ref{fig:redback_slsn}; left panel), the model captures well the rise (apart from the $i$ band, which does not have data during the rise), peak, and decline in all five filters up to $30$ days, after which the model declines more slowly than the data. The model fails to fit the first real detection in the $c$ band and a possible small bump at $\sim -30$ days that is visible in the $g$ and $r$ bands. The latter is not unexpected since this model can explain only a smooth light curve (see discussion in \citealt{Omand2024a}). Similarly, in \xgc\ (Fig.~\ref{fig:redback_slsn}; right panel), the model fits well the SN multiband light curve both in the rise and the decline up to $\sim 60$ days, after which the model declines more quickly than the data in the $r$ and $c$ bands. In addition, the model does not capture the first three data points in the $r$ band, which, as is discussed in Sect.~\ref{sec:lc_properties}, could potentially be a pre-bump often seen in the light curves of SLSNe-I. A possible explanation for the early bumps could be interaction with extended material \citep{Piro2015} or magnetar-driven shock breakout \citep{Kasen2017}. 

The \texttt{general magnetar-driven supernova} model uses the initial magnetar spin-down luminosity, $L_{0}$, and the magnetar spin-down time, $t_{\rm SD}$, as input parameters instead of the initial magnetar spin
period in millisecond, $P_{0,\rm ms} (= P/10^{3}~\rm s)$, the magnetic field, $B_{14} (= B/10^{14}~\rm G)$, and the NS mass, $M_{\rm NS}$, used in previous magnetar models \citep[e.g.,][]{Nicholl2017}. To recover these parameters, we used the scalings

\begin{equation}
    E_{\rm rot}=\frac{n-1}{2}L_{0}t_{\rm SD} \, \rm erg
\end{equation}
and
\begin{equation}
    E_{\rm rot}=2.6 \times 10^{52} P_{0,\rm ms}^{-2} \,\rm erg
.\end{equation}
Assuming a $1.4~\rm M_\odot$ NS with the same equation of state as in \cite{Nicholl2017}, we found the rotational energy of the magnetar, $E_{\rm rot}$, to be $1.1 \times 10^{52}~\rm erg$ for \xga\ and $E_{\rm rot} = 3.3 \times 10^{52}~\rm erg$ for \xgc, while $P_{0,\rm ms}$ is estimated to be $1.6 \pm 0.1$ and $0.9 \pm 0.2$ for \xga\ and \xgc, respectively. The estimated $P_{0,\rm ms}$ values in \xga\  and \xgc\ are close to the so-called mass-shedding limit, the limit at which the centrifugal force throws mass off the surface of the magnetar \citep[e.g.,][]{Metzger2015,Watts2016}.

The ejecta masses of $7.0~\rm M_\odot$ and $9.3~\rm M_\odot$ for \xga\ and \xgc, respectively, resulting from the \texttt{general magnetar-driven supernova} model, are consistent with the findings of \cite{Chen2023b}, who found that the median ejecta mass is $5.1^{+4.0}_{-2.4}~\rm M_\odot$. However, we note that this comparison is limited by the different physics included in the model used in this paper and the model of \cite{Nicholl2017} used in the paper of \cite{Chen2023b}. In addition, the model estimates the explosion dates, $t_{\rm exp}$, of \xga\ and \xgc\ to be 13 and 1 days, respectively, which are earlier than the values found in Sect.~\ref{sec:lc_properties}. These small discrepancies are not unreasonable given that both SNe were luminous already at the time of the first detection, and thus our approach in Sect.~\ref{sec:lc_properties} constrains the time of first light rather than the explosion date.

Since the ejecta velocity, $v_{\rm ej}$, in the \texttt{general magnetar-driven supernova} model is not constant and the ejecta evolve dynamically as a function of time, we calculated the diffusion timescale, $t_{\rm diff}$,

\begin{equation} 
    t_{\rm diff} = 9.8 \times 10^{5} \, \Bigg(\frac{M_{\rm ej}}{1 \, \rm M_\odot}\Bigg)^{1/2} \, \Bigg(\frac{v_{\rm ej}}{10^4 \, \rm km~s^{-1}}\Bigg)^{-1/2} \, \Bigg(\frac{\kappa_{\rm }}{0.1 \, \rm cm^{2}~g^{-1}}\Bigg)^{1/2}~\rm s
,\end{equation}
using the ejecta velocity at the peak. The $v_{\rm ej}$ at the peak derived from the most likely model in \program{redback} is $10446~\rm km~s^{-1}$ for \xga\ and $8304~\rm km~s^{-1}$ for \xgc, and thus the $t_{\rm diff}$ is estimated to be $18$ days and $24$ for \xga\ and \xgc, respectively. We note that the ejecta velocities calculated in \program{redback} do not have the same physical meaning as the line velocities extracted from the spectra in Sect.~\ref{sec:velocity}. By computing the ratio of $t_{\rm SD}/t_{\rm diff}$, we could determine the fraction of the spin-down luminosity converted to kinetic energy accelerating the ejecta \citep{Suzuki2021,Sarin2022}. The ratios of the two timescales are $7.6$ for \xga\ and $6.6$ for \xgc. These two ratios show that the radiated and kinetic energy could be possibly dominated by magnetar spin-down  \citep{Suzuki2021,Omand2024a}. We note that a discussion of alternative power-mechanism scenarios is included in Sect.~\ref{sec:power}.

\subsection{Imaging polarimetry} \label{sec:pola}

The polarization degrees obtained on \xgc, and reported in Table~\ref{tab:pol_results} are all very low ($<0.5 \%$), and all constrained to about or less than 2$\sigma$. The final values shown in bold (Column 9) were obtained after bias correction following the equation given in \cite{wang1997}:
\begin{equation} 
\label{eqn:pdeb}
    P = (P_{\rm obs} - \sigma_{P}^{2}/P_{\rm obs}) \times h(P_{\rm obs} - \sigma_{P}),
\end{equation}
where $h$ is the Heaviside function, $P_{\rm obs}$ is the observed polarization, and $\sigma_{P}$ is the 1$\sigma$ error.

The debiased measurements obtained on \xgc\ displayed in the last column of Table~\ref{tab:pol_results} have been obtained without making any MW polarization correction. The first important point to notice is that they all show that the percentage of polarization is consistently low and does not seem to vary with time. The second important point is that these measurements are all consistent with the low level of Galactic polarization expected in that region of the sky. The Galactic extinction along the line of sight at the coordinates of \xgc\ is such that $E(B-V)=0.061$ \citep[][]{Schlafly2011}. Following \cite{serkowski1975}, this means that, in case of a magnetic field perfectly lying on the plane of the sky, the empirical upper limit on the optical degree of polarization produced by dichroic absorption by magnetically aligned Galactic dust grains should be $P_{\rm max}= 9 \times E(B-V)=0.55 \%$. A look at the measurements of the two closest known polarized stars in the vicinity of \xgc, HD\,58784 with $P_{\rm V}=0.65 \pm 0.2 \%$ and HD\,57291 with $P_{\rm V}=0.37 \pm 0.2 \%$, support this statement. These data were retrieved from the compiled catalog of optical polarization measurements by \cite{Heiles2000}. 

The polarization angles displayed in Table~\ref{tab:pol_results} show polarization angles that differ by about 90 degrees between the two epochs. The constraints on the polarization angles are low S/N, but as an ultimate test we estimated the variations in the polarization degree obtained between each epoch in each filter. This was done using the values of the Stokes parameters before debiasing the data. Any IP and ZPA corrected Stokes parameter on \xgc\ should be the sum of a constant contribution from the MW and from the host galaxy (if any) added to a possibly variable contribution from the SN. Therefore a differential measurement between two epochs should assess the degree of variation in polarization associated to the SN. In the $R$-band the differential is, $\Delta P(R) = 0.29 \pm 0.16 \%$, while in the $V$-band it is, $\Delta P(V) = 0.52 \pm 0.23 \%$. 

We conclude that the estimates given in Table~\ref{tab:pol_results} are likely estimates of the MW polarization contribution and that any contribution that could be associated with \xgc\ and its host galaxy should only be a few tenths of a percent. This low level rejects any detection of jet activities. If the level of polarization associated with \xgc\ changed between the two epochs, it should only be a fraction of a percent; in other words, very low. This low level of variation in the polarization level refutes the idea that there is any strong change in the shape of the photosphere of the SN between the two epochs. 

All these results seem consistent with the statistical results obtained by \cite{Pursiainen2023} on a sample of 16 SLSNe. In this work, the data obtained before maximum light indicate nearly spherical photospheres. No clear relation is found between the polarimetry and spectral phase after maximum light, and an increasing polarization degree is measured only on a subsample of four SLSNe that have irregular light curve shapes on decline. The light curve decline of \xgc\ looks smooth and regular (see Fig.~\ref{fig:redback_slsn}, right) at the phases when polarimetry was obtained (+26.1 days and +60.1 days). If any strong CSM interaction with the ejecta of \xgc\ happened during these two phases, it appears that it did not aﬀect the symmetry of the system. We point out, however, that these results are not indicative of the lack of CSM playing a role in powering the observed light curve of the events.

\section{Spectroscopy} \label{sec:spectroscopy}

\subsection{Redshift} \label{sec:redshift}

\begin{figure*}[!ht]
     \centering
          \begin{subfigure}[b]{0.47\textwidth}
         \centering
         \includegraphics[width=\textwidth]{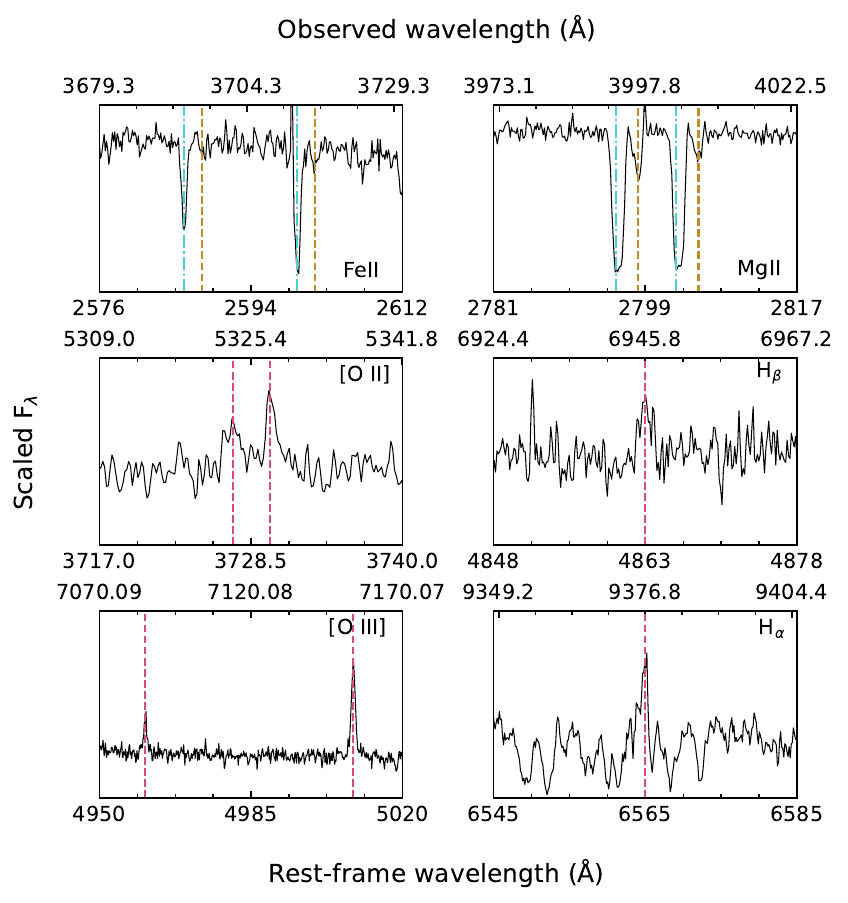}
         \caption{}
     \end{subfigure}
     \begin{subfigure}[b]{0.49\textwidth}
         \centering
         \includegraphics[width=\textwidth]{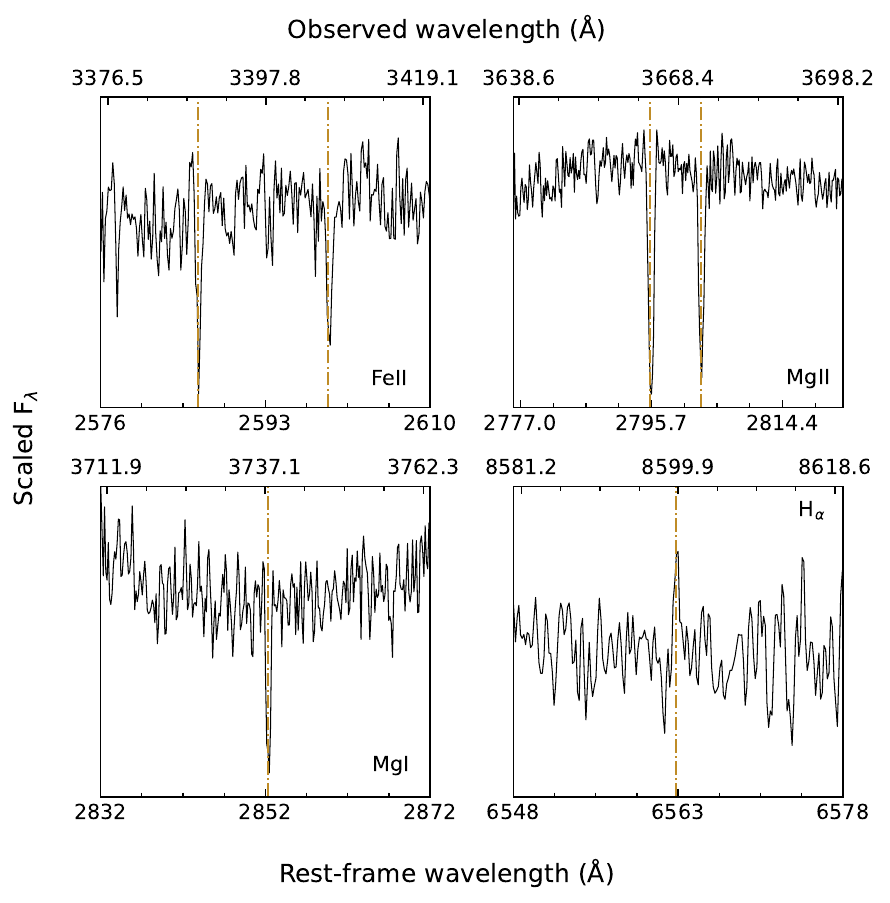}
         \caption{}
     \end{subfigure}
     
    \caption{Host galaxy absorption and emission lines in the X-shooter spectrum of \xga\ (panel a) $-8.3$ days and \xgc\ (panel b)  $+21.9$ days after maximum light. The vertical lines illustrate the various redshift values of the galaxy lines. Panel a: In the spectrum of \xga\,, the emission lines (dashed red lines) give a consistent redshift of $z = 0.4287$, whereas the strong (dashed blue line) and weak (dashed gold line) absorption lines indicate redshifts of $z = 0.4283$ and $z = 0.4296$, respectively. Panel b: The host galaxy lines (dashed gold line) of \xgc\ agree on a redshift of $z = 0.3103$.}
    \label{fig:redshift}
\end{figure*}

To estimate the precise redshifts of \xga\ and \xgc, we examined the X-shooter spectra of \xga\ at $-8.3$ days and \xgc\ at $+21.9$ days after maximum light and identified emission and absorption lines from the interstellar medium and \ion{H}{ii} regions in the host galaxy \citep[e.g.,][]{Vreeswijk2014,Leloudas2015}. Figure~\ref{fig:redshift} shows the galaxy lines that appear in the spectra of \xga\ and \xgc\ and that we used for the redshift determination.

We identified in the spectrum of \xga\ the galaxy's narrow absorption \ion{Fe}{II} doublet $\lambda\lambda2586,2600$ and \ion{Mg}{II} doublet $\lambda\large2796, 2803$, and the galaxy's narrow emission [\ion{O}{II}] doublet $\lambda\lambda3727, 3729$, H$\beta$ $\lambda 4861$, forbidden [\ion{O}{III}] doublet $\lambda\lambda4959,5007$ and narrow H${\alpha}$ $\lambda6563$. We found two galactic \ion{Fe}{II} and \ion{Mg}{II} absorption systems in the spectrum of \xga, with the stronger lines at a redshift of $z = 0.4283 \pm 0.0002$ and the weaker lines at $z = 0.4296 \pm 0.0002$. In addition, the host's emission lines are consistent with a redshift of $z = 0.4287 \pm 0.0001$. Throughout the paper, we chose the redshift of SN\,2020xga, as the higher value, and hence we assumed $z = 0.4296$. We further discuss this implication in Sect.~\ref{sec:host_galaxy}.

In the case of \xgc, the host galaxy is not detected in the images since it falls below the sensitivity limits of the surveys (see Table~\ref{tab:phot:host}); however, host galaxy lines are detected in the X-shooter spectrum of \xgc. This is not unprecedented since it has been also seen in other SLSNe-I \citep[e.g.,][]{Vreeswijk2014,Chen2015,Leloudas2015}. The host galaxy lines displayed in the spectrum of SN\,2022gxc are the narrow \ion{Fe}{II} doublet $\lambda\lambda2586,2600$, the \ion{Mg}{II} doublet $\lambda\lambda2796, 2803$, the \ion{Mg}{I} $\lambda2852$ and H${\alpha}$ $\lambda6563$. These lines support a redshift of $z = 0.3103 \pm 0.0001$ for \xgc. 

\subsection{Spectroscopic sequence} \label{sec:line_id}

\begin{figure*}[!ht]
     \centering
          \begin{subfigure}[b]{0.5\textwidth}
         \centering
         \includegraphics[width=\textwidth]{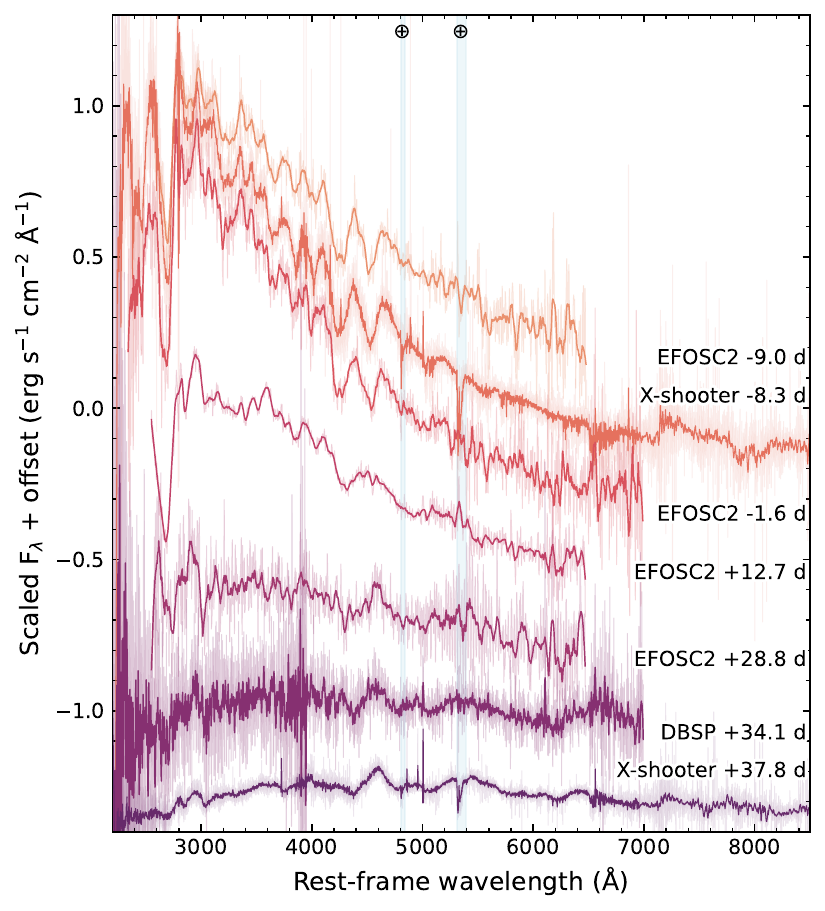}
         \caption{}

     \end{subfigure}
     \begin{subfigure}[b]{0.49\textwidth}
         \centering
         \includegraphics[width=\textwidth]{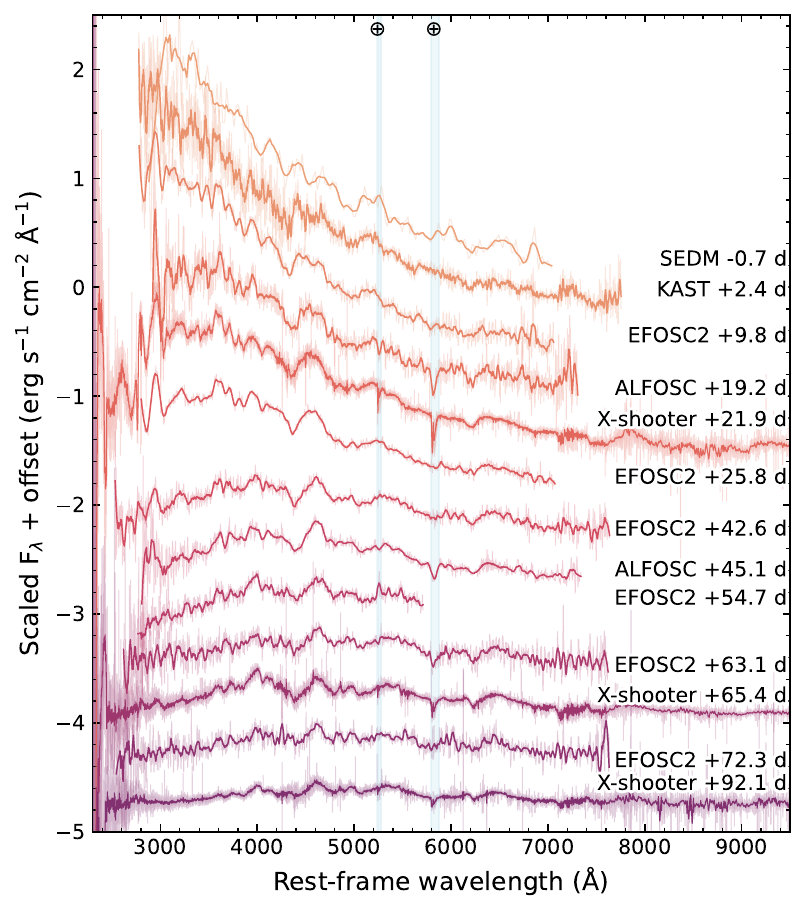}
         \caption{}

     \end{subfigure}

    \caption{Spectral sequence of \xga\ (panel a) from $-9$ to $+37.8$ rest-frame days and \xgc\ (panel b) from $-0.7$ to $+92.1$ $g$-band rest-frame days. An offset in flux was applied for illustration purposes. The spectroscopic measurements have undergone absolute flux calibration to align with the photometric data. The spectra are corrected for MW extinction and are smoothed using a Savitzky-Golay filter. The original data are presented in lighter colors. Regions of strong atmospheric absorption are blue-shaded.}
    \label{fig:spectral_evolution}
\end{figure*}

Figure~\ref{fig:spectral_evolution} depicts the spectral evolution from $-9.0$ to $+37.8$ rest-frame days past maximum brightness of \xga\ and from $-0.7$ to $+92.1$ rest-frame days of \xgc\ from 2500~\AA\ up to $\sim$ 10\,000~\AA. All spectra were taken during the photospheric phase. As the ejecta cools down, the spectra of both \xga\ and \xgc\ begin to resemble those of typical type Ic SNe, which is expected for SLSNe-I \citep{GalYam2019b}.

\begin{figure}[!ht]
   \centering
  \includegraphics[width=0.5\textwidth]{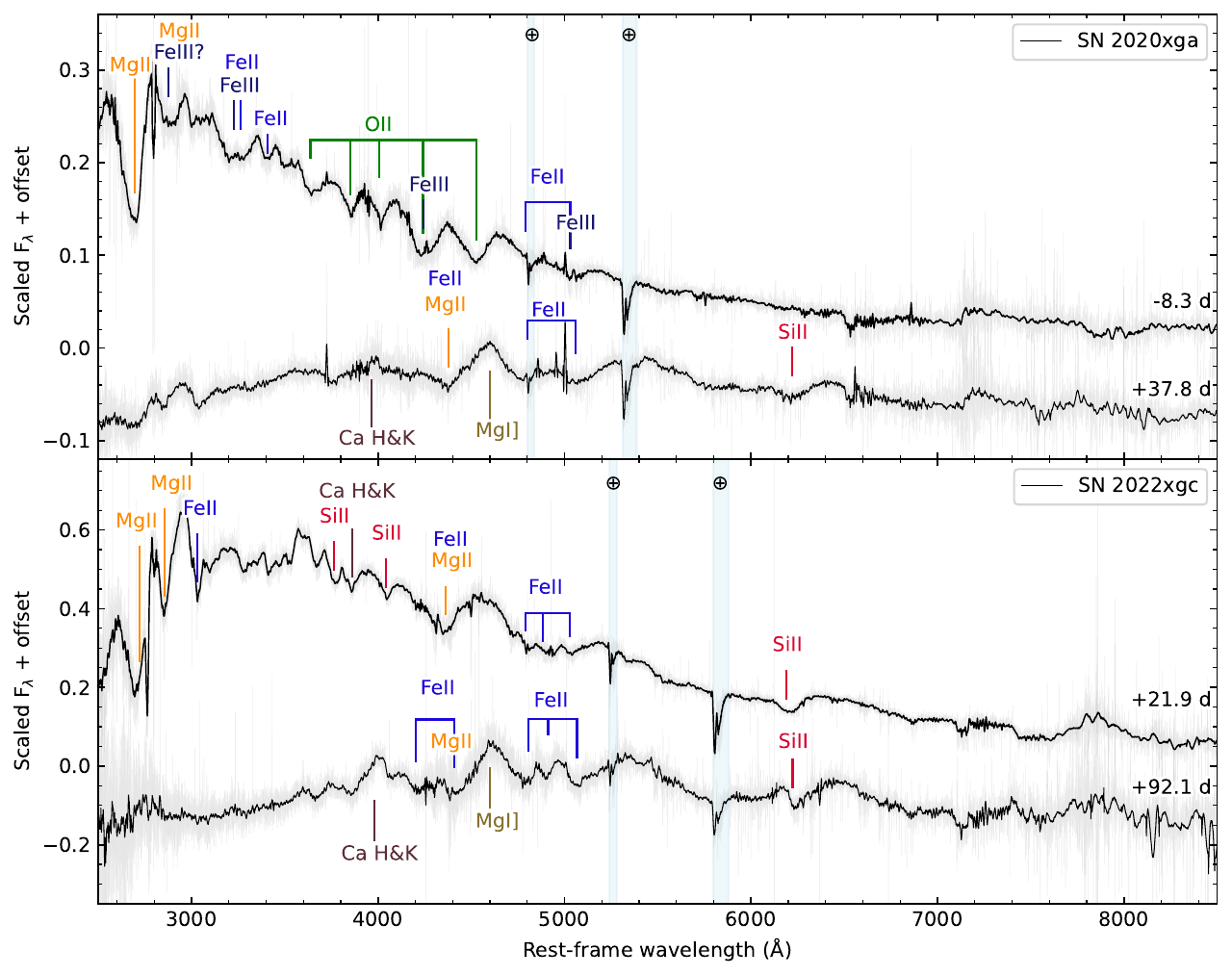}
   \caption{X-shooter spectra of \xga\ (upper panel) at $-8.3$ and $+37.8$ days and \xgc\ (lower panel) at $+21.9$ and $+92.1$ days after maximum light. The spectra are corrected for MW extinction and are smoothed using a Savitzky-Golay filter. The original spectra are shown in lighter gray. The most conspicuous features are labeled. Uncertain line identifications are denoted with question marks. The ions beneath the spectrum are shown at the rest wavelength, whilst those above have been shifted to match the absorption component. The light blue regions represent the telluric absorptions.}
      \label{fig:line_id}%
\end{figure}

To identify the spectral lines in \xga\ and \xgc\ we used the medium-resolution X-shooter spectra due to their high S/N. For \xga, the spectra at $-8.3$ and $+37.8$ days were utilized, and for \xgc, we used those at $+21.9$ and $+92.1$ days. The line identification was done by comparing the spectra with well-studied SLSNe-I from the literature \citep[e.g.,][]{Quimby2011,Inserra2013,Nicholl2015b,Quimby2018,GalYam2019a}, by modeling the earlier spectra at $-8.3$ days and $+21.9$ days (for \xga\ and \xgc, respectively) using the synthesis code SYN++ \citep{Thomas2011} and finally by searching the National Institute of Standards and Technology (NIST; \citealt{Kramida2022}) atomic spectra database for lines above a certain strength, similar to what was done in \citet{GalYam2019a}. Figure~\ref{fig:line_id} depicts the X-shooter spectra of \xga\ and \xgc\ along with the most prominent features blueshifted by 6000 -- 8000~km~s$^{-1}$ to match the absorption lines (see Sect.~\ref{sec:velocity}). The \ion{Ca}{H \& K} and \ion{Mg}{I]} lines in the spectra at $+37.8$ days and $+92.1$ days are shown at zero rest-frame velocity. The SYN++ modeling of the $-8.3$ day phase spectrum of \xga\ and the $+21.9$ day phase spectrum of \xgc\ can be found in Fig.~\ref{fig:syn++}, while Table~\ref{tab:syn} collects the best-fit parameter values obtained by the modeling. 

In the early spectrum of \xga, SYN++ tentatively identify the strong W-shape feature between 3500 -- 5000~\AA\ with the \ion{O}{II} $\lambda 4358$ and $\lambda 4651$ that characterize the spectra of numerous SLSNe-I. A small contribution of \ion{C}{II} might be present at the troughs of 4300 and 4550~\AA. Comparison with other SLSNe-I showed that the absorption trough at 4300~\AA\ is most likely a blend of \ion{Fe}{III} $\lambda 4432$ and \ion{O}{II} $\lambda 4357$. Redward of the \ion{O}{II} lines, the \ion{Fe}{II} $\lambda4923$  and $\lambda5169$ are present, but the \ion{Fe}{II} $\lambda4923$ falls into the telluric band and the \ion{Fe}{II} $\lambda5169$ is likely mixed with \ion{Fe}{III} $\lambda5129$ \citep{Liu2017}. Above 5000~\AA\, the early spectrum of \xga\ does not show any obvious feature. Blueward of 3500~\AA\ a number of features are visible, but owing to severe blending the identification is challenging. The trough at 2670~\AA\ is associated with \ion{Mg}{II}, as is seen in other SLSNe-I, and the trough at $\sim$ 2880~\AA\ has been observed in a number of SLSNe \citep[e.g.,][]{Vreeswijk2014,Quimby2018,Gkini2024} and has been suggested by a few studies \citep{Dessart2012,Mazzali2016,Quimby2018,Gkini2024} to have some contribution from \ion{Ti}{III}, \ion{Fe}{III}, \ion{Si}{III}, \ion{C}{II} and \ion{Mg}{II}. Searching NIST, we discovered that the absorption at 3200~\AA\ could be attributed to \ion{Fe}{II} $\lambda3325$ and \ion{Fe}{III} $\lambda3305$. The feature at 3410~\AA\ may be related with \ion{Fe}{II} $\lambda3500$. 

In the spectrum of \xgc\ at $+21.9$ days the major ions that are securely identified by SYN++ are \ion{Fe}{II}, \ion{Si}{II} and \ion{Ca}{II}. Comparison with other SLSNe-I revealed that the absorption trough at 2670~\AA\ is due to \ion{Mg}{II}. The absorption component at 2880~\AA\ is stronger than what is seen in \xga\ and similar to the one observed in the SLSN-I SN\,2020zbf \citep{Gkini2024}. As was previously stated, this component is likely a contribution of multiple elements, including \ion{Mg}{II}. Searching the NIST, we identified some plausible contribution of \ion{Fe}{II} between 3000 and 3600~\AA, although the high level of blending makes this identification dubious.  Additional \ion{Mg}{II} $\lambda 4481$ may be present in the absorption feature at 4300~\AA. Absorption from \ion{O}{II} between 3500 -- 5000~\AA, as is seen in many SLSN-I spectra around the peak, is not present. Similarly to the case of \xga, we were unable to identify any line beyond 6500~\AA\ in the spectrum of \xgc\ owing to low S/N.

The spectra of \xga\ at $+37.8$ days and \xgc\ at $+92.1$ days resemble the spectra of a SN Ic at maximum light \citep{Pastorello2010,Quimby2011}. Both objects show \ion{Ca}{II} $\lambda\lambda 3966,3934$ (though in \xga\ the emission line is weak), \ion{Mg}{II} $\lambda 4481$, \ion{Mg}{I}] $\lambda 4571$,  and strong \ion{Fe}{II} lines between 4000 and 5200~\AA\ blueshifted by 7500 km s$^{-1}$ (see Sect.~\ref{sec:velocity}) to match the absorption component. In \xgc, the strong absorption trough at 6230~\AA\ is connected with \ion{Si}{II} $\lambda6355$, whereas in \xga, \ion{Si}{II} may contribute to the weak absorption component at 6230~\AA. The contribution of the \ion{O}{I} triplet $\lambda\lambda 7772,7774,7775$ in the spectra of \xga\ and \xgc\ might be visible at 7580~\AA, however owing to the low S/N, this identification is uncertain.

\subsection{Ejecta velocities} \label{sec:velocity}

\begin{figure}[!ht]
   \centering
  \includegraphics[width=0.5\textwidth]{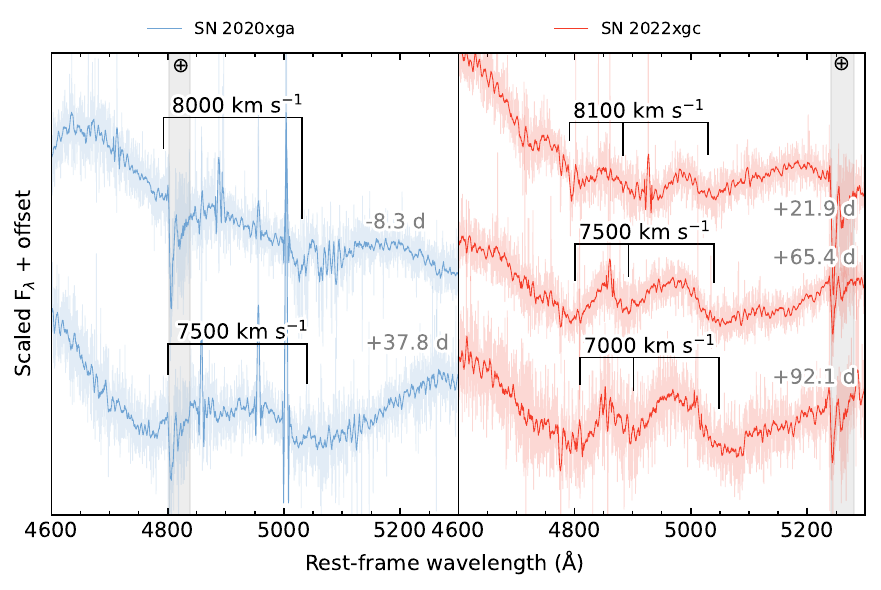}
   \caption{\ion{Fe}{II} triplet $\lambda\lambda4923,5018,5169$ region of \xga\ (left panel) at $-8.3$ and $+37.8$ day spectra and \xgc\ at $+21.9$,$+65.5$ and $+92.1$ days spectra. A normalization and an arbitrary offset has been applied for illustration purposes. The spectra have been smoothed using the Savitzky-Golay filter and the original data are shown in lighter colors. The absorption features that correspond to the blueshift of the \ion{Fe}{II} lines are denoted along with the velocities.}
      \label{fig:velocity_evolution}%
\end{figure}

The ejecta velocities of SLSNe-I and their evolution can be measured from the \ion{O}{II} absorption lines at 3500 -- 5000~\AA\ at early phases \citep{Quimby2018,GalYam2019a,GalYam2019b} and from the \ion{Fe}{II} triplet $\lambda\lambda4923,5018,5169$ \citep{Branch2002,Nicholl2016b,Modjaz2016,Liu2017}. In \xga, the \ion{O}{II} lines with the most noticeable characteristic, the W-shape, are present from $-9.0$ to $-1.6$ days. The absorption troughs of \ion{O}{II} are shifted by $-8000$~km~s$^{-1}$ and the velocity remains constant throughout the seven-day period. \cite{Chen2023b}, studying a sample of 77 SLSNe-I, estimates the median \ion{O}{II} velocity of the ZTF sample to be 9700~km~s$^{-1}$. To report a dispersion in this value we bootstrapped the 41 SLSNe-I from the ZTF sample with \ion{O}{II} velocities within $\pm 30$ days post maximum light and propagated the measurement uncertainties with a Monte Carlo simulation. This process resulted in a median velocity of the ZTF-I sample of $9794^{+3106}_{-2804}$~km~s$^{-1}$. Our estimated value of $8000$~km~s$^{-1}$ for \xga\ is lower than the median but it is in the range of the velocities of SLSNe-I. In \xgc, the \ion{O}{II} lines are not clearly visible in the early spectra, therefore we cannot determine the ejecta velocity using the \ion{O}{II} ion.

\begin{figure}[!ht]
\centering
  \includegraphics[width=0.5\textwidth]{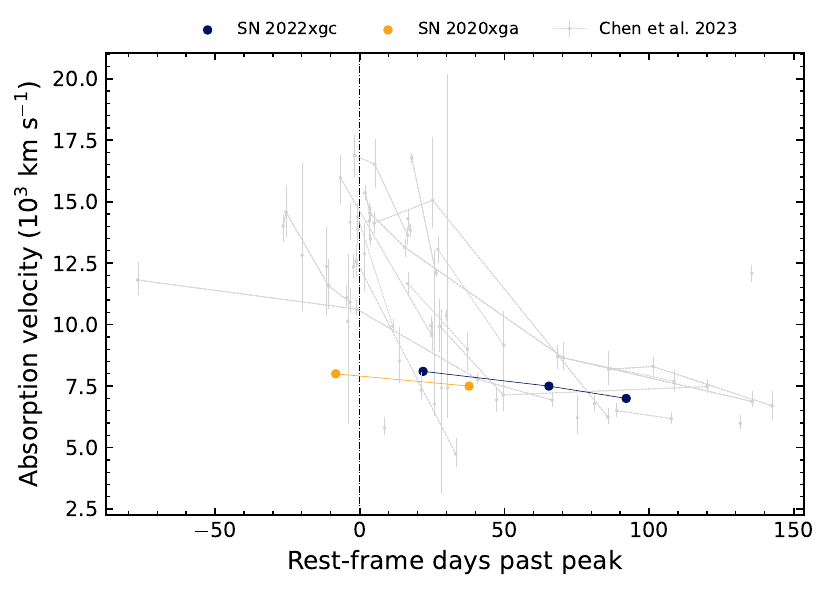}
   \caption{\ion{Fe}{II} ejecta velocities of \xga\ and \xgc\ as measured from the X-shooter spectra as a function of time. The velocity evolution of the ZTF SLSN-I sample is shown in gray for comparison. The vertical dash-dotted line illustrates the phase of the maximum light.}
      \label{fig:velocity_track}%
\end{figure}

The second method of measuring the ejecta velocity and following its evolution is to use the \ion{Fe}{II} triplet as a tracer \citep{Branch2002,Nicholl2016b,Modjaz2016,Liu2017}. In Fig.~\ref{fig:velocity_evolution}, a zoomed-in view of the \ion{Fe}{II} triplet region at $-8.3$ and $+37.8$ days post-maximum for \xga\ and at $+21.9$, $+65.4$, and $+92.1$ days post-maximum for \xgc\ is shown; we used the high-quality X-shooter spectra, since the low S/N of the low-resolution spectra prevent us from tracking the velocity evolution. In \xga, two absorption lines are visible, which we identified as \ion{Fe}{II} $\lambda4923$ and \ion{Fe}{II} $\lambda5169$. Since the \ion{Fe}{II} $\lambda4923$ suffer from telluric absorption, we utilized the \ion{Fe}{II} $\lambda5169$ line to estimate the velocities. In the X-shooter spectrum at $-8.3$ days after the peak, the marked absorption components match well with the \ion{Fe}{II} triplet blueshifted by $\sim 8000$~km~s$^{-1}$ despite the \ion{Fe}{II} $\lambda4923$ being veiled by tellurics. We note that this value may be underestimated due to contamination with \ion{Fe}{III} $\lambda5129$. The resulting velocity from the \ion{Fe}{II} lines is in agreement with the velocity estimated from the \ion{O}{II} ion and the absorption components of other identified elements, such as \ion{Fe}{III}. The strong features at 4770 and 5045~\AA\ in the spectrum at $+37.8$ days suggest that the \ion{Fe}{II} $\lambda4923$ and $\lambda5169$ may be blueshifted by $\sim 7500$~km~s$^{-1}$, which would result in a relatively constant velocity over a period of 50 days seen also in other SLSNe-I \citep{Nicholl2013,Nicholl2016b,Nicholl2016a,Liu2017}.

In Fig.~\ref{fig:velocity_evolution}, in the spectrum of \xgc\ we resolved three absorption lines that we identify as \ion{Fe}{II} $\lambda\lambda4924, 5018,5169$. At $+21.9$~days the troughs match with a ejecta velocity of $\sim 8100$~km~s$^{-1}$, though the mismatch of the absorption at 4940~\AA\ can be explained by a blend of \ion{Fe}{II} with other ions. The triplet is better resolved in the spectra at $+65.4$ and $+92.1$ days and the velocity decreases by $\sim 1000$~km~s$^{-1}$ within 70 days. 

To compare these values with the ZTF sample, we plot in Fig.~\ref{fig:velocity_track} the velocity evolution of 38 ZTF SLSNe-I with \ion{Fe}{II} velocities, together with the velocity evolution of \xga\ and \xgc. The value of $\sim 8000$~km~s$^{-1}$ measured from the pre-peak spectrum of \xga\ is lower than what we find in the ZTF sample, but not unprecedented since there are at least two SLSNe within $\pm 20$ days after the peak in the ZTF sample with \ion{Fe}{II} velocities close to $8000$~km~s$^{-1}$. In \xgc, the first measurement of the velocity is derived from the spectrum at $+21.9$ days following the peak. At this phase, the estimated value of $8100$~km~s$^{-1}$ is within the velocity range of the ZTF sample at similar phases but lower than the bulk of the population. Both objects appear to evolve slower than the ZTF sample.

\subsection{Comparison with other hydrogen-poor superluminous supernovae}

\begin{figure*}[!ht]
\centering
  \includegraphics[width=1\textwidth]{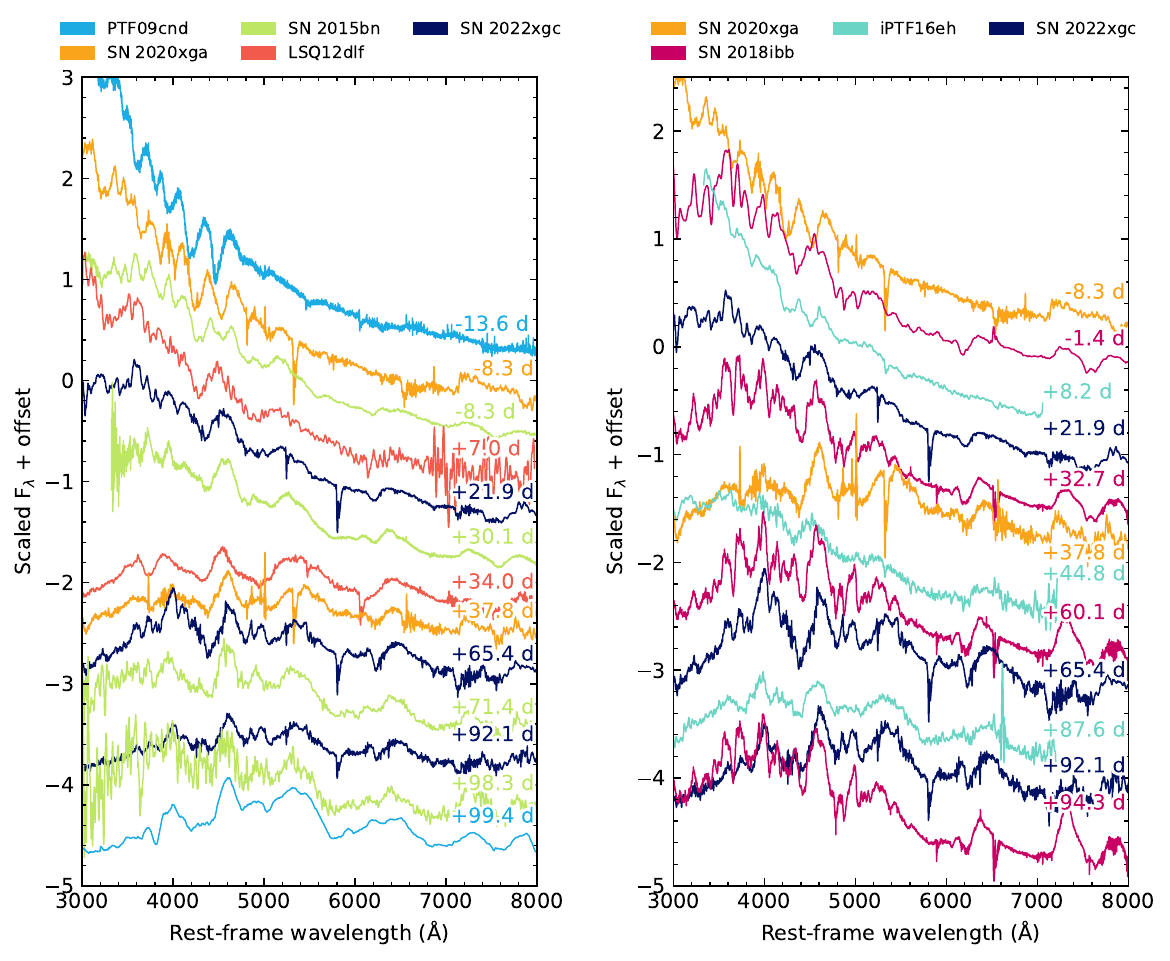}
   \caption{Spectral comparison of \xga\ and \xgc\ with SLSNe-I from the literature. Left: Comparison of  \xga\ and \xgc\ spectra with typical well-studied SLSNe-I at similar epochs. Right: \xga\ and \xgc\ spectra in comparison with SLSNe-I that display the second narrow \ion{Mg}{II} absorption system. All spectra are corrected for MW extinction and have been smoothed using a Savitzky-Golay filter.}
      \label{fig:spectral_comparison}%
\end{figure*}

In Sect.~\ref{sec:lc_properties}, we compared the photometric properties of \xga\ and \xgc\ with the homogeneous ZTF sample \citep{Chen2023a} and found that the light curve characteristics of \xga\ and \xgc\ are either average or span across the entire distribution, aside being very bright. However, \xga\ and \xgc\ display a spectroscopic signature rarely seen in SLSN-I spectra, and thus we seek to compare the spectra of \xga\ and \xgc\ with those of typical SLSNe-I.

In Fig.~\ref{fig:spectral_comparison} (left), we compare the X-shooter spectra of \xga\ and \xgc\ at $-8.3$ and $+37.8$, and $+21.9$, $+65.4$ and $+92.1$ days after the peak, respectively, with a sample of well-studied SNe from the literature including PTF09cnd \citep{Quimby2011,Quimby2018}, LSQ12dlf \citep{Nicholl2014} and SN\,2015bn \citep{Nicholl2016a}. As was previously indicated, the pre-peak spectrum of \xga\ shows a strong \ion{O}{II} series, similar to the one found in PTF09cnd, albeit the absorption components in PTF09cnd are shifted to higher velocities compared to the ones in \xga. The \ion{O}{II} lines are also seen in the spectrum of SN\,2015bn at the same phase as the pre-peak spectrum of \xga, although they are weaker than for \xga\ and PTF09cnd. The \ion{O}{II} lines are not clearly seen in the spectra of \xgc\ and LSQ12dlf during the early photospheric phase, which could be explained by the fact that the conditions for \ion{O}{II} excitation may not satisfied \citep[e.g.,][]{Mazzali2016,Dessart2019,Konyves2022,Saito2024}. 

As the ejecta cool, the spectra of \xga\ and \xgc\ become similar to those of type Ic SNe at maximum, as anticipated for the typical SLSNe-I. Overall, the general shape of the spectra of \xga\ and \xgc\ both pre- and post-peak are similar to the spectra of typical SLSNe-I and do not present any unusual spectral properties aside the second narrow \ion{Mg}{II} absorption system in the UV part of the spectrum, which is further discussed in Sect.~\ref{sec:mgii}. We note that in the spectra of \xga\ and \xgc\ more features are resolved than often seen in
spectra of typical SLSNe-I, due to the high signal and good resolution of the X-shooter data. 

In Fig.~\ref{fig:spectral_comparison} (right), we compare the X-shooter spectra of \xga\ and \xgc\ with the spectra of SN\,2018ibb and iPTF16eh at similar phases. \xga\ is the only object in this class that displays strong \ion{O}{II} lines; the \ion{O}{II} W-shape appears to be present in iPTF16eh but weaker than in \xga. In the spectra of \xgc\ and SN\,2018ibb the \ion{Fe}{II} triplet is resolved though the \ion{Fe}{II} $\lambda5018$ in the spectrum of \xgc\ at $+21.9$ days does not align with the absorption component. In contrast to \xga\ and iPTF16eh, which do not exhibit any obvious feature in the red part of the optical spectrum in the spectra near peak, both \xgc\ and SN\,2018ibb show strong \ion{Si}{II} $\lambda6355$. As the temperature drops and elements from deeper inside are revealed, the spectra of all these objects become similar to those of standard type Ic SNe. However, SN\,2018ibb \citep{Schulze2024} develops features such as [\ion{O}{II}], [\ion{O}{III}], and [\ion{Ca}{II}] at the early photospheric phase that are not common for SNe. The spectroscopic peculiarity along with the outstanding light curve timescales make SN\,2018ibb a unique SLSN-I, which stands out even among the SLSNe-I that display the rare signature of the second \ion{Mg}{II} system \citep{Schulze2024}. 

\section{Circumstellar material shell around \xga\ and \xgc} \label{sec:mgii}

\subsection{Modeling of the \ion{Mg}{II} absorption lines} 

\begin{figure}[!ht]
   \centering
  \includegraphics[width=0.5\textwidth]{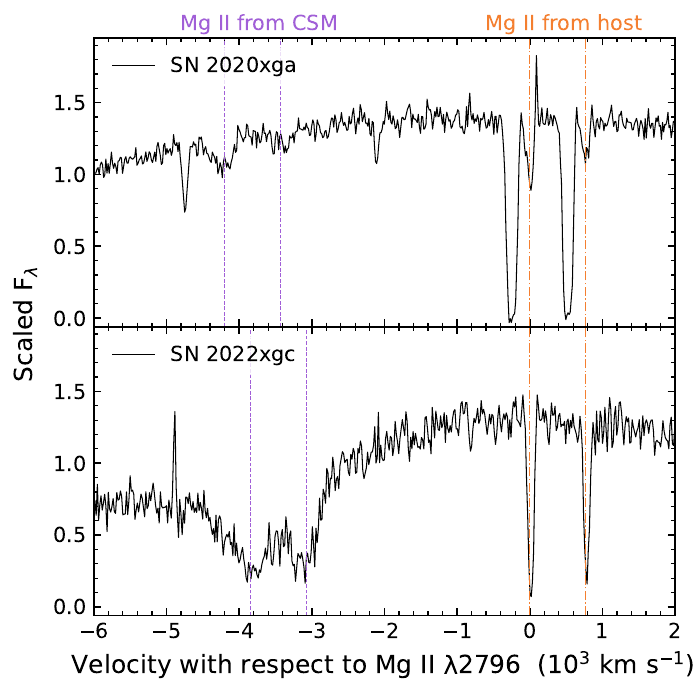}
   \caption{X-shooter spectra of \xga\ at $-8.3$ days (top panel) and \xgc\ at +21.9 days (bottom panel). The spectra show resolved, narrow absorption lines from the host ISM (marked by the vertical dashed orange lines) and a blueshifted absorption line system (marked by the vertical dashed purple lines) from a CSM shell expelled shortly before the SN explosion.}
 \label{fig:mgii}
\end{figure}

The X-shooter spectra of \xga\ and \xgc\ at $-8.3$ and $+21.9$ days, respectively, show two \ion{Mg}{II} absorption systems. The positive identification of the second system as \ion{Mg}{II} is supported by the absence of other transitions from the SN ejecta and the characteristic separation between the \ion{Mg}{II} doublet components. In Fig.~\ref{fig:mgii} a zoom-in region around the \ion{Mg}{II} lines at $\sim 2800$~\AA\ of \xga\ and \xgc\ is presented in velocity space. The narrow \ion{Mg}{II} systems at zero rest-frame velocity originate from the ISM of the host galaxies and are used for the determination of the SN redshift (see Sect.~\ref{sec:redshift}). The blueshifted \ion{Mg}{II} systems are significantly broader ($270~\rm km~s^{-1}$ for \xga\ and $500~\rm km~s^{-1}$ for \xgc) than expected for the ISM in dwarf host galaxies \citep{Kruhler2015,Arabsalmani2018} but also narrower compared to the SN features (> 1000 $\rm km~s^{-1}$). This supports the hypothesis that these systems arise from fast-moving absorbing gas that is not part of the ejecta or the host galaxy. The only objects that showed such lines are iPTF16eh and SN\,2018ibb and they have been associated with the existence of a rapidly expanding CSM shell expelled a few years before the SN explosion \citep{Lunnan2018}. We rule out the possibility that the \ion{Mg}{II} absorption reflects a peculiar composition of the ejecta, as this would require unusually high Mg abundances, in which case O and Ne lines would also be present in the spectra \citep{Woosley2002}.

\begin{figure}[!ht]
     \centering
          \begin{subfigure}[b]{0.55\textwidth}
         \centering
         \includegraphics[width=\textwidth]{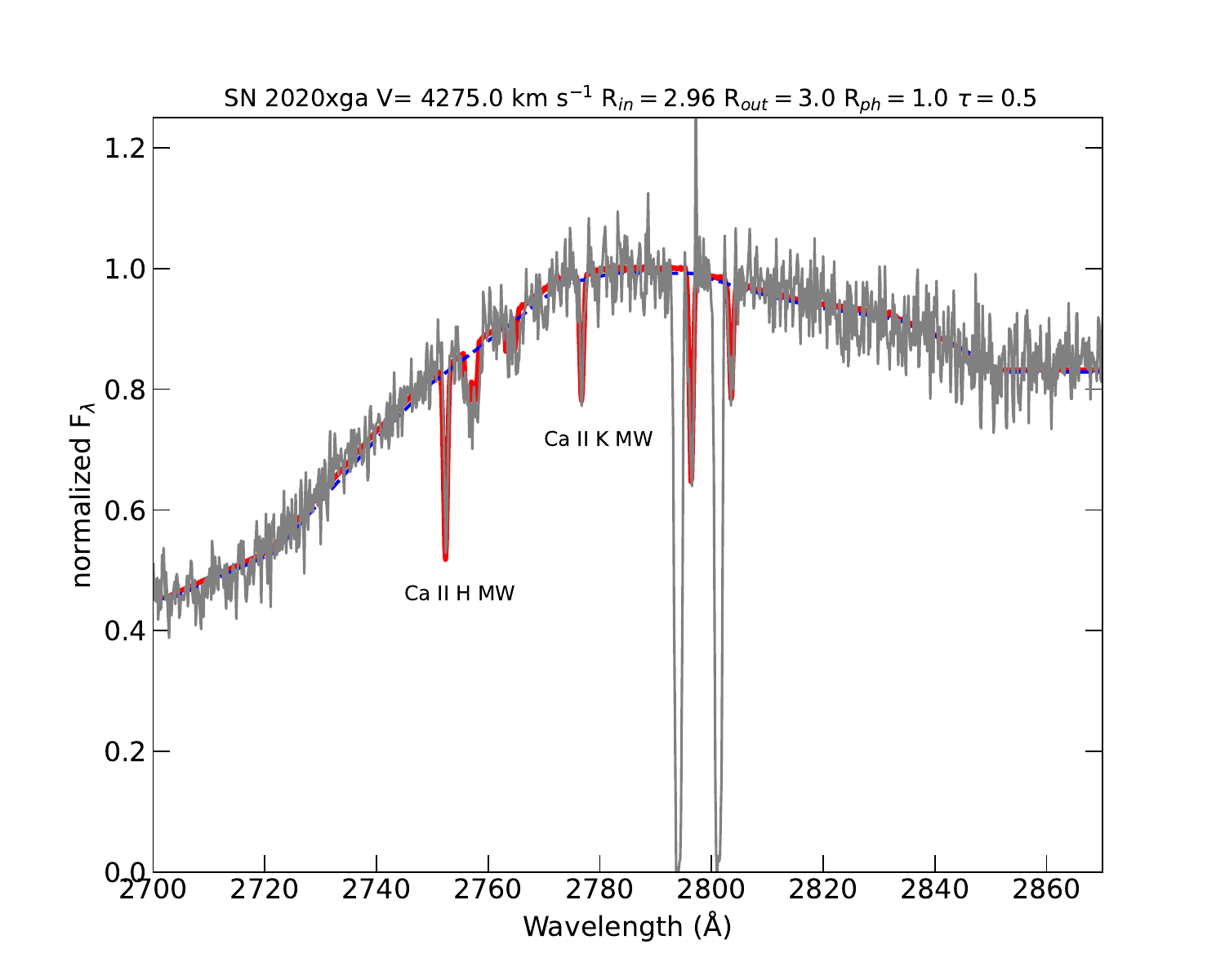}
     \end{subfigure}
     \begin{subfigure}[b]{0.55\textwidth}
         \centering
         \includegraphics[width=\textwidth]{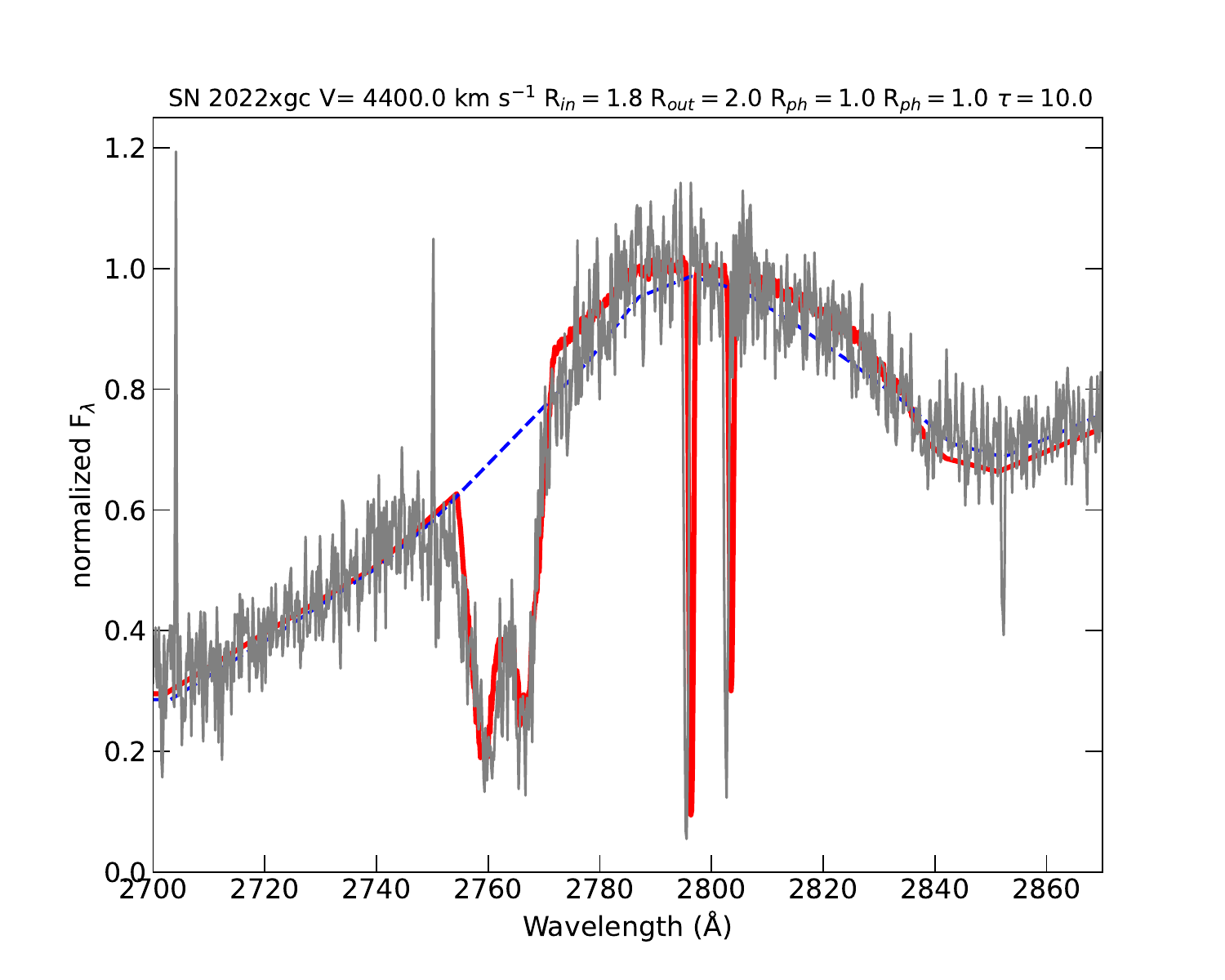}
     \end{subfigure}

    \caption{Modeling on the \ion{Mg}{II} doublets originating from the CSM shell (broad features) and the ISM of the host galaxy (narrow features) for \xga\ (top) and \xgc\ (bottom). The observed spectrum in the 2800~\AA\ region is presented in gray and the best model fit is shown in red. The dashed blue line illustrates the SN continuum. For \xga\ we have marked the \ion{Ca}{H \& K} lines from the MW.}
    \label{fig:MgII_modelling}
\end{figure}

To estimate the distance and the thickness of the CSM shell, we modeled the line profile of the \ion{Mg}{II} doublet using  a modified Monte Carlo scattering code.  While the line profile from a spherically symmetric shell for a single scattering line can be calculated analytically \citep{Fransson1984}, the doublet nature of the \ion{Mg}{II} lines will cause scattering from the blue component by the red component that affects the line profiles. We therefore used a Monte Carlo code to model the line profile, based on the Monte Carlo code in \cite{Fransson2014a} and \cite{Taddia2020}, used for type IIn SNe.  We assumed that the photons around 2800~\AA\ produced in the SN ``photosphere'' are scattered by an expanding spherical shell with an inner radius of $R_{\rm in}$ and an outer radius of $R_{\rm out}$.  The scattering is assumed to be coherent and isotropic in the frame of the expanding shell.  In reality, the background ``continuum'' is produced by lines from the expanding ejecta, most likely dominated by the broad Mg II lines (Sect. \ref{sec:spectroscopy}). We assumed that this continuum is given by the observed spectrum at velocities larger than that of the shell, and adjusted the background spectrum so that the total spectrum, ``continuum''  plus the scattered emission from the shell agreed with the observations. At the observed absorption velocity of the shell, the ``continuum'' flux was simply interpolated. Since there is no indication of thermal emission from the shell  \citep[contrary to the case of iPTF\,16eh,][]{Lunnan2018}, we neglected collisional excitation for the line profile, as well as other NLTE effects. Given that our primary interest is in the kinematics of the shell, which is only observed in the \ion{Mg}{II} lines, a more detailed NLTE treatment is not justified. We did not include scattering by the thermal motions of the electrons, as the electron scattering depth of the shell should be small, and there are no indications of smooth, extended wings of the lines in our best S/N spectrum (Fig. \ref{fig:MgII_modelling}), as is discussed in \cite{Hillier1991}.

We assumed that the expansion of the shell is homologous with $V= V_{\rm max}(r/R_{\rm out})$, which is reasonable for a time-limited eruption, such as PPI or LBV eruptions. This configuration gives rise to the second \ion{Mg}{II} system from the shell, which we observe as an emission and/or absorption lines on top of the SN photospheric flux. The homologous assumption is expected to break down when the fast moving ejecta catches up with the slower shell and hydrodynamically interacts. This is expected to either heat the shell to X-ray temperatures and ionize Mg almost completely,  or, if it would radiatively cool, produce flat topped emission lines. There is no evidence for either case.

We assumed that the CSM is mainly characterized by an optical depth $\tau$ of the shell. In the Sobolev approximation \citep{Sobolev1957} this is for a homologous expansion given by 
\begin{equation}
    \tau=\frac{g_2 \lambda^3 A_{21} n_{1} t}{8 \pi g_1 } \ .
\end{equation}
Here, $A_{21} = 2.8 \times 10^8$ s$^{-1}$ is the transition rate between the upper level 2 and lower level 1, $n_{1}$ is the number density in the ground state, $\lambda$ the wavelength, $g_1=2$ and $g_2$ the statistical weights of the lower and upper levels, respectively, and $t$ the time from the eruption. For the 2795.5 \AA \ line $g_2=4$ and for the 2802.7 \AA \ line $g_2=2$. The relative depth of the blue and red \ion{Mg}{II} doublet can be used to determine the optical depth. In the optically thin limit this ratio is 2.0, decreasing as the optical depth increases. 

The shell is assumed to have a constant density, or, more precisely, a constant optical depth. Since the shells we are considering are geometrically thin, variations in density, such as a wind-like distribution where $\rho \propto r^{-2}$, are expected to have only a minor effect. Although this is a simplification, it is reasonable given the uncertainty in the details of the ejection process, such as the mass loss history during the eruption. In particular, the inner and outer radii of the shell should not be sensitive to this assumption.

The velocity range of the absorption from the shell gives a direct measure of the radial extent of the shell. For a homologous expansion of the shell and for a single line the minimum, $V_{\rm blue}$ and maximum, $V_{\rm red}$, velocity of the absorption is given by 
\begin{equation}
\begin{split}
    V_{\rm blue} &  = -V_{\rm max} 
    \\
    V_{\rm red} &  = -V_{\rm max} \left(\frac{R_{\rm in}}{R_{\rm out}}\right)^\alpha \left[1 - \left(\frac{R_{\rm phot}}{R_{\rm in}}\right)^2\right]^{1/2}
\end{split}   
\label{eq:vel_br}
.\end{equation}
For homologous expansion, $\alpha=1$, and for a constant velocity shell, $\alpha=0$. Determining the maximum and minimum velocities of the absorption we can therefore estimate the $R_{\rm in}$ and $R_{\rm out}$ of the shell relative to the photospheric radius. We note that the minimum velocity of the shell is not $-V_{\rm red}$, but $V_{\rm in} = R_{\rm in}/t$. While we have here assumed homologous expansion, a shell with constant velocity only differs marginally in a slightly lower velocity unless the shell is very broad. The absorption line profile, however, differs from the homologous, lacking a flat bottom at the minimum absorption and instead being more V-shaped \citep[see e.g., Figs.~4 and 5 in][]{Fransson1984}. The observed absorption profiles for \xgc\ (Fig. \ref{fig:mgii}) argue against a constant velocity shell, but is compatible with a homologous expansion. This is also the case for \xga \ (Fig. \ref{fig:mgii}) and SN\,2018ibb (see Sect.~\ref{sec:mgii_comp}), although the S/N makes this conclusion marginal.  

For homologous expansion, Eq.~\ref{eq:vel_br}  can be inverted to show that 
\begin{equation}
\frac{R_{\rm in}}{R_{\rm out}} =   
\left[ \left(\frac{V_{\rm red}}{V_{\rm blue}}\right)^2 + \left(\frac{R_{\rm phot}}{R_{\rm out}}\right)^2 \right]^{1/2}
\label{eq:rinout}
.\end{equation}
The relative thickness of the shell, and therefore the depth of the absorption, decreases as the radius of the shell increases. The depth of the absorption therefore provides a constraint on the radius of the shell.

Figure~\ref{fig:MgII_modelling} shows the observed data for \xga\ and \xgc\ around the 2800~\AA\ region along with the modeled \ion{Mg}{II} line profiles. For \xga\  there is good agreement between the model and the observations for $R_{\rm in} = 2.96 \ R_{\rm phot}$ and $R_{\rm out} = 3.00 \ R_{\rm phot}$, where $R_{\rm phot}$ is the photospheric radius, and $V_{\rm max} = 4275~\rm km~s^{-1}$. Using the blackbody radius we derived in Sect.~\ref{sec:blackbody} from the given observed spectrum, we estimate that the CSM shell is located at $ 1.29 \pm 0.01 \times 10^{16}~\rm cm$ and extends to $1.31 \pm 0.01 \times 10^{16}~\rm cm$. The emission for the CSM shell is weak and consistent with the SN continuum. Given the derived distance of the shell and the estimated maximum velocity of $4275$~km~s$^{-1}$, the shell was expelled $10.5^{+0.3}_{-0.2}$ months before the core collapse, assuming a constant shell velocity as is seen in the case of SN\,2018ibb.

In \xgc\ the best-fit model is for a CSM shell with properties of $R_{\rm in} = 1.8 \ R_{\rm phot}$, $R_{\rm out} = 2.0 \ R_{\rm phot}$,  $V_{\rm max} = 4400~\rm km~s^{-1}$. This leads to a shell extending from $7.96 \pm 0.03  \times 10^{15}\rm cm$ to $8.84 \pm 0.03 \times 10^{15}~\rm cm$. Contrary to \xga, the emission from the CSM shell in \xgc\ is well above the SN continuum (dashed red line) and contributes significantly to the shape of the output spectra around the 2800~\AA\ region. This is a result of the considerably broader CSM shell, which scatter a larger fraction of the photons outside the absorption lines. Utilizing the maximum velocity of $4400~\rm km~s^{-1}$ calculated by the absorption minima of the broad \ion{Mg}{II} lines and assuming that the shell has not been decelerated, we estimate the time of CSM ejection to be $5.1^{+0.3}_{-0.4}$ months before the SN explosion. We caution that the error bars in the time of ejection reflect only statistical uncertainties and do not account for any systematic errors that may arise from uncertainties in the explosion date. However, uncertainties of a few days in the explosion date will not affect significantly the results.

The relative depth of the blue and red \ion{Mg}{II} doublet in \xga\ indicate an optical depth of $\tau \approx 0.5$. Using $\Delta r = (V_{\rm out}-V_{\rm in}) t$ for a homologous shell one finds for the column density of \ion{Mg}{II}, $N=n_1 \Delta r$, 
\begin{equation}
    N=2.2 \times 10^{14} \left(\frac{V_{\rm max}}{10^3 ~\rm km~s^{-1}}\right) \left(1 - \frac{R_{\rm in}}{R_{\rm out}}\right) \tau \  \ \rm cm^{-2} \ ,
\label{eq:density}
\end{equation}  
which is also valid for large $\tau$. For \xga, we find $N(\ion{Mg}{II}) \sim 4.7 \times 10^{12} \rm cm^{-2}$. The uncertainty in $\tau$,
 and therefore $N(\ion{Mg}{II})$, is at least a factor of two.  In the case of \xgc \ the lines are saturated with $\tau > 5$ and we only find a lower limit of $N(\ion{Mg}{II})  > 4.7 \times 10^{14} \rm cm^{-2}$.

In both \xga\ and \xgc, the X-shooter spectra obtained at later epochs are very noisy in the UV part of the spectrum that we are interested in and any potential \ion{Mg}{II} lines from the CSM shell cannot be resolved. Thus, we are not able to track the evolution of the \ion{Mg}{II} line profiles, which could provide a hint about the geometry of the shell. We note that for this part of the analysis we used only the spectra obtained through our X-shooter program due to the high S/N and the particular sensitivity of the instrument in the bluer wavelengths.

\subsection{Comparison with similar hydrogen-poor superluminous supernovae} \label{sec:mgii_comp}

\begin{figure}[!ht]
   \centering
  \includegraphics[width=0.5\textwidth]{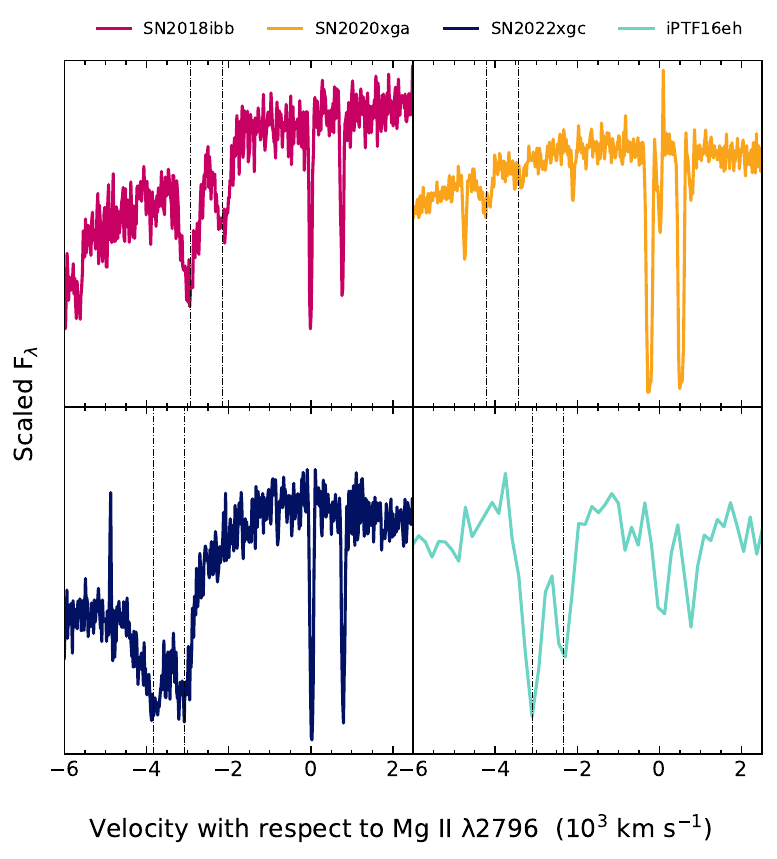}
   \caption{Comparison of the broad \ion{Mg}{II} lines originating from the CSM shells for \xga, \xgc, SN\,2018ibb and iPTF\,16eh. The spectra are corrected for MW extinction. The vertical black lines mark the velocity of the absorption minima of the secondary \ion{Mg}{II} lines.}
      \label{fig:mgii_comp}
\end{figure}

Figure~\ref{fig:mgii_comp} shows a zoom-in of the \ion{Mg}{II} region of all the four objects that have shown the high velocity CSM absorption systems. The \ion{Mg}{II} doublets originating from the CSM have various shapes and are shifted by various velocities reflecting the diversity of the CSM shells located around the SLSNe. In iPTF16eh the broad \ion{Mg}{II} lines are blueshifted by $3300~\rm km~s^{-1}$ and in SN\,2018ibb by $2918~\rm km~s^{-1}$ indicating shells moving somewhat slower but still comparable to the ones in \xga\ and \xgc. The \ion{Mg}{II} doublet in \xga\ is shallower compared to the rest of the objects, which supports the idea that a CSM in \xga\ is placed at larger radius with respect to its photosphere in comparison to the other three objects showing this signature in their spectra. On the other hand, the deep and blended \ion{Mg}{II} lines in \xgc\ and iPTF16eh points toward broader CSM shells in comparison to \xga\ and SN\,2018ibb.

\begin{figure}[!ht]
   \centering
  \includegraphics[width=0.55\textwidth]{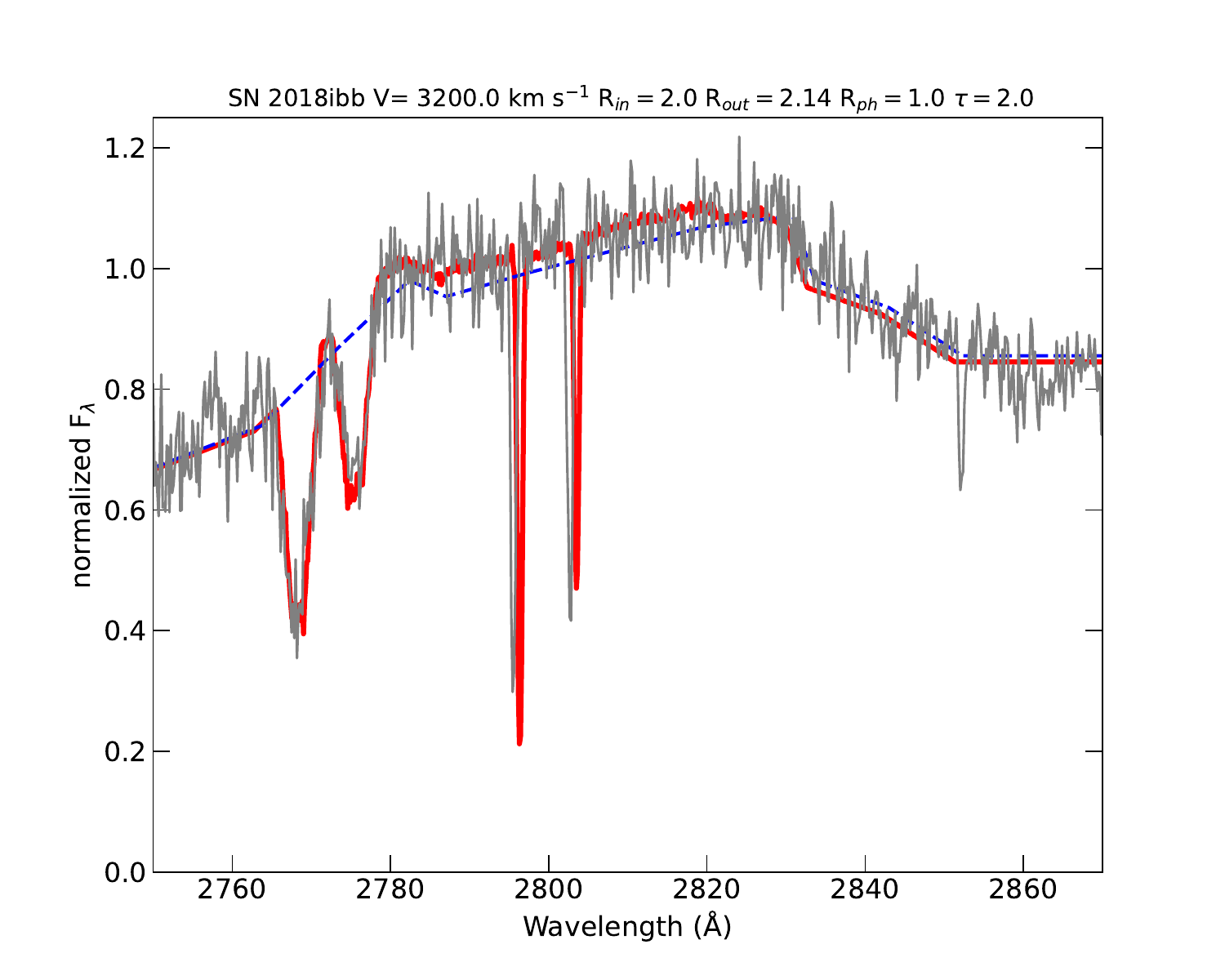}
   \caption{Modeling on the \ion{Mg}{II} doublets originated from the CSM shell (broad features) and the ISM of the host galaxy (narrow features) for SN\,2018ibb. The observed spectrum at the 2800~\AA\ region is presented in gray and the best model fit in red. The dashed blue line illustrates the SN continuum.}
      \label{fig:mgii_2018ibb}%
\end{figure}

Motivated by the above analysis for \xga\ and \xgc, we modeled the \ion{Mg}{II} doublet line profile of SN\,2018ibb \citep{Schulze2024} using the X-shooter spectrum at $+32.7$ days after maximum to quantify the properties of the CSM shell (see Fig.~\ref{fig:mgii_2018ibb}). The best fit model is for a CSM shell with $R_{\rm in} = 2.00 \ R_{\rm phot}$, $R_{\rm out} = 2.14 \ R_{\rm phot}$ and $V_{\rm max} = 3200~\rm km~s^{-1}$. Using the photospheric-radius values reported in \cite{Schulze2024}, we calculated a shell extended from $1.00 \pm 0.05 \times 10^{16}~\rm cm$ out to $1.07 \pm 0.05 \times 10^{16}~\rm cm$. Similarly to \xgc, the emission from the shell contributes significantly to the spectrum of SN\,2018ibb around the 2800~\AA\ region. Using  Eq.~\ref{eq:density} we estimated the lower limit to the column density of \ion{Mg}{II} in the CSM shell of SN\,2018ibb to be $N(\ion{Mg}{II}) \sim 9.2 \times 10^{13} \rm cm^{-2}$, in agreement with the value derived in \cite{Schulze2024}. Finally, we estimated the time of the CSM expelling to be $< 9$ months before the core collapse given the high uncertainty of the explosion date in SN\,2018ibb.

In iPTF16eh (\citealt{Lunnan2018}, see their Fig.~1), an intermediate-width \ion{Mg}{II} emission appeared at $\sim 2800$~\AA\ approximately 100 days after explosion and was persistent for more than 200 days. During this timescale, the line centroid was shifting from $-1600~\rm km~s^{-1}$ to $2900~\rm km~s^{-1}$, while the FWHM remained constant. This time- and wavelength-dependant emission line was associated with a resonance scattering light echo from a CSM shell. In the case of iPTF16eh, due to the low-resolution spectrum, the \ion{Mg}{II} scattered shell emission was modeled instead to estimate the size and the thickness of the CSM. \cite{Lunnan2018} find that the shell is located at $\sim 3.37 \times 10^{17}~\rm cm$, extended out to $\sim 3.55 \times 10^{17}~\rm cm$ and was expelled $\sim 30$ years before explosion. The CSM radius of iPTF16eh is almost an order of magnitude higher than the one derived from the modeling of \xga\ and \xgc. The evolution of the emerging \ion{Mg}{II} emission line in iPTF16eh was used as diagnostic for the geometry of the shell, which according to \cite{Lunnan2018} is found to be roughly spherical. Motivated by this study, \cite{Schulze2024} analyzed the spectra of SN\,2018ibb between $+230$ and $+378$ days after the peak and detected \ion{Mg}{II} in emission. However, due to heavy rebinning of the data, it is uncertain whether this emission line is connected with light echo from the CSM shell. In \xga\ and \xgc, we do not observe any variable \ion{Mg}{II} in emission; thus, we cannot derive any conclusion about the geometry of the shell.

\section{Host galaxy properties} \label{sec:host_galaxy}

\subsection{Host galaxy of \xga}

In Fig.~\ref{fig:redshift} (left panel), two \ion{Mg}{II} and \ion{Fe}{II} absorption systems are resolved in the host galaxy of \xga. As is discussed in Sect.~\ref{sec:redshift}, the stronger system is at $z = 0.4283$ and the weaker one at $z = 0.4296$, with a velocity separation of $\sim 260~\rm km~s^{-1}$. These two absorbing systems could arise either as a consequence of gas cloud motions (infall or outflows of gas) in the host galaxy or from a neighboring galaxy intervening in the line of sight \citep[e.g.,][]{Ledoux2006,Chen2012,Moller2013,Friis2015}. Since the emission lines from the star-forming regions are centered on a redshift in between the absorption components ($z = 0.4287$) and the velocity separation of $\sim 260~\rm km~s^{-1}$ is the typical difference in rotational velocity in a galaxy \citep{Galbany2016}, the scenario of mapping different regions of a single host galaxy is more likely \citep{Friis2015}.

The determination of the SN redshift in this case in not straightforward. \cite{Vreeswijk2014} showed that the absorbing gas responsible for the narrow lines in the spectrum of the SLSN iPTF\,13ajg was produced by gas in the ISM located at least 50 pc from the SN. In addition, \cite{Friis2015}, which presented a similar picture to \xga\ in terms of the lines of the GRB 121024A host galaxy, found that the absorbing clouds are not probing the actual GRB environment, but rather gas that has been photoionized by the GRB out to hundreds of parsecs. Although there are some distinctions between GRB and SLSN host galaxies \citep[e.g.,][]{Vreeswijk2014,Lunnan2014,Orum2020}, both are tracers of star formation, and therefore the assumption that the SN redshift is same as the redshift of the emission lines would not be unreasonable. Alternatively, to break the redshift degeneracy we would need high-resolution imaging and spectroscopy of the host galaxy to resolve the morphology of the galaxy and to infer the SN location and measure the redshift of that location. However, since analyzing the host environment of \xga\ is out of the scope of this paper and the redshift values are different only in the third digit, we take as the redshift of the SN the highest value. We note that considering the redshift of the emission lines or the strong \ion{Mg}{II} system would not affect our analysis and would not change significantly the results inferred from the CSM modeling.

\subsection{Stellar population synthesis modeling}

\begin{figure}[!ht]
   \centering
  \includegraphics[width=0.5\textwidth]{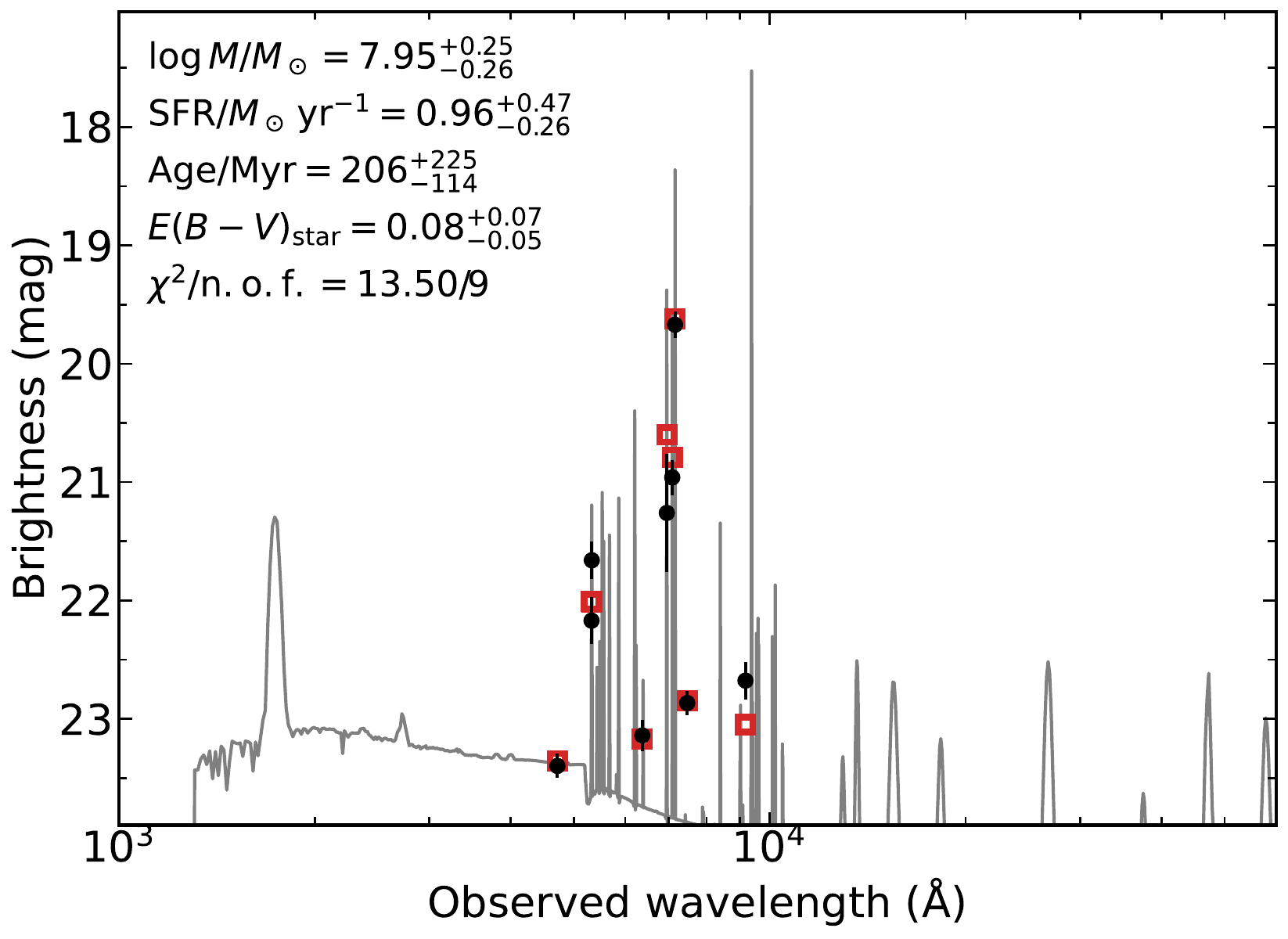}
   \caption{Spectral energy distribution of the host galaxy of SN\,2020xga (black data points). The solid line displays the best-fitting model of the SED. The red squares represent the model-predicted magnitudes. The fitting parameters are shown in the upper-left corner. The abbreviation `n.o.f.' stands for the number of filters.}
      \label{fig:sed}%
\end{figure}

To infer the mass and SFR of the host of \xga, we modeled the observed SED built from the broadband photometry (Table~\ref{tab:phot:host}) and the measured emission lines (Table~\ref{tab:line_fluxes}) with the software package \program{Prospector} \citep{Johnson2021a} version 1.1.\footnote{\program{Prospector} uses the \program{Flexible Stellar Population Synthesis} [\program{FSPS}] code \citep{Conroy2009a} to generate the underlying physical model and \program{python-fsps} \citep{ForemanMackey2014a} to interface with \program{FSPS} in \program{python}. The \program{FSPS} code also accounts for the contribution from the diffuse gas based on the \program{Cloudy} models from \citep{Byler2017a}. We use the dynamic nested sampling package \program{dynesty} \citep{Speagle2020a} to sample the posterior probability.} We assumed a Chabrier initial mass function (IMF; \citealt{Chabrier2003a}) and approximated the star formation history (SFH) by a linearly increasing SFH at early times followed by an exponential decline at late times (functional form $t \times \exp\left(-t/t_{1/e}\right)$, where $t$ is the age of the SFH episode and $t_{1/e}$ is the $e$-folding timescale). The model was attenuated with the Calzetti model \citep{Calzetti2000a}. The priors of the model parameters were set to be identical to the ones used by \citet{Schulze2021a}. The
observed SED is adequately described by a galaxy model with a stellar mass of log $M_{*}/ M_{\odot} = 7.95^{+0.25}_{-0.26}$ and SFR
of $0.96^{+0.47}_{-0.26}~\rm M_{\odot}~{\rm yr}^{-1}$ (gray curve in Fig.~\ref{fig:sed}),
leading to a specific SFR of $10^{-8}~\rm yr^{-1}$. The values derived from the modeling of \xga\ are consistent with what is seen in the host galaxies of other SLSNe-I \citep{Perley2016,Angus2019,Schulze2021a}.

\subsection{Emission line diagnostics}

Using the narrow emission lines from the galaxy present in the X-shooter spectra of \xga\ at $+37.8$ days and \xgc\ at $+21.9$ days, we can derive some properties of the two SLSN hosts regarding their metallicity, SFR, and host galaxy extinction. After calibrating the spectra to the photometry and correcting for MW extinction, we measured the fluxes of the emission lines utilizing the python package \program{LiMe} \citep{Fernandez2024}. The resulting flux values are listed in Table~\ref{tab:line_fluxes}. 

\begin{table} [!ht]
\centering
\small
\caption{Observed host galaxy emission line fluxes of \xga\ and \xgc.}
\label{tab:line_fluxes}
\begin{tabular}{ccc}
\hline
\hline

 & SN\,2020xga & SN\,2022xgc\\
\hline
Line         & Flux   & Flux    \\
           &  ($10^{-17}~{\rm erg~s}^{-1}{\rm cm}^{-2}$) & ($10^{-17}~{\rm erg~s}^{-1}{\rm cm}^{-2}$) \\ \hline
$[$\ion{S}{II}$]~\lambda6732$ & $0.91   \pm 0.27$ & -- \\
$[$\ion{S}{II}$]~\lambda6718$ & $0.64   \pm 0.23$ ($<0.69$) & -- \\
$[$\ion{N}{II}$]~\lambda6584$  &$ < 1.23$ &   -- \\ 
H$\alpha~\lambda6563$ & $4.29   \pm 0.48$ & $0.37 \pm 0.12$ \\
$[$\ion{O}{III}$]~\lambda5007$ & $7.42\pm 0.30$ &  -- \\ 
$[$\ion{O}{III}$]~\lambda4959$ & $2.31 \pm 0.23$&  --  \\
H$\beta~\lambda4861$& $ 1.87  \pm 0.70$ ($<2.1$)  & -- \\
$[$\ion{O}{II}$]~\lambda3729$ & $3.06 \pm 0.30$ &  --  \\
$[$\ion{O}{II}$]~\lambda3727$ & $1.91 \pm 0.28$ &  --  \\ \hline
\end{tabular}
\tablefoot{The fluxes are corrected for MW extinction. For lines detected with a significance of < 3$\sigma$, we also report the < 3$\sigma$ upper limits.}
\end{table}

We used the Balmer decrement to measure the host galaxy extinction of \xga\, and found a value of ${\rm H}\alpha / {\rm H}\beta = 2.3 \pm 0.9$, which is consistent within the uncertainty with the theoretical ratio of 2.87 for no extinction (assuming Case B recombination and a temperature of 10,000~K; \citealt{Osterbrock2006}); thus we did not apply any host galaxy extinction correction to the \xga\ photometry. In the spectrum of \xgc\ only the H$\alpha$ line was detected, thus we cannot estimate the extinction of \xgc's host. 

The SFR can be derived from the luminosity of  H$\alpha$ emission line using the \cite{Kennicutt1998} relation and the relation from \cite{Madau2014} to convert from the Salpeter to the Chabrier IMF in the \cite{Kennicutt1998}. This gives a SFR of $0.24 \pm 0.03~\rm M_{\odot}~{\rm yr}^{-1}$ for \xga\ and $ 0.009 \pm 0.003 ~\rm M_{\odot}~yr^{-1}$ for \xgc\ assuming zero host extinction. The value for \xga\ is lower than what is inferred from the SED modeling but not unprecedented compared to what is seen among SLSN host galaxies, while the value for \xgc\ points toward a ultra-faint dwarf galaxy \citep[e.g.,][]{Perley2016,Angus2016}.

Since we do not detect the auroral line [\ion{O}{III}] $\lambda4363$ in the spectrum of \xga, we are limited to strong-line metallicity diagnostics, such as R23, O3N2, or N2 (\citealt{Pagel1979,Pettini2004}; see discussion in \citealt{Kewley2008}). The O3N2 and N2 indexes require the [\ion{N}{II}] line flux that evaded detection. Thus, we used the R23 index with the calibrations of \cite{Jiang2019}, and the non-detection of [\ion{N}{II}] to break the degeneracy. We calculated for \xga\ $12 + \log({\rm O/H}) = 7.96^{+0.22}_{-0.18}$~dex. Taking the solar value to be $12+\log({\rm O/H})_{\odot} = 8.69$~dex \citep{Asplund2021}, this corresponds to a metallicity $Z = 0.2~Z_{\odot}$. The metallicity of \xga\ is in agreement with the values reported for other SLSN-I host galaxies ($< 0.5~Z_{\odot}$; e.g., \citealt{Lunnan2014,Leloudas2015,Perley2016,Chen2017b,Schulze2018}). We do not derive any metallicity measurement for the host of \xgc\ due to the lack of detection of emission lines originating from the host galaxy in the spectrum of \xgc.

\section{Discussion} \label{sec:discussion}

\subsection{The origin of the circumstellar material}

In Sect.~\ref{sec:mgii}, we modeled the broad \ion{Mg}{II} lines in the spectra of \xgc\ and \xga\ originating from CSM shells located around the SNe. The question that we need to address is what process led to the ejection of the material that generated the CSM at this distance and these velocities. Our findings indicate that this material was expelled less than a year before the core collapse in the final stages of stellar evolution, supporting a scenario of eruptive mass loss, since line-driven winds have much longer timescales than the evolutionary timescales at the post-MS phase.

One scenario to explain the detected CSM shells is the giant eruptions of LBV stars. The best observed example is the “Great Eruption” of $\eta$ Carinae in 1843 in which $12 - 20~\rm M_{\odot}$ (or more; \citealt{Smith2003}) of material was ejected within a decade, moving at velocities between $650 - 6000~\rm km~s^{-1}$  \citep{Davidson2001,Currie2002,Smith2002,Smith2004,Smith2008}. The geometry of ejecta-nebulae in LBV eruptions can be very anisotropic, as is observed in $\eta$ Carinae. Although the causes of LBV eruptions are still unknown, several theories have been proposed \citep[e.g.,][]{Davidson1987,Owocki2004,Smith2003,Smith2006,Smith2008,Woosley2017,Akashi2020,Cheng2024}.

For iPTF16eh, \cite{Lunnan2018} did not exclude an LBV-like eruption as a possibility for the formation of the CSM shell located around the SN. However, the spherically detached shell seen in iPTF16eh is not consistent with the asymmetric CSM structure may be expected in this type of eruption (see Fig.~1 in \citealt{Smith2008}). Since in the spectra of \xgc\ and \xga\ the shell velocities of $\sim 4000~\rm km~s^{-1}$ are consistent with an LBV-related eruption and the geometry of the shells cannot be constrained (see Sect.~\ref{sec:mgii}), previous massive ejection such as that in an LBV cannot be ruled out.

The LBV eruptions have also been discussed that could be driven by the PPI mechanism \citep{Woosley2017}, which occurs for a He core in the mass range of $30 - 65~\rm M_{\odot}$ \citep{Woosley2007}. For the PPI, the primary parameter determining the duration between the shell ejections and the ultimate core collapse, as well as the expelled mass and their kinetic energy, is the mass of the He core $M_{\rm He}$ \citep{Woosley2017,Leung2019,Marchant2019}. The time interval could range from a few hours up to 10\,000~years. For $M_{\rm He} < 40 - 42~\rm M_{\odot}$, the energy released is insufficient to unbind significant amounts of material, resulting in multiple weak pulses \citep{Woosley2017,Renzo2020}, whereas for $M_{\rm He} > 50~\rm M_{\odot}$, the time delay between strong pulses becomes long, forming shells at distances greater than the photospheric radius of the SN \citep{Woosley2017}. We note that the PPI scenario is addressed in relation to the mechanism responsible for forming the observed CSM shells, rather than as a powering mechanism for \xga\ and \xgc\ (see discussion in Sect.~\ref{sec:power}).

For \xga, the CSM was estimated to be expelled $\sim$ 11 months before the SN explosion. This value corresponds to that of a pure \ion{He}{} core of $51~\rm M_{\odot}$ with no rotation and zero metallicity in the study of \cite{Woosley2017}. Exploring the models of \cite{Woosley2017} for the blue supergiant progenitors developed using the \program{KEPLER} code \citep{Weaver1978,Weaver1993,Woosley2002}, we found that the model best describing the observed CSM properties of \xga\ is the B105. Assuming that the material in the B105 model has been ejected in one strong pulse $\sim 2$ years before the core collapse, at the time of the detection, the shell is moving at $\sim 2700~\rm km~s^{-1}$ at a radius of about $1.8 \times 10^{16}~\rm cm$. The shell velocity derived from \cite{Woosley2017} model is substantially lower than our prediction, but its distance is comparable to our estimated value for \xga. The model B100, on the other hand, has slightly higher velocity, but at the moment of detection, the shell is four times closer to the SN than in our model. We note that the exploration of these models focuses on the properties of the CSM shell resulting from their ejection, and we do not intend to directly link these models to the light curve properties of \xga\ and \xgc.

The discrepancy between the observed properties of the CSM and the predicted CSM properties from \cite{Woosley2017} models may stem from our assumption that the material was ejected in a single pulse, which is unlikely for stars of these masses (see Table 1 in \citealt{Woosley2017} and \citealt{Marchant2019}). However, assuming that the models must have at least one pulse on a time-scale equivalent to the ejection time of the CSM shell we observed, we could set a lower limit on the stellar mass. Thus, in \xga\ given that the shell was ejected $\sim 11$ months before the core collapse, the mass of the progenitor should be $> 51~\rm M_{\odot}$ for zero metallicity, within this framework. The lower limit corresponds to the case in which the CSM was formed in the first pulses. An upper limit is harder to set because it is closely related to the energetics of the PPI, the number of pulses and the timescales between the pulses; however, if restricted by the PPI regime it must be $< 62~\rm M_{\odot}$.

A similar comparison can be done with the H-free low metallicity models developed by \cite{Marchant2019} with the \program{mesa} software \citep{Paxton2011,Paxton2013,Paxton2015,Paxton2018}. In this study, the limit goes down to $47~\rm M_{\odot}$ presumably due to the fact that the models of \cite{Marchant2019} consider mass loss by stellar winds. This value is consistent with the modeling of \cite{Renzo2020} (see their Table C1), and the shell velocities resulting from their simulations can be similar to the ones found in the spectrum of \xga. 

A comparable analysis for \xgc\ indicates that the predicted ejection timeframe of $\sim$ 5 months corresponds to the model with $> 50~\rm M_{\odot}$ of pure He core \citep{Woosley2017} for zero metallicity, and $> 46~\rm M_{\odot}$ when low metallicity is taken into consideration. However, given the variations in the shell velocities in \xga\ and \xgc\, and therefore in the energetics, the properties of the progenitor stars of \xga\ and \xgc\ are likely to be different. According to \cite{Renzo2020}, even a difference of only $0.2~\rm M_{\odot}$ across the models may result in distinct PPI characteristics.

As is discussed in \cite{Leung2019}, there are substantial quantitative differences for the same He core mass between the models of \cite{Woosley2017}, \cite{Marchant2019} and by extension \cite{Renzo2020}, including ejected masses and time intervals between the pulses that can be traced back to the treatment of shocks and convection. We note that none of the aforementioned models have been constructed to match the observable CSM properties of \xga\ and \xgc\ and that tuning is required to obtain  better estimates for the shell properties; however, a qualitative comparison could provide some limits for the mass of the progenitor star that undergoes PPI before the final SN explosion. The threshold for the He-core progenitor mass could be further reduced if rotation is considered, as the induced chemical mixing could enable the formation of the required He-core in lower-mass stars compared to nonrotating models \citep{Chatzopoulos2012,Woosley2017}.

The discussion of SN\,2018ibb in the context of PPI is challenging as it is considered the best candidate of a PISN \citep{Schulze2024}. PISN
can be the end fate for stars with \ion{He}{} cores between $65 - 130~\rm M_{\odot}$ \citep{Heger2002}, where the energy produced by the pair creation obliterates the entire star. There are no models of PI SNe that also show eruptions similar to the PPI SNe, although this needs further investigations. However, \cite{Schulze2024} supports the presence of a CSM shell surrounding SN\,2018ibb based on the spectroscopic signs of interaction between the ejecta and the shell. The presence of CSM in SN\,2018ibb was also supported by our modeling in Sect.~\ref{sec:mgii}. The formation of CSM shells has been discussed by \cite{Schulze2024} in the context of an LBV-like eruption analogous to the one in $\eta$ Carinae. 

To strengthen the hypothesis of eruptive mass loss in \xga\ and \xgc, we analyzed the available ATLAS and ZTF forced photometry between 400 days pre-explosion and the time of explosion, paying particular attention to the estimated timeframes of the mass ejections to see if there is any detection. In \xgc\, the time of eruption falls into the gap owing to the solar conjunction, therefore we cannot detect any light from the putative blast. In the case of \xga,  we were unable to find any meaningful detection at $\sim 10$ months before explosion, but merely upper limits, implying that the eruption was probably fainter than the detection limits. \cite{Strotjohann2021} searched systematically for precursor eruptions hundreds of days before the explosion in a sample of SNe at $z<0.14$. Their sample include one SLSN-II with $z \approx 0.20$ in which a precursor is detected; however, \cite{Strotjohann2021} mentioned that at this progenitor distance the activity is more likely associated with AGN rather than stellar flares. Thus, it is not surprising that we do not detect any activity in the light curve prior to the explosion of \xga\ and \xgc\ given the distances of these two objects ($z = 0.4296$ and $z = 0.3103$).

The analysis of the four SLSNe  that exhibit the broad \ion{Mg}{II} absorption system, revealing the existence of a CSM shell expelled less than a year before the core collapse, has demonstrated that determining the expelling mechanism is challenging because various mechanisms could result in the formation of CSM at such distances. One possible hint to distinguish the various eruptive mass loss mechanisms could be to constrain the geometry of the CSM shells. 
In the future, improved modeling of the very final moments of massive stars will be necessary to comprehend the star's late activity and the mechanisms that lead to the formation of the CSM shells.  

\subsection{When the ejecta will interact with the circumstellar material shell}

As is discussed throughout the paper, the second \ion{Mg}{II} absorption system in the spectra of \xga\ and \xgc\ is explained by resonance-line scattering of the SLSN background photospheric emission by a rapidly expanding CSM shell located at a few $10^{16}~\rm cm$. To calculate the time of interaction $t_{\rm int}$ of the ejecta with the fast moving material, we assume that the inner radius of the shell is equal to the distance covered by the outer layer of ejecta. This is given by

\begin{equation} \label{eq:t_int}
    t_{\rm int} = \frac{V_{\rm in}~t_{\rm erup}}{V_{\rm ej,max} - V_{\rm in}} 
,\end{equation}
where $V_{\rm ej,max}$ is the velocity of the outer layer of the ejecta, $t_{\rm erup}$ is the time of eruption before explosion, and $V_{\rm in}$ is the minimum velocity of the shell (see definition in Sect.~\ref{sec:mgii}). From the spectra of \xga, we measure a maximum ejecta velocity of $\sim 13000~\rm km~s^{-1}$ using the \ion{O}{II} lines. Using the predicted value for $R_{\rm in}$ in Sect.~\ref{sec:mgii} for the spectra at $-8.3$ days, we calculated the $V_{\rm in} = 4210~\rm km~s^{-1}$. Thus, from Eq.~\ref{eq:t_int} we computed that the ejecta will collide with the CSM $153^{+4}_{-3}$ days after explosion and $109^{+8}_{-9}$ days after maximum, provided that the rise of the light curve from the first light is $44^{+7}_{-8}$ days. We do not have any spectroscopic or photometric observation at the time of collision to detect any possible interaction signature in \xga.

Following the same technique for \xgc, and measuring a maximum velocity of $11500~\rm km~s^{-1}$ from the \ion{Fe}{II} triplet and $V_{\rm in} = 3960~\rm km~s^{-1}$, we estimated a collision time of $80^{+5}_{-6}$ days after the explosion. Given that the rise from the time of first light is $59^{+12}_{-9}$ days, the ejecta will interact with the CSM $21^{+10}_{-13}$ days after maximum. This suggests that when we obtained the X-shooter spectra $+21.9$ days after the peak, the ejecta had more likely already begun interacting with the CSM shell. In addition, in Fig.~\ref{fig:temperature} the blackbody radius of \xgc\ at $+55$ days after the peak is $1 \times 10^{16}~\rm cm$, similar to the radius of the CSM, strengthening the hypothesis that the ejecta have interacted with the CSM at the time of our observations. We do not see any emerging emission line in the spectra of \xgc\, that could have offered information on the CSM interaction; however, this could be due to the fact that the spectral lines are highly sensitive to the density profiles of the ejecta and the CSM \citep[e.g.,][]{Fransson1984,Chevalier2017,Chatzopoulos2012}. Furthermore, the polarimetry of \xgc\ does not reveal any evidence of asymmetry that may come from the interaction of the ejecta with the CSM shell, which we anticipate to have happened during the period of time covered by the polarimetry ($+26.1$ days and $+60.1$ days after the peak). The effect of the CSM interaction might start becoming visible at around $80$ post-peak days at which the light curve of \xgc\ (see Fig.~\ref{fig:LCs}) shows a potential flattening in the $gcr$ filters. We should highlight that the computed time of interaction for \xga\ and \xgc\ should be taken with caution since it is highly dependent on the (uncertain) maximum velocity and the explosion date. 

Motivated by the above discussion for \xga\ and \xgc, we conducted the same calculations for SN\,2018ibb to check whether our predicted values are consistent with the timescales stated in \cite{Schulze2024}. Using the maximum velocity reported in \cite{Schulze2024} of $12500~\rm km~s^{-1}$ and the estimated $R_{\rm in}$, we found that the ejecta of SN\,2018ibb is predicted to interact with the CSM $\sim 90$ days after explosion. The rise time of the SN\,2018ibb is $> 93$ days, implying the interaction occurred during the rise. The light curve of SN\,2018ibb shows bumps and undulations in various bands post peak. Furthermore, the [\ion{Ca}{II}] emission line, which was already present in the spectrum of SN\,2018ibb at $-1.4$ days, and the [\ion{O}{II}] and [\ion{O}{III}] emission lines appear approximately $+30$ days after maximum light, has been associated with the CSM shell. These timescales agree with our predicted interaction time. 

\subsection{The composition and mass of the circumstellar material shells}

The determination of the chemical composition of the CSM shells in \xga\ and \xgc\ is hampered by a lack of observations. In \xga\, we do not have observational data at our estimated time of interaction, while for \xgc\, even though we found that the ejecta had already started interacting with the CSM shell at the time we obtained the spectra, there are no spectroscopic interaction signatures such as narrow emission lines. This does not necessarily imply that the CSM of \xgc\ is H-, He-, and O-free; rather, the evidence of interaction has yet to be revealed. 

The \ion{Mg}{II} doublet lines that we observed from the CSM are one of the few resonance lines of abundant elements in the observed part of the spectrum along with the \ion{Ca}{II} $\lambda\lambda3968.5,4226.7$, the \ion{Na}{I} $\lambda\lambda5890.0,5895.9$ and the \ion{K}{I} $\lambda\lambda7664.9,7699.0$ \citep{Morton2003}. Due to the fact that \ion{Na}{I} and \ion{K}{I} have a low ionization potential, and Ca is less abundant and has a lower ionization potential than Mg, these lines are ionized by the SN continuum. Thus, it is not surprising that we observe only the \ion{Mg}{II} doublet from the CSM. The far-UV rest frame below $\sim 2000$~\AA\ contains a large number of resonance lines, as well as intercombination lines of abundant ions such as \ion{C}{II-IV}, \ion{N}{II-V}, \ion{O}{I-IV}, which could be used for a more constraining abundance determination. This would, however, require either observations with HST or a SLSN at considerably higher redshift.

Assuming that the CSM shell of \xga\ and \xgc\ are H-dominated and using the same method as in \cite{Lunnan2018}, we could put an upper limit on the H mass of the shell by utilizing the luminosity of H$\alpha$. We assumed that the CSM H$\alpha$ peak is blueshifted from the host H$\alpha$ by the velocity of the CSM measured from the \ion{Mg}{II} doublet and that the line profile of the CSM H$\alpha$ is similar to the estimated \ion{Mg}{II} emission one extended to the maximum velocity of the CSM \ion{Mg}{II} absorption. We found that $L_{\rm H\alpha} < 1.8 \times 10^{38}~\rm erg~s^{-1}$ and $L_{\rm H\alpha} < 1.2 \times 10^{39}~\rm erg~s^{-1}$ for \xga\ and \xgc\,, respectively. This yields a limit of $M_{\rm shell} < 0.03f^{0.5}~\rm M_{\odot}$ for \xga\ and $M_{\rm shell} < 0.02f^{0.5}~\rm M_{\odot}$ for \xgc\ where $f$ is the filling factor. These CSM masses are substantially lower than the values for ejected masses expected in \cite{Woosley2017}, \cite{Marchant2019}, and \cite{Renzo2020}, indicating that the shell is most likely not dominated by H and that the H envelope must have been lost before this eruption. This is also evident in SN\,2018ibb in which the absence of H and He lines throughout the entire spectral evolution of SN\,2018ibb and the existence of O lines suggests that the CSM shell must be O-dominated and any H and He must reside at much larger radii \citep{Schulze2024}.

\subsection{What powers \xga\ and \xgc} \label{sec:power} 

In Sect.~\ref{sec:lc_modeling}, we modeled the light curves of \xga\ and \xgc\ with the \program{redback} software, assuming they are powered by a magnetar. The modeling resulted in very rapidly spinning magnetars ($1.6$ and $0.9$~ms, respectively) with ejecta masses of $7-9~\rm M_{\odot}$, consistent with what is seen in SLSNe-I \citep{Chen2023b,Gomez2024}. However, the spin period in the case of \xga\ and \xgc\ approaches the mass-shedding limit \citep{Watts2016}, which suggests that in order to explain the very high luminosities observed in \xga\ and \xgc\ with the magnetar model, the magnetar must produce energy near to its upper limits without being destabilized. 

A remaining question is how a star more massive than $\sim 50~\rm M_{\odot}$ can explode with $\sim 9~\rm M_{\odot}$ of ejecta. The stars massive enough to satisfy the PPI conditions form Fe cores, which tend to collapse rather than explode \citep[e.g.,][]{Heger2002}. However, \cite{Woosley2007} discuss a possibility in which a massive star (with initial mass of $\sim 95~\rm M_{\odot}$) with mild rotation and magnetic torques may have enough angular momentum in its Fe core to form a NS with a period of $2~\rm ms$. Furthermore, \cite{Woosley2017} investigated the explosions of numerous PPI models, including the model of a $50~\rm M_{\odot}$ He-core, and discovered that the light curve produced is compatible with observations of some of the brightest SLSNe-I and therefore, in accordance with the observables of \xga\ and \xgc. In the case of the formation of a NS, \cite{Lunnan2018} argue that the magnetic field must be $>10^{15}~\rm G$ initially to allow the star to explode, but then it must decay to $<10^{14}~\rm G$ to power the light curve for longer timescales. This scenario would require a high level of fine-tuning in the rotation of the star since it is expected to have a key part in the star's final death and consequently in the SN explosion. 

Alternative scenarios include \xga\ and \xgc\ being PPISNe, or powered by fallback accretion into a black hole or CSM interaction. The observed radiated energies of $> 1.8 \times 10^{51}~\rm erg$ and $> 0.9 \times 10^{51}~\rm erg$ for \xga\ and \xgc, respectively, cannot be reproduced by the PPISN scenario because the maximum energy of the pure PPISN models is $\sim 5 \times 10^{50}~\rm erg$ unless there is a contribution from a magnetar \citep{Woosley2017}. The fallback scenario proposes that the star only partially explodes, resulting in a weak SN explosion and fallback accretion on the equatorial plane \citep[e.g.,][]{Dexter2013,Woosley2017,Moriya2018}. While the fallback material is accreted onto the black hole, the resulting outflows can come in the form of disk winds or jets and they may interact with previously ejected material, causing a significant impact in the SLSN-I light curves. We fit the light curves of \xga\ and \xgc\ with the fallback model from \cite{Moriya2018}. Assuming an 100\% accretion efficiency, we would need $5~\rm M_{\odot}$ of accreted mass to power both \xga\ and \xgc. Assuming a more realistic efficiency of $10^{-3}$ \citep{Moriya2018} would require an accreted mass of $5000~\rm M_{\odot}$, which we deem unrealistic. The efficiency could be on the order of 10\% if a jet is formed \citep{Gilkis2016}, implying an accretion mass of $50~\rm M_{\odot}$. This would be consistent with the ``missing’' ejecta in the PPI scenario of \xga\ and \xgc. However, there is no evidence of a jet in our data (see Sect.~\ref{sec:pola} for \xgc). Nevertheless, given the low ejecta mass inferred from the relatively short rise times of the light curves of \xga\ and \xgc, the fallback accretion scenario cannot be ruled out as a plausible powering mechanism.

Finally, we explored the scenario in which \xga\ and \xgc\ are powered by the interaction of the ejecta with CSM. We modeled the light curves using \program{redback} and the \texttt{CSM+Ni} \citep{Chatzopoulos2012} model, and found that the ejecta and CSM masses range from $1-3~\rm M_{\odot}$ and $35-50~\rm M_{\odot}$, respectively. While these findings may seem unphysical, they are not unprecedented, as semianalytic models \citep{Chatzopoulos2012} and hydrodynamic simulations \citep{Moriya2013, Moriya2018, Sorokina2016} have shown to yield conflicting results. The development of consistent radiation-hydrodynamic simulations is necessary to properly model the light curves of SLSNe-I, similar to what has be done in \cite{Dessart2015}. However, considering that these stars expel material prior to their explosion, and particularly in the case of \xgc\, the presence of the three early $r$-band data points, along with the fact that the ejecta is estimated to have interacted with the CSM during our observations, we conclude that CSM interaction could contribute to the light curves of \xga\ and \xgc.

Determining the powering mechanism of the light curves of SLSNe-I is still an open question and especially in the high mass ranges in which the “explodability” is highly uncertain (e.g., \citealt{Ertl2016,Coughlin2018}; see discussion in \citealt{Renzo2020}). Our observations put limitations on any progenitor model by requiring a significant mass ejections less than a few years before explosion.

\section{Conclusions} \label{sec:conclusions}

In this work, we provide optical observations of the H-poor SLSNe-I \xga\ and \xgc\ covering $-44$ to $+59$ and $-59$ to $+110$ days after maximum light, respectively. Our key findings are as follows:

\begin{itemize}
    \item \xga\ and \xgc\ are among the most luminous SLSNe-I, with $M_{\rm g} = -22.3~\rm mag$ and $M_{\rm g} = -22.0~\rm mag$, respectively.
    \item The spectra of \xga\ and \xgc\ show a second \ion{Mg}{II} absorption system coming from the CSM surrounding the SNe, but are otherwise similar to the spectra of normal SLSNe-I.
    \item The modeling of the narrow \ion{Mg}{II} results in a CSM shell for \xga\ located at $\sim 1.3 \times 10^{16}~\rm cm$ and for \xgc\ located at $\sim 0.8 \times 10^{16}~\rm cm$ moving at maximum velocity of $4275~\rm km~s^{-1}$ and $4400~\rm km~s^{-1}$, respectively.
    \item The CSM shells of \xga\ and \xgc\ were expelled less than a year before the core collapse as a result of eruptive-mass loss in the form of LBV-like eruptions or PPI.
    \item The light curve modeling of \xga\ and \xgc\ is consistent with magnetar-powered SNe with an ejecta mass of about $\sim 7-9~\rm M_{\odot}$ with magnetars near the mass-shedding limit.
    \item The PPI scenario suggests He cores $>50~\rm M_{\odot}$, which is incompatible with the findings of the light curve modeling; hence, alternatives such as fallback accretion and CSM interaction are discussed.
    \item The host galaxy properties of \xga\  and \xgc\ are similar to the ones of typical SLSN-I host galaxies and point toward dwarf galaxies. 
    
\end{itemize}

In this paper, we focus on the extensive analysis of two objects that show the spectroscopic signature of the presence of a CSM shell expelled less than a year before the core collapse. The analysis of the whole high-quality X-shooter spectral sample constraining the fraction of SLSNe that show evidence of eruptive mass loss and establishing observational limitations will be addressed in a future work. The discovery of these objects can provide an insight into the late stages of stellar evolution as well as a better knowledge of SLSN-I progenitors.

\section*{Data availability} \label{data}

The photometric data for \xga\ and \xgc\ are available in electronic form at the CDS, accessible via anonymous FTP at \href{cdsarc.u-strasbg.fr}{cdsarc.u-strasbg.fr} (130.79.128.5) or through \href{http://cdsweb.u-strasbg.fr/cgi-bin/qcat?J/A+A/}{http://cdsweb.u-strasbg.fr/cgi-bin/qcat?J/A+A/}. All spectra of \xga\ and \xgc\ have been uploaded to the WISeREP\footnote{\url{https://www.wiserep.org}} archive \citep{Yaron2012a}. The corner plots of \xga\ and \xgc\ resulting from the \program{redback} modeling are available at  \href{https://zenodo.org/records/14565605}{https://zenodo.org/records/14565605}.

\bibliographystyle{aa}
\bibliography{references}
%
%

\begin{appendix}

\section{Acknowledgments}
Based on observations obtained with the Samuel Oschin Telescope 48-inch and the 60-inch Telescope at the Palomar Observatory as part of the Zwicky Transient Facility project. ZTF is supported by the National Science Foundation under Grants No. AST-1440341 and AST-2034437 and a collaboration including Caltech, IPAC, the Weizmann Institute of Science, the Oskar Klein Center at Stockholm University, the University of Maryland, Deutsches Elektronen-Synchrotron and Humboldt University, the TANGO Consortium of Taiwan, the University of Wisconsin at Milwaukee, Trinity College Dublin, Lawrence Livermore National Laboratories, IN2P3, University of Warwick, Ruhr University Bochum, Northwestern University and former partners the University of Washington, Los Alamos National Laboratories, and Lawrence Berkeley National Laboratories. Operations are conducted by COO, IPAC, and UW.
SED Machine is based upon work supported by the National Science Foundation under Grant No. 1106171.
The ZTF forced-photometry service was funded under the Heising-Simons Foundation grant \#12540303 (PI: Graham).
The Gordon and Betty Moore Foundation, through both the Data-Driven Investigator Program and a dedicated grant, provided critical funding for SkyPortal.
Based on observations collected at the European Organisation for Astronomical Research in the Southern Hemisphere, Chile, as part of ePESSTO+ (the advanced Public ESO Spectroscopic Survey for Transient Objects Survey). ePESSTO+ observations were obtained under ESO programs ID 106.216C and 108.220C.
Some of the observations with the Las Cumbres Observatory data have been obtained via OPTICON proposals and as part of the Global Supernova Project. The OPTICON project has received funding from the European Union's Horizon 2020 research and innovation programme under grant agreement No 730890.
This work has made use of data from the Asteroid Terrestrial-impact Last Alert System (ATLAS) project. ATLAS is primarily funded to search for near earth asteroids through NASA grants NN12AR55G, 80NSSC18K0284, and 80NSSC18K1575; byproducts of the NEO search include images and catalogs from the survey area. The ATLAS science products have been made possible through the contributions of the University of Hawaii Institute for Astronomy, the Queen’s University Belfast, the Space Telescope Science Institute, and the South African Astronomical Observatory.
Partially based on observations made with the Nordic Optical Telescope, owned in collaboration by the University of Turku and Aarhus University, and operated jointly by Aarhus University, the University of Turku and the University of Oslo, representing Denmark, Finland and Norway, the University of Iceland and Stockholm University at the Observatorio del Roque de los Muchachos, La Palma, Spain, of the Instituto de Astrofisica de Canarias.
The Liverpool Telescope is operated on the island of La Palma by Liverpool John Moores University in the Spanish Observatorio del Roque de los Muchachos of the Instituto de Astrofisica de Canarias with financial support from the UK Science and Technology Facilities Council. Based on observa- tions made with the Italian Telescopio Nazionale Galileo (TNG) operated on the island of La Palma by the Fundación Galileo Galilei of the INAF (Istituto Nazionale di Astrofisica) at the Spanish Observatorio del Roque de los Muchachos of the Instituto de Astrofisica de Canarias.

RL is supported by the European Research Council (ERC) under the European Union’s Horizon Europe research and innovation programme (grant agreement No. 10104229 - TransPIre).
FP acknowledges support from the Spanish Ministerio de Ciencia, Innovaci\'{o}n y Universidades (MICINN) under grant numbers PID2022-141915NB-C21.
SS is partially supported by LBNL Subcontract 7707915.
NS and AS are supported by the Knut and Alice Wallenberg foundation through the “Gravity Meets Light” project.
RKT is supported by the NKFIH/OTKA FK-134432 and the NKFIH/OTKA K-142534 grant of the National Research, Development and Innovation (NRDI) Office of Hungary.
MN is supported by the European Research Council (ERC) under the European Union’s Horizon 2020 research and innovation programme (grant agreement No.~948381) and by UK Space Agency Grant No.~ST/Y000692/1.
JPA is supported by ANID, Millennium Science Initiative, ICN12\_009.
T.E.M.B. acknowledges financial support from the Spanish Ministerio de Ciencia e Innovaci\'on (MCIN), the Agencia  Estatal de Investigaci\'on (AEI) 10.13039/501100011033, and the  European Union Next Generation EU/PRTR funds under the 2021 Juan de la Cierva program FJC2021-047124-I and the PID2020-115253GA-I00 HOSTFLOWS project, from Centro Superior de Investigaciones Cient\'ificas (CSIC) under the PIE project 20215AT016, and the program Unidad de Excelencia Mar\'ia de Maeztu CEX2020-001058-M.
MB and TP acknowledge the financial support from the Slovenian Research Agency (grants I0-0033, P1-0031, J1-8136, J1-2460 and Z1-1853) and the Young Researchers program. 
MK acknowledges financial support from MICINN (Spain) through the programme Juan de la Cierva-Incorporación [JC2022-049447-I] and from AGAUR, CSIC, MCIN and AEI 10.13039/501100011033 under projects PID2023-151307NB-I00, PIE 20215AT016, CEX2020-001058-M, and 2021-SGR-01270.
CPG acknowledges financial support from the Secretary of Universities
and Research (Government of Catalonia) and by the Horizon 2020 Research
and Innovation Programme of the European Union under the Marie
Sk\l{}odowska-Curie and the Beatriu de Pin\'os 2021 BP 00168 programme,
from the Spanish Ministerio de Ciencia e Innovaci\'on (MCIN) and the
Agencia Estatal de Investigaci\'on (AEI) 10.13039/501100011033 under the
PID2020-115253GA-I00 HOSTFLOWS project, and the program Unidad de
Excelencia Mar\'ia de Maeztu CEX2020-001058-M.
L.G. acknowledges financial support from AGAUR, CSIC, MCIN and AEI 10.13039/501100011033 under projects PID2023-151307NB-I00, PIE 20215AT016, CEX2020-001058-M, and 2021-SGR-01270.
T.-W.C. acknowledges the Yushan Fellow Program by the Ministry of Education, Taiwan for the financial support (MOE-111-YSFMS-0008-001-P1).
MF is supported by a Royal Society - Science Foundation Ireland University Research Fellowship.
This work makes use of data from the Las Cumbres Observatory global network of telescopes.  The LCO group is supported by NSF grants AST-1911151 and AST-1911225.
IPF acknowledges financial support from the Spanish Agencia
Estatal de Investigaci\'{o} n del Ministerio de Ciencia e Innovaci\'{o} n
(AEI–MCINN) under grant PID2022-137779OB-C44.
AS acknowledges the Warwick Astrophysics PhD prize scholarship made possible thanks to a generous philanthropic donation.

\onecolumn
    \section{Spectroscopic data} \label{app:spectra}

\begin{table*}[!ht]

\centering
\caption{\xga\ spectroscopic observations.}\label{tab:2020xga_spectra}
\begin{tabular*}{.98\linewidth}[!ht]{@{\extracolsep{\fill}}ccccccc}
\hline
\hline
UT date & MJD & Phase\tablefootmark{\scriptsize a}  & Telescope + & Exposure & Disperser & Wavelength range\\
 & (days) & (days) & Instrument & (s) & & (\AA)\\
\hline
20201106 & 59159.6 & -9 & NTT + EFOSC2 & 1500 & Gr\#13 & 3650 – 9250 \\
20201107 & 59160.6 & -8.3 & VLT + X-shooter & 3600 & -- & 3000 – 24800  \\
20201116 & 59169.8 & -1.9 & NTT + EFOSC2 & 2700 & Gr\#11 + Gr\#16 & 3345 – 9995\\
20201117 & 59170.7 & -1.3 &      NTT + EFOSC2 &  2700 & Gr\#11 & 3345 – 7470\\
20201207 & 59190.6 & +12.7 &      NTT + EFOSC2 &  5400 & Gr\#13 & 3650 – 9250\\
20201230 & 59213.6 & +28.8 &      NTT + EFOSC2 &  5400 & Gr\#13 & 3650 – 9250\\
20210107 & 59221.2 & +34.1  & P200 + DBSP & 3600  & 600/316 & 3500 – 10000  \\
20210110 & 59224.6 & +36.4 & VLT + X-shooter & 4800 & -- & 3000 – 24800  \\
20210114 & 59228.6 & +39.3 & VLT + X-shooter & 4800 & -- & 3000 – 24800  \\
\hline
\end{tabular*}
\tablefoot{\tablefoottext{a}{Rest-frame relative to the $g$-band maximum (MJD 59172.5).}}
\end{table*}

\begin{table*}[!ht]

\centering
\caption{SN\,2022xgc spectroscopic observations.}\label{tab:2022xgc_spectra}
\begin{tabular*}{.98\linewidth}[!ht]{@{\extracolsep{\fill}}ccccccc}
\hline
\hline
UT date & MJD & Phase\tablefootmark{\scriptsize a}  & Telescope + & Exposure & Disperser & Wavelength range\\
 & (days) & & Instrument & (s) & & (\AA)\\
\hline
20221113 & 59896.0 & -4.5 & NOT + ALFOSC & 3344 & Gr\#4 & 3900 -- 9600 \\
20221114 &  59897.8 & -3.7  & P60 + SEDm & 2250  & -- & 3950 -- 9200 \\
20221118 &  59901.3 & -0.7 & P60 + SEDm & 2250  & -- & 3950 -- 9200 \\ 
20221122 &  59905.0 & +2.4 & Lick + KAST & 3600 & 600/4310 + 300/7500 & 3500 – 10500 \\
20221201 & 59914.8 & +9.8 & NTT + EFOSC2 & 900 & Gr\#13 & 3650 – 9250 \\
20221214 & 59927.0  & +19.2 & NOT + ALFOSC & 3600 & Gr\#4 & 3900 -- 9600 \\
20221207 & 59930.6 & +21.9 & VLT + X-shooter & 3600 & -- & 3000 – 24800  \\
20221222 & 59935.7 & +25.8 &      NTT + EFOSC2 &  2700 & Gr\#13 & 3650 – 9250 \\
20230113 & 59957.8 & +42.6 &      NTT + EFOSC2 &  2700 & Gr\#11 + Gr\#16 & 3345 – 9995\\
20230117 & 59961.0 & +45.1 & NOT + ALFOSC & 4000  & Gr\#4 & 3900 -- 9600 \\
20230129 & 59973.6 & +54.7 &    NTT + EFOSC2 &  2700 & Gr\#11 & 3345 – 7470\\
20230209 & 59987.6 & +61.3 &      NTT + EFOSC2 &  2700 & Gr\#11 + Gr\#16 & 3345 – 9995\\
20230212 & 59987.6 & +65.4 & VLT + X-shooter & 4800 & -- & 3000 – 24800  \\
20230221 & 59996.7 & +72.3 &      NTT + EFOSC2 &  2700 & Gr\#11 + Gr\#16 & 3345 – 9995\\
20230319 & 60022.5 & +92.1 & VLT + X-shooter & 3600 & -- & 3000 – 24800  \\
\hline
\end{tabular*}
\tablefoot{\tablefoottext{a}{Rest-frame relative to the rest-frame g-band maximum (MJD 59901.9).}}
\end{table*}

\FloatBarrier

\onecolumn
\section{Polarimetry data} \label{app:pol}

The log of the polarimetry obtained on \xgc\ using ALFOSC on the NOT discussed in Sect.~\ref{pol_data} is presented in Table~\ref{tab:log_polarimetry}. The results discussed in Sects.~\ref{pol_data} and ~\ref{sec:pola} are given in Table~\ref{tab:pol_results}.

\begin{table*} [!ht]
        \centering
        \caption{Observations log of the imaging polarimetry observations. N.A. means Not Available.}
        \label{tab:log_polarimetry}
        \begin{tabular}{lllcc} 
          \hline
          UT Time  & Object & Exp. Time & Filter & Seeing \\
           &  &  [s] & & [$\arcsec$] \\
          \hline
          2022-12-15 00:00:24  &  \xgc\       & $ 2 \times (4 \times 420) $ & V & N.A. \\          
          2022-12-15 01:03:00  &  \xgc\       & $ 2 \times (4 \times 420) $ & R & N.A. \\          
          2022-12-14 22:03:00  &  HD\,14069   & $  2 \times (4 \times 1)  $ & V & N.A. \\
          2022-12-14 22:05:48  &  HD\,14069   & $  2 \times (4 \times 1)  $ & R & N.A.  \\
          2022-12-14 22:10:11  &  HD\,251204  & $  4 \times (4 \times 3)  $ & V & N.A.  \\
          2022-12-14 22:12:44 &  HD\,251204   & $  4 \times (4 \times 3)  $ & R & N.A.  \\           
          \hline
          2023-01-18 00:57:12 &  \xgc\        & $ 4 \times (4 \times 200) $ & V & 1.1 \\          
          2023-01-18 00:00:02 &  \xgc\        & $ 4 \times (4 \times 200) $ & R & 1.0 \\          
          2023-01-17 19:43:50 &  HD\,14069    & $ 2 \times (4 \times 2)   $ & V & 1.2 \\
          2023-01-17 19:46:05 &  HD\,14069    & $ 2 \times (4 \times 2)   $ & R & 1.2 \\
          2023-01-17 19:49:44 &  BD+59\,389   & $ 2 \times (4 \times 1.2) $ & V & 1.2 \\
          2023-01-17 19:51:41 &  BD+59\,389   & $ 2 \times (4 \times 1.5) $ & R & 1.2 \\           
          \hline
        \end{tabular}
\end{table*}

\begin{table*}[!ht]
\centering
\caption{Polarimetry results on SN\,2022xgc obtained in the R-band and V-band Bessel filters. 
        $^{\rm(a)}$: Stokes parameters, $\overline{Q}$ and $\overline{U}$, directly obtained
          from the ALFOSC data frames Extraordinary and Ordinary images without applying
          any further corrections. $^{\rm(b)}$: instrumental
          polarization corrected. $^{\rm(c)}$: instrumental
          polarization corrected and polarization angle corrected. 
          $^{\rm(d)}$: instrumental
          polarization corrected, polarization angle corrected and bias corrected.}     \label{tab:pol_results}

\begin{tabular}{lllccccccc} 
\hline
\hline
Date & Source &filter & $\overline{Q}^{\rm (a)}$ & $\overline{U}^{\rm(a)}$ & $P [\%]^{\rm(a)} $ & $P [\%]^{\rm(b,c)} $ & $\theta [^{\circ}]^{\rm(c)}$ & $P_{\rm deb} [\%]^{\rm(d)}$ \\
\hline
2022-12-14 & HD 14069 & R      & 0.06  & 0.12  &  0.13 $\pm$ 0.04  & --  & -- & --  \\
2022-12-14        & HD 251204 & R   & -2.53 & 4.10  & 4.82 $\pm$ 0.03            &  4.75 $\pm$ 0.05 & -- & --  \\
2022-12-14        & SN\,2022xgc & R & 0.11  & -0.07 & 0.13 $\pm$ 0.10 & 0.19 $\pm$ 0.11 &  46.85 $\pm$ 16.75 & 0.13 $\pm$ 0.11   \\
2022-12-14        & HD 14069 &      V & 0.02 & 0.16  &  0.16 $\pm$ 0.06  & -- & --  & --   \\
2022-12-14       & HD 251204 &   V & -3.12 & 3.89 & 4.99 $\pm$ 0.05            & 4.87 $\pm$ 0.08  & -- & --  \\
2022-12-14        & SN\,2022xgc & V & 0.13 & -0.03 & 0.14 $\pm$ 0.12            & 0.22 $\pm$ 0.14  &  53.05 $\pm$ 17.63 & 0.13 $\pm$ 0.14  \\
2023-01-17 & HD 14069 &      R & 0.09 & 0.06  &  0.11 $\pm$ 0.05 & -- & --  & --  \\
2023-01-17        & BD$+$59\,389 & R  & 6.39 & 1.40  & 6.55 $\pm$ 0.04 &  6.44 $\pm$ 0.06 & -- & --  \\
2023-01-17        & SN\,2022xgc & R & 0.11 & 0.16  & 0.19 $\pm$ 0.10  & 0.10 $\pm$ 0.11  & 132.89 $\pm$ 33.47 & 0.10 $\pm$ 0.11 \\
2023-01-17        & HD 14069 &     V  & -0.07 & 0.16 &  0.17 $\pm$ 0.04 & -- & --  & --  \\
2023-01-17        & BD$+$59\,389 & V  & 6.23 & 2.22  & 6.62 $\pm$ 0.08 &  6.63 $\pm$ 0.09 & -- & --  \\
2023-01-17        & SN\,2022xgc & V & 0.34 & 0.34  & 0.48 $\pm$ 0.18  & 0.44 $\pm$ 0.18 &  100.92 $\pm$ 11.95 & 0.37 $\pm$ 0.18  \\
\hline
\end{tabular}
\end{table*}

\FloatBarrier

\onecolumn

\section{SYNOW+ results} \label{app:synow}

\begin{figure*} [!ht]
\centering
\includegraphics[width=8cm]{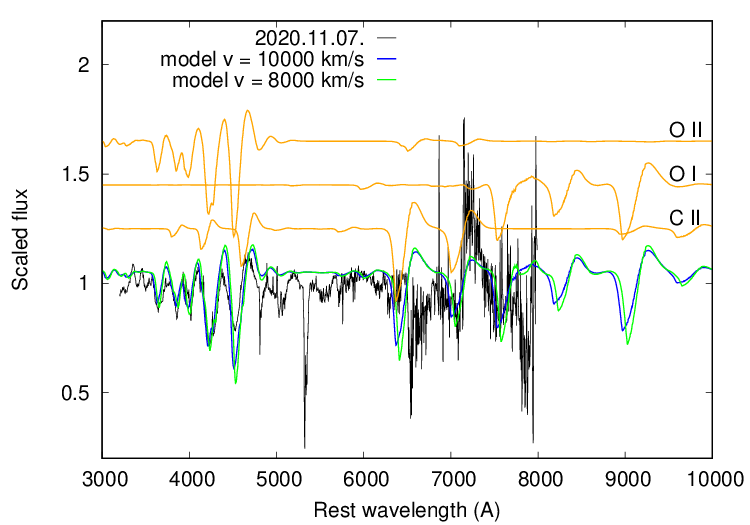}
\includegraphics[width=8cm]{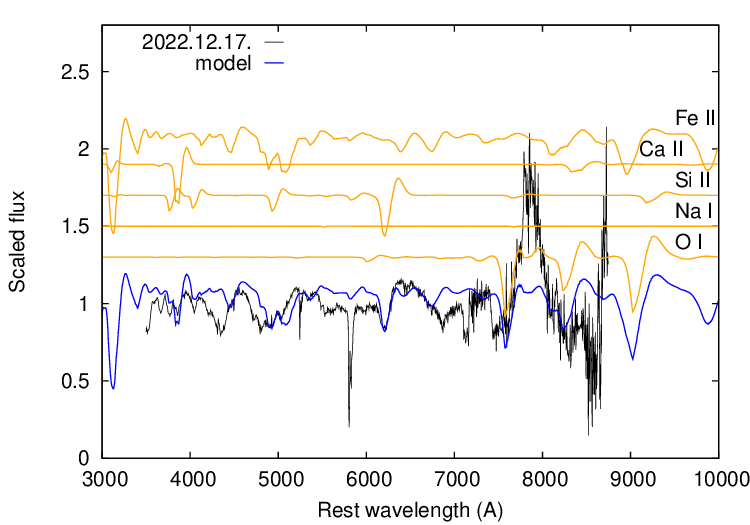}
\caption{The SYN++ modeling of one of the spectra of SN~2020xga (left) and SN~2022xgc (right). The observed spectra (black lines) are corrected for interstellar reddening and redshift, and continuum-normalized for clarification. The best-fit model obtained using SYN++ is shown with blue and green colors, while the contribution of the single ions to the model are plotted with orange, and shifted vertically.}
\label{fig:syn++}
\end{figure*}

\begin{table*}[!ht]

\centering
\caption{Best-fit parameter values of the SYN++ modeling of the spectra of SN~2020xga and SN~2022xgc. The fit parameters are the following: the velocity at the photosphere ($v_{\rm phot}$ [$10^3$ km s$^{-1}$]),  the photospheric temperature ( $T_{\rm phot}$ [1000 K]), the optical depth of each ion  ($\log\tau$ [--]), the inner velocity of the line forming region ($v_{\rm min}$ [$10^3$ km s$^{-1}$]), the outer velocity of the line forming region ($v_{\rm max}$ [$10^3$ km s$^{-1}$]), the scale hight of the optical depth (aux [$10^3$ km s$^{-1}$]), and the excitation temperature of each ion ($T_{\rm exc}$ [1000 K]). }\label{tab:syn}
\begin{tabular*}{.98\linewidth}[!ht]{@{\extracolsep{\fill}}lccccccc}
\hline
\hline
Ions & C II &  O I &  O II & Na I & Si II  & Ca II & Fe II \\
\hline
  \multicolumn{8}{c}{SN~2020xga (2020.11.07)  $v_{\rm phot}$~=~8.0; $T_{\rm phot}$=14.0} \\
\hline
$\log\tau$ & -0.4 & 0.1 & -1.2 & --& --& --& -- \\
$v_{\rm min}$ & 8.0 & 8.0 & 8.0& --&--&--&--\\
$v_{\rm max}$ & 30.0 & 30.0 & 30.0 &--&--&--&--\\
aux & 1.0 & 1.0 & 1.0 &--&--&--&-- \\
$T_{\rm exc}$ & 14.0 & 14.0 & 14.0 &--&--&--&--\\
\hline
 \multicolumn{8}{c}{SN~2022xgc (2022.12.17) $v_{\rm phot}$~=~8.0; $T_{\rm phot}$=13.0} \\
\hline
$\log\tau$ &--& 0.2&--&3.2 &0.1 &0.2&-0.1\\
$v_{\rm min}$ &--&8.0&--&20.0&8.0 &8.0&8.0\\
$v_{\rm max}$ &--&30.0&--& 30.0 &30.0&30.0&30.0\\
aux &--&1.0&--&1.0 &1.0&1.0&1.0\\
$T_{\rm exc}$ &--&13.0&--&13.0&13.0 &13.0&13.0\\
\hline
\hline
\end{tabular*}
\end{table*}

\end{appendix}

\end{document}